\documentclass[sn-mathphys,Numbered,iicol]{sn-jnl}

\usepackage{graphicx}\usepackage{multirow}\usepackage{amsmath,amssymb,amsfonts}\usepackage{amsthm}\usepackage{mathrsfs}\usepackage[title]{appendix}\usepackage{xcolor}\usepackage{textcomp}\usepackage{manyfoot}\usepackage{booktabs}\usepackage{algorithm}\usepackage{algorithmicx}\usepackage{algpseudocode}\usepackage{listings}\usepackage{xcolor}
\usepackage{subcaption}
\usepackage{tikz}
\usetikzlibrary{backgrounds}
\usetikzlibrary{decorations.pathreplacing,calligraphy}
\usepackage{placeins}               \usepackage{textcomp}
\usepackage{upgreek}                \usepackage[switch, mathlines]{lineno}

\usepackage{tikz,pgfplots,pgfplotstable}
\usepackage{svg}
\usetikzlibrary{intersections}
\pgfplotsset{compat=newest}
\usetikzlibrary{positioning}
\usetikzlibrary{arrows}
\usetikzlibrary{arrows.meta}
\usepgflibrary{arrows.meta}
\usetikzlibrary{patterns}
\usepackage{contour}
\usepackage{anyfontsize}
\definecolor{AliceBlue}{rgb}{0.94,0.97,1.00}
\definecolor{AntiqueWhite1}{rgb}{1.00,0.94,0.86}
\definecolor{AntiqueWhite2}{rgb}{0.93,0.87,0.80}
\definecolor{AntiqueWhite3}{rgb}{0.80,0.75,0.69}
\definecolor{AntiqueWhite4}{rgb}{0.55,0.51,0.47}
\definecolor{AntiqueWhite}{rgb}{0.98,0.92,0.84}
\definecolor{BlanchedAlmond}{rgb}{1.00,0.92,0.80}
\definecolor{BlueViolet}{rgb}{0.54,0.17,0.89}
\definecolor{CadetBlue1}{rgb}{0.60,0.96,1.00}
\definecolor{CadetBlue2}{rgb}{0.56,0.90,0.93}
\definecolor{CadetBlue3}{rgb}{0.48,0.77,0.80}
\definecolor{CadetBlue4}{rgb}{0.33,0.53,0.55}
\definecolor{CadetBlue}{rgb}{0.37,0.62,0.63}
\definecolor{CornflowerBlue}{rgb}{0.39,0.58,0.93}
\definecolor{DarkBlue}{rgb}{0.00,0.00,0.55}
\definecolor{DarkCyan}{rgb}{0.00,0.55,0.55}
\definecolor{DarkGoldenrod1}{rgb}{1.00,0.73,0.06}
\definecolor{DarkGoldenrod2}{rgb}{0.93,0.68,0.05}
\definecolor{DarkGoldenrod3}{rgb}{0.80,0.58,0.05}
\definecolor{DarkGoldenrod4}{rgb}{0.55,0.40,0.03}
\definecolor{DarkGoldenrod}{rgb}{0.72,0.53,0.04}
\definecolor{DarkGray}{rgb}{0.66,0.66,0.66}
\definecolor{DarkGreen}{rgb}{0.00,0.39,0.00}
\definecolor{DarkGrey}{rgb}{0.66,0.66,0.66}
\definecolor{DarkKhaki}{rgb}{0.74,0.72,0.42}
\definecolor{DarkMagenta}{rgb}{0.55,0.00,0.55}
\definecolor{DarkOliveGreen1}{rgb}{0.79,1.00,0.44}
\definecolor{DarkOliveGreen2}{rgb}{0.74,0.93,0.41}
\definecolor{DarkOliveGreen3}{rgb}{0.64,0.80,0.35}
\definecolor{DarkOliveGreen4}{rgb}{0.43,0.55,0.24}
\definecolor{DarkOliveGreen}{rgb}{0.33,0.42,0.18}
\definecolor{DarkOrange1}{rgb}{1.00,0.50,0.00}
\definecolor{DarkOrange2}{rgb}{0.93,0.46,0.00}
\definecolor{DarkOrange3}{rgb}{0.80,0.40,0.00}
\definecolor{DarkOrange4}{rgb}{0.55,0.27,0.00}
\definecolor{DarkOrange}{rgb}{1.00,0.55,0.00}
\definecolor{DarkOrchid1}{rgb}{0.75,0.24,1.00}
\definecolor{DarkOrchid2}{rgb}{0.70,0.23,0.93}
\definecolor{DarkOrchid3}{rgb}{0.60,0.20,0.80}
\definecolor{DarkOrchid4}{rgb}{0.41,0.13,0.55}
\definecolor{DarkOrchid}{rgb}{0.60,0.20,0.80}
\definecolor{DarkRed}{rgb}{0.55,0.00,0.00}
\definecolor{DarkSalmon}{rgb}{0.91,0.59,0.48}
\definecolor{DarkSeaGreen1}{rgb}{0.76,1.00,0.76}
\definecolor{DarkSeaGreen2}{rgb}{0.71,0.93,0.71}
\definecolor{DarkSeaGreen3}{rgb}{0.61,0.80,0.61}
\definecolor{DarkSeaGreen4}{rgb}{0.41,0.55,0.41}
\definecolor{DarkSeaGreen}{rgb}{0.56,0.74,0.56}
\definecolor{DarkSlateBlue}{rgb}{0.28,0.24,0.55}
\definecolor{DarkSlateGray1}{rgb}{0.59,1.00,1.00}
\definecolor{DarkSlateGray2}{rgb}{0.55,0.93,0.93}
\definecolor{DarkSlateGray3}{rgb}{0.47,0.80,0.80}
\definecolor{DarkSlateGray4}{rgb}{0.32,0.55,0.55}
\definecolor{DarkSlateGray}{rgb}{0.18,0.31,0.31}
\definecolor{DarkSlateGrey}{rgb}{0.18,0.31,0.31}
\definecolor{DarkTurquoise}{rgb}{0.00,0.81,0.82}
\definecolor{DarkViolet}{rgb}{0.58,0.00,0.83}
\definecolor{DeepPink1}{rgb}{1.00,0.08,0.58}
\definecolor{DeepPink2}{rgb}{0.93,0.07,0.54}
\definecolor{DeepPink3}{rgb}{0.80,0.06,0.46}
\definecolor{DeepPink4}{rgb}{0.55,0.04,0.31}
\definecolor{DeepPink}{rgb}{1.00,0.08,0.58}
\definecolor{DeepSkyBlue1}{rgb}{0.00,0.75,1.00}
\definecolor{DeepSkyBlue2}{rgb}{0.00,0.70,0.93}
\definecolor{DeepSkyBlue3}{rgb}{0.00,0.60,0.80}
\definecolor{DeepSkyBlue4}{rgb}{0.00,0.41,0.55}
\definecolor{DeepSkyBlue}{rgb}{0.00,0.75,1.00}
\definecolor{DimGray}{rgb}{0.41,0.41,0.41}
\definecolor{DimGrey}{rgb}{0.41,0.41,0.41}
\definecolor{DodgerBlue1}{rgb}{0.12,0.56,1.00}
\definecolor{DodgerBlue2}{rgb}{0.11,0.53,0.93}
\definecolor{DodgerBlue3}{rgb}{0.09,0.45,0.80}
\definecolor{DodgerBlue4}{rgb}{0.06,0.31,0.55}
\definecolor{DodgerBlue}{rgb}{0.12,0.56,1.00}
\definecolor{FloralWhite}{rgb}{1.00,0.98,0.94}
\definecolor{ForestGreen}{rgb}{0.13,0.55,0.13}
\definecolor{GhostWhite}{rgb}{0.97,0.97,1.00}
\definecolor{GreenYellow}{rgb}{0.68,1.00,0.18}
\definecolor{HotPink1}{rgb}{1.00,0.43,0.71}
\definecolor{HotPink2}{rgb}{0.93,0.42,0.65}
\definecolor{HotPink3}{rgb}{0.80,0.38,0.56}
\definecolor{HotPink4}{rgb}{0.55,0.23,0.38}
\definecolor{HotPink}{rgb}{1.00,0.41,0.71}
\definecolor{IndianRed1}{rgb}{1.00,0.42,0.42}
\definecolor{IndianRed2}{rgb}{0.93,0.39,0.39}
\definecolor{IndianRed3}{rgb}{0.80,0.33,0.33}
\definecolor{IndianRed4}{rgb}{0.55,0.23,0.23}
\definecolor{IndianRed}{rgb}{0.80,0.36,0.36}
\definecolor{LavenderBlush1}{rgb}{1.00,0.94,0.96}
\definecolor{LavenderBlush2}{rgb}{0.93,0.88,0.90}
\definecolor{LavenderBlush3}{rgb}{0.80,0.76,0.77}
\definecolor{LavenderBlush4}{rgb}{0.55,0.51,0.53}
\definecolor{LavenderBlush}{rgb}{1.00,0.94,0.96}
\definecolor{LawnGreen}{rgb}{0.49,0.99,0.00}
\definecolor{LemonChiffon1}{rgb}{1.00,0.98,0.80}
\definecolor{LemonChiffon2}{rgb}{0.93,0.91,0.75}
\definecolor{LemonChiffon3}{rgb}{0.80,0.79,0.65}
\definecolor{LemonChiffon4}{rgb}{0.55,0.54,0.44}
\definecolor{LemonChiffon}{rgb}{1.00,0.98,0.80}
\definecolor{LightBlue1}{rgb}{0.75,0.94,1.00}
\definecolor{LightBlue2}{rgb}{0.70,0.87,0.93}
\definecolor{LightBlue3}{rgb}{0.60,0.75,0.80}
\definecolor{LightBlue4}{rgb}{0.41,0.51,0.55}
\definecolor{LightBlue}{rgb}{0.68,0.85,0.90}
\definecolor{LightCoral}{rgb}{0.94,0.50,0.50}
\definecolor{LightCyan1}{rgb}{0.88,1.00,1.00}
\definecolor{LightCyan2}{rgb}{0.82,0.93,0.93}
\definecolor{LightCyan3}{rgb}{0.71,0.80,0.80}
\definecolor{LightCyan4}{rgb}{0.48,0.55,0.55}
\definecolor{LightCyan}{rgb}{0.88,1.00,1.00}
\definecolor{LightGoldenrod1}{rgb}{1.00,0.93,0.55}
\definecolor{LightGoldenrod2}{rgb}{0.93,0.86,0.51}
\definecolor{LightGoldenrod3}{rgb}{0.80,0.75,0.44}
\definecolor{LightGoldenrod4}{rgb}{0.55,0.51,0.30}
\definecolor{LightGoldenrodYellow}{rgb}{0.98,0.98,0.82}
\definecolor{LightGoldenrod}{rgb}{0.93,0.87,0.51}
\definecolor{LightGray}{rgb}{0.83,0.83,0.83}
\definecolor{LightGreen}{rgb}{0.56,0.93,0.56}
\definecolor{LightGrey}{rgb}{0.83,0.83,0.83}
\definecolor{LightPink1}{rgb}{1.00,0.68,0.73}
\definecolor{LightPink2}{rgb}{0.93,0.64,0.68}
\definecolor{LightPink3}{rgb}{0.80,0.55,0.58}
\definecolor{LightPink4}{rgb}{0.55,0.37,0.40}
\definecolor{LightPink}{rgb}{1.00,0.71,0.76}
\definecolor{LightSalmon1}{rgb}{1.00,0.63,0.48}
\definecolor{LightSalmon2}{rgb}{0.93,0.58,0.45}
\definecolor{LightSalmon3}{rgb}{0.80,0.51,0.38}
\definecolor{LightSalmon4}{rgb}{0.55,0.34,0.26}
\definecolor{LightSalmon}{rgb}{1.00,0.63,0.48}
\definecolor{LightSeaGreen}{rgb}{0.13,0.70,0.67}
\definecolor{LightSkyBlue1}{rgb}{0.69,0.89,1.00}
\definecolor{LightSkyBlue2}{rgb}{0.64,0.83,0.93}
\definecolor{LightSkyBlue3}{rgb}{0.55,0.71,0.80}
\definecolor{LightSkyBlue4}{rgb}{0.38,0.48,0.55}
\definecolor{LightSkyBlue}{rgb}{0.53,0.81,0.98}
\definecolor{LightSlateBlue}{rgb}{0.52,0.44,1.00}
\definecolor{LightSlateGray}{rgb}{0.47,0.53,0.60}
\definecolor{LightSlateGrey}{rgb}{0.47,0.53,0.60}
\definecolor{LightSteelBlue1}{rgb}{0.79,0.88,1.00}
\definecolor{LightSteelBlue2}{rgb}{0.74,0.82,0.93}
\definecolor{LightSteelBlue3}{rgb}{0.64,0.71,0.80}
\definecolor{LightSteelBlue4}{rgb}{0.43,0.48,0.55}
\definecolor{LightSteelBlue}{rgb}{0.69,0.77,0.87}
\definecolor{LightYellow1}{rgb}{1.00,1.00,0.88}
\definecolor{LightYellow2}{rgb}{0.93,0.93,0.82}
\definecolor{LightYellow3}{rgb}{0.80,0.80,0.71}
\definecolor{LightYellow4}{rgb}{0.55,0.55,0.48}
\definecolor{LightYellow}{rgb}{1.00,1.00,0.88}
\definecolor{LimeGreen}{rgb}{0.20,0.80,0.20}
\definecolor{MediumAquamarine}{rgb}{0.40,0.80,0.67}
\definecolor{MediumBlue}{rgb}{0.00,0.00,0.80}
\definecolor{MediumOrchid1}{rgb}{0.88,0.40,1.00}
\definecolor{MediumOrchid2}{rgb}{0.82,0.37,0.93}
\definecolor{MediumOrchid3}{rgb}{0.71,0.32,0.80}
\definecolor{MediumOrchid4}{rgb}{0.48,0.22,0.55}
\definecolor{MediumOrchid}{rgb}{0.73,0.33,0.83}
\definecolor{MediumPurple1}{rgb}{0.67,0.51,1.00}
\definecolor{MediumPurple2}{rgb}{0.62,0.47,0.93}
\definecolor{MediumPurple3}{rgb}{0.54,0.41,0.80}
\definecolor{MediumPurple4}{rgb}{0.36,0.28,0.55}
\definecolor{MediumPurple}{rgb}{0.58,0.44,0.86}
\definecolor{MediumSeaGreen}{rgb}{0.24,0.70,0.44}
\definecolor{MediumSlateBlue}{rgb}{0.48,0.41,0.93}
\definecolor{MediumSpringGreen}{rgb}{0.00,0.98,0.60}
\definecolor{MediumTurquoise}{rgb}{0.28,0.82,0.80}
\definecolor{MediumVioletRed}{rgb}{0.78,0.08,0.52}
\definecolor{MidnightBlue}{rgb}{0.10,0.10,0.44}
\definecolor{MintCream}{rgb}{0.96,1.00,0.98}
\definecolor{MistyRose1}{rgb}{1.00,0.89,0.88}
\definecolor{MistyRose2}{rgb}{0.93,0.84,0.82}
\definecolor{MistyRose3}{rgb}{0.80,0.72,0.71}
\definecolor{MistyRose4}{rgb}{0.55,0.49,0.48}
\definecolor{MistyRose}{rgb}{1.00,0.89,0.88}
\definecolor{NavajoWhite1}{rgb}{1.00,0.87,0.68}
\definecolor{NavajoWhite2}{rgb}{0.93,0.81,0.63}
\definecolor{NavajoWhite3}{rgb}{0.80,0.70,0.55}
\definecolor{NavajoWhite4}{rgb}{0.55,0.47,0.37}
\definecolor{NavajoWhite}{rgb}{1.00,0.87,0.68}
\definecolor{NavyBlue}{rgb}{0.00,0.00,0.50}
\definecolor{OldLace}{rgb}{0.99,0.96,0.90}
\definecolor{OliveDrab1}{rgb}{0.75,1.00,0.24}
\definecolor{OliveDrab2}{rgb}{0.70,0.93,0.23}
\definecolor{OliveDrab3}{rgb}{0.60,0.80,0.20}
\definecolor{OliveDrab4}{rgb}{0.41,0.55,0.13}
\definecolor{OliveDrab}{rgb}{0.42,0.56,0.14}
\definecolor{OrangeRed1}{rgb}{1.00,0.27,0.00}
\definecolor{OrangeRed2}{rgb}{0.93,0.25,0.00}
\definecolor{OrangeRed3}{rgb}{0.80,0.22,0.00}
\definecolor{OrangeRed4}{rgb}{0.55,0.15,0.00}
\definecolor{OrangeRed}{rgb}{1.00,0.27,0.00}
\definecolor{PaleGoldenrod}{rgb}{0.93,0.91,0.67}
\definecolor{PaleGreen1}{rgb}{0.60,1.00,0.60}
\definecolor{PaleGreen2}{rgb}{0.56,0.93,0.56}
\definecolor{PaleGreen3}{rgb}{0.49,0.80,0.49}
\definecolor{PaleGreen4}{rgb}{0.33,0.55,0.33}
\definecolor{PaleGreen}{rgb}{0.60,0.98,0.60}
\definecolor{PaleTurquoise1}{rgb}{0.73,1.00,1.00}
\definecolor{PaleTurquoise2}{rgb}{0.68,0.93,0.93}
\definecolor{PaleTurquoise3}{rgb}{0.59,0.80,0.80}
\definecolor{PaleTurquoise4}{rgb}{0.40,0.55,0.55}
\definecolor{PaleTurquoise}{rgb}{0.69,0.93,0.93}
\definecolor{PaleVioletRed1}{rgb}{1.00,0.51,0.67}
\definecolor{PaleVioletRed2}{rgb}{0.93,0.47,0.62}
\definecolor{PaleVioletRed3}{rgb}{0.80,0.41,0.54}
\definecolor{PaleVioletRed4}{rgb}{0.55,0.28,0.36}
\definecolor{PaleVioletRed}{rgb}{0.86,0.44,0.58}
\definecolor{PapayaWhip}{rgb}{1.00,0.94,0.84}
\definecolor{PeachPuff1}{rgb}{1.00,0.85,0.73}
\definecolor{PeachPuff2}{rgb}{0.93,0.80,0.68}
\definecolor{PeachPuff3}{rgb}{0.80,0.69,0.58}
\definecolor{PeachPuff4}{rgb}{0.55,0.47,0.40}
\definecolor{PeachPuff}{rgb}{1.00,0.85,0.73}
\definecolor{PowderBlue}{rgb}{0.69,0.88,0.90}
\definecolor{RosyBrown1}{rgb}{1.00,0.76,0.76}
\definecolor{RosyBrown2}{rgb}{0.93,0.71,0.71}
\definecolor{RosyBrown3}{rgb}{0.80,0.61,0.61}
\definecolor{RosyBrown4}{rgb}{0.55,0.41,0.41}
\definecolor{RosyBrown}{rgb}{0.74,0.56,0.56}
\definecolor{RoyalBlue1}{rgb}{0.28,0.46,1.00}
\definecolor{RoyalBlue2}{rgb}{0.26,0.43,0.93}
\definecolor{RoyalBlue3}{rgb}{0.23,0.37,0.80}
\definecolor{RoyalBlue4}{rgb}{0.15,0.25,0.55}
\definecolor{RoyalBlue}{rgb}{0.25,0.41,0.88}
\definecolor{SaddleBrown}{rgb}{0.55,0.27,0.07}
\definecolor{SandyBrown}{rgb}{0.96,0.64,0.38}
\definecolor{SeaGreen1}{rgb}{0.33,1.00,0.62}
\definecolor{SeaGreen2}{rgb}{0.31,0.93,0.58}
\definecolor{SeaGreen3}{rgb}{0.26,0.80,0.50}
\definecolor{SeaGreen4}{rgb}{0.18,0.55,0.34}
\definecolor{SeaGreen}{rgb}{0.18,0.55,0.34}
\definecolor{SkyBlue1}{rgb}{0.53,0.81,1.00}
\definecolor{SkyBlue2}{rgb}{0.49,0.75,0.93}
\definecolor{SkyBlue3}{rgb}{0.42,0.65,0.80}
\definecolor{SkyBlue4}{rgb}{0.29,0.44,0.55}
\definecolor{SkyBlue}{rgb}{0.53,0.81,0.92}
\definecolor{SlateBlue1}{rgb}{0.51,0.44,1.00}
\definecolor{SlateBlue2}{rgb}{0.48,0.40,0.93}
\definecolor{SlateBlue3}{rgb}{0.41,0.35,0.80}
\definecolor{SlateBlue4}{rgb}{0.28,0.24,0.55}
\definecolor{SlateBlue}{rgb}{0.42,0.35,0.80}
\definecolor{SlateGray1}{rgb}{0.78,0.89,1.00}
\definecolor{SlateGray2}{rgb}{0.73,0.83,0.93}
\definecolor{SlateGray3}{rgb}{0.62,0.71,0.80}
\definecolor{SlateGray4}{rgb}{0.42,0.48,0.55}
\definecolor{SlateGray}{rgb}{0.44,0.50,0.56}
\definecolor{SlateGrey}{rgb}{0.44,0.50,0.56}
\definecolor{SpringGreen1}{rgb}{0.00,1.00,0.50}
\definecolor{SpringGreen2}{rgb}{0.00,0.93,0.46}
\definecolor{SpringGreen3}{rgb}{0.00,0.80,0.40}
\definecolor{SpringGreen4}{rgb}{0.00,0.55,0.27}
\definecolor{SpringGreen}{rgb}{0.00,1.00,0.50}
\definecolor{SteelBlue1}{rgb}{0.39,0.72,1.00}
\definecolor{SteelBlue2}{rgb}{0.36,0.67,0.93}
\definecolor{SteelBlue3}{rgb}{0.31,0.58,0.80}
\definecolor{SteelBlue4}{rgb}{0.21,0.39,0.55}
\definecolor{SteelBlue}{rgb}{0.27,0.51,0.71}
\definecolor{VioletRed1}{rgb}{1.00,0.24,0.59}
\definecolor{VioletRed2}{rgb}{0.93,0.23,0.55}
\definecolor{VioletRed3}{rgb}{0.80,0.20,0.47}
\definecolor{VioletRed4}{rgb}{0.55,0.13,0.32}
\definecolor{VioletRed}{rgb}{0.82,0.13,0.56}
\definecolor{WhiteSmoke}{rgb}{0.96,0.96,0.96}
\definecolor{YellowGreen}{rgb}{0.60,0.80,0.20}
\definecolor{aliceblue}{rgb}{0.94,0.97,1.00}
\definecolor{antiquewhite}{rgb}{0.98,0.92,0.84}
\definecolor{aquamarine1}{rgb}{0.50,1.00,0.83}
\definecolor{aquamarine2}{rgb}{0.46,0.93,0.78}
\definecolor{aquamarine3}{rgb}{0.40,0.80,0.67}
\definecolor{aquamarine4}{rgb}{0.27,0.55,0.45}
\definecolor{aquamarine}{rgb}{0.50,1.00,0.83}
\definecolor{azure1}{rgb}{0.94,1.00,1.00}
\definecolor{azure2}{rgb}{0.88,0.93,0.93}
\definecolor{azure3}{rgb}{0.76,0.80,0.80}
\definecolor{azure4}{rgb}{0.51,0.55,0.55}
\definecolor{azure}{rgb}{0.94,1.00,1.00}
\definecolor{beige}{rgb}{0.96,0.96,0.86}
\definecolor{bisque1}{rgb}{1.00,0.89,0.77}
\definecolor{bisque2}{rgb}{0.93,0.84,0.72}
\definecolor{bisque3}{rgb}{0.80,0.72,0.62}
\definecolor{bisque4}{rgb}{0.55,0.49,0.42}
\definecolor{bisque}{rgb}{1.00,0.89,0.77}
\definecolor{black}{rgb}{0.00,0.00,0.00}
\definecolor{blanchedalmond}{rgb}{1.00,0.92,0.80}
\definecolor{blue1}{rgb}{0.00,0.00,1.00}
\definecolor{blue2}{rgb}{0.00,0.00,0.93}
\definecolor{blue3}{rgb}{0.00,0.00,0.80}
\definecolor{blue4}{rgb}{0.00,0.00,0.55}
\definecolor{blueviolet}{rgb}{0.54,0.17,0.89}
\definecolor{blue}{rgb}{0.00,0.00,1.00}
\definecolor{brown1}{rgb}{1.00,0.25,0.25}
\definecolor{brown2}{rgb}{0.93,0.23,0.23}
\definecolor{brown3}{rgb}{0.80,0.20,0.20}
\definecolor{brown4}{rgb}{0.55,0.14,0.14}
\definecolor{brown}{rgb}{0.65,0.16,0.16}
\definecolor{burlywood1}{rgb}{1.00,0.83,0.61}
\definecolor{burlywood2}{rgb}{0.93,0.77,0.57}
\definecolor{burlywood3}{rgb}{0.80,0.67,0.49}
\definecolor{burlywood4}{rgb}{0.55,0.45,0.33}
\definecolor{burlywood}{rgb}{0.87,0.72,0.53}
\definecolor{cadetblue}{rgb}{0.37,0.62,0.63}
\definecolor{chartreuse1}{rgb}{0.50,1.00,0.00}
\definecolor{chartreuse2}{rgb}{0.46,0.93,0.00}
\definecolor{chartreuse3}{rgb}{0.40,0.80,0.00}
\definecolor{chartreuse4}{rgb}{0.27,0.55,0.00}
\definecolor{chartreuse}{rgb}{0.50,1.00,0.00}
\definecolor{chocolate1}{rgb}{1.00,0.50,0.14}
\definecolor{chocolate2}{rgb}{0.93,0.46,0.13}
\definecolor{chocolate3}{rgb}{0.80,0.40,0.11}
\definecolor{chocolate4}{rgb}{0.55,0.27,0.07}
\definecolor{chocolate}{rgb}{0.82,0.41,0.12}
\definecolor{coral1}{rgb}{1.00,0.45,0.34}
\definecolor{coral2}{rgb}{0.93,0.42,0.31}
\definecolor{coral3}{rgb}{0.80,0.36,0.27}
\definecolor{coral4}{rgb}{0.55,0.24,0.18}
\definecolor{coral}{rgb}{1.00,0.50,0.31}
\definecolor{cornflowerblue}{rgb}{0.39,0.58,0.93}
\definecolor{cornsilk1}{rgb}{1.00,0.97,0.86}
\definecolor{cornsilk2}{rgb}{0.93,0.91,0.80}
\definecolor{cornsilk3}{rgb}{0.80,0.78,0.69}
\definecolor{cornsilk4}{rgb}{0.55,0.53,0.47}
\definecolor{cornsilk}{rgb}{1.00,0.97,0.86}
\definecolor{cyan1}{rgb}{0.00,1.00,1.00}
\definecolor{cyan2}{rgb}{0.00,0.93,0.93}
\definecolor{cyan3}{rgb}{0.00,0.80,0.80}
\definecolor{cyan4}{rgb}{0.00,0.55,0.55}
\definecolor{cyan}{rgb}{0.00,1.00,1.00}
\definecolor{darkblue}{rgb}{0.00,0.00,0.55}
\definecolor{darkcyan}{rgb}{0.00,0.55,0.55}
\definecolor{darkgoldenrod}{rgb}{0.72,0.53,0.04}
\definecolor{darkgray}{rgb}{0.66,0.66,0.66}
\definecolor{darkgreen}{rgb}{0.00,0.39,0.00}
\definecolor{darkgrey}{rgb}{0.66,0.66,0.66}
\definecolor{darkkhaki}{rgb}{0.74,0.72,0.42}
\definecolor{darkmagenta}{rgb}{0.55,0.00,0.55}
\definecolor{darkolive}{rgb}{0.33,0.42,0.18}
\definecolor{darkorange}{rgb}{1.00,0.55,0.00}
\definecolor{darkorchid}{rgb}{0.60,0.20,0.80}
\definecolor{darkred}{rgb}{0.55,0.00,0.00}
\definecolor{darksalmon}{rgb}{0.91,0.59,0.48}
\definecolor{darksea}{rgb}{0.56,0.74,0.56}
\definecolor{darkslate}{rgb}{0.18,0.31,0.31}
\definecolor{darkslate}{rgb}{0.18,0.31,0.31}
\definecolor{darkslate}{rgb}{0.28,0.24,0.55}
\definecolor{darkturquoise}{rgb}{0.00,0.81,0.82}
\definecolor{darkviolet}{rgb}{0.58,0.00,0.83}
\definecolor{deeppink}{rgb}{1.00,0.08,0.58}
\definecolor{deepsky}{rgb}{0.00,0.75,1.00}
\definecolor{dimgray}{rgb}{0.41,0.41,0.41}
\definecolor{dimgrey}{rgb}{0.41,0.41,0.41}
\definecolor{dodgerblue}{rgb}{0.12,0.56,1.00}
\definecolor{firebrick1}{rgb}{1.00,0.19,0.19}
\definecolor{firebrick2}{rgb}{0.93,0.17,0.17}
\definecolor{firebrick3}{rgb}{0.80,0.15,0.15}
\definecolor{firebrick4}{rgb}{0.55,0.10,0.10}
\definecolor{firebrick}{rgb}{0.70,0.13,0.13}
\definecolor{floralwhite}{rgb}{1.00,0.98,0.94}
\definecolor{forestgreen}{rgb}{0.13,0.55,0.13}
\definecolor{gainsboro}{rgb}{0.86,0.86,0.86}
\definecolor{ghostwhite}{rgb}{0.97,0.97,1.00}
\definecolor{gold1}{rgb}{1.00,0.84,0.00}
\definecolor{gold2}{rgb}{0.93,0.79,0.00}
\definecolor{gold3}{rgb}{0.80,0.68,0.00}
\definecolor{gold4}{rgb}{0.55,0.46,0.00}
\definecolor{goldenrod1}{rgb}{1.00,0.76,0.15}
\definecolor{goldenrod2}{rgb}{0.93,0.71,0.13}
\definecolor{goldenrod3}{rgb}{0.80,0.61,0.11}
\definecolor{goldenrod4}{rgb}{0.55,0.41,0.08}
\definecolor{goldenrod}{rgb}{0.85,0.65,0.13}
\definecolor{gold}{rgb}{1.00,0.84,0.00}
\definecolor{gray0}{rgb}{0.00,0.00,0.00}
\definecolor{gray100}{rgb}{1.00,1.00,1.00}
\definecolor{gray10}{rgb}{0.10,0.10,0.10}
\definecolor{gray11}{rgb}{0.11,0.11,0.11}
\definecolor{gray12}{rgb}{0.12,0.12,0.12}
\definecolor{gray13}{rgb}{0.13,0.13,0.13}
\definecolor{gray14}{rgb}{0.14,0.14,0.14}
\definecolor{gray15}{rgb}{0.15,0.15,0.15}
\definecolor{gray16}{rgb}{0.16,0.16,0.16}
\definecolor{gray17}{rgb}{0.17,0.17,0.17}
\definecolor{gray18}{rgb}{0.18,0.18,0.18}
\definecolor{gray19}{rgb}{0.19,0.19,0.19}
\definecolor{gray1}{rgb}{0.01,0.01,0.01}
\definecolor{gray20}{rgb}{0.20,0.20,0.20}
\definecolor{gray21}{rgb}{0.21,0.21,0.21}
\definecolor{gray22}{rgb}{0.22,0.22,0.22}
\definecolor{gray23}{rgb}{0.23,0.23,0.23}
\definecolor{gray24}{rgb}{0.24,0.24,0.24}
\definecolor{gray25}{rgb}{0.25,0.25,0.25}
\definecolor{gray26}{rgb}{0.26,0.26,0.26}
\definecolor{gray27}{rgb}{0.27,0.27,0.27}
\definecolor{gray28}{rgb}{0.28,0.28,0.28}
\definecolor{gray29}{rgb}{0.29,0.29,0.29}
\definecolor{gray2}{rgb}{0.02,0.02,0.02}
\definecolor{gray30}{rgb}{0.30,0.30,0.30}
\definecolor{gray31}{rgb}{0.31,0.31,0.31}
\definecolor{gray32}{rgb}{0.32,0.32,0.32}
\definecolor{gray33}{rgb}{0.33,0.33,0.33}
\definecolor{gray34}{rgb}{0.34,0.34,0.34}
\definecolor{gray35}{rgb}{0.35,0.35,0.35}
\definecolor{gray36}{rgb}{0.36,0.36,0.36}
\definecolor{gray37}{rgb}{0.37,0.37,0.37}
\definecolor{gray38}{rgb}{0.38,0.38,0.38}
\definecolor{gray39}{rgb}{0.39,0.39,0.39}
\definecolor{gray3}{rgb}{0.03,0.03,0.03}
\definecolor{gray40}{rgb}{0.40,0.40,0.40}
\definecolor{gray41}{rgb}{0.41,0.41,0.41}
\definecolor{gray42}{rgb}{0.42,0.42,0.42}
\definecolor{gray43}{rgb}{0.43,0.43,0.43}
\definecolor{gray44}{rgb}{0.44,0.44,0.44}
\definecolor{gray45}{rgb}{0.45,0.45,0.45}
\definecolor{gray46}{rgb}{0.46,0.46,0.46}
\definecolor{gray47}{rgb}{0.47,0.47,0.47}
\definecolor{gray48}{rgb}{0.48,0.48,0.48}
\definecolor{gray49}{rgb}{0.49,0.49,0.49}
\definecolor{gray4}{rgb}{0.04,0.04,0.04}
\definecolor{gray50}{rgb}{0.50,0.50,0.50}
\definecolor{gray51}{rgb}{0.51,0.51,0.51}
\definecolor{gray52}{rgb}{0.52,0.52,0.52}
\definecolor{gray53}{rgb}{0.53,0.53,0.53}
\definecolor{gray54}{rgb}{0.54,0.54,0.54}
\definecolor{gray55}{rgb}{0.55,0.55,0.55}
\definecolor{gray56}{rgb}{0.56,0.56,0.56}
\definecolor{gray57}{rgb}{0.57,0.57,0.57}
\definecolor{gray58}{rgb}{0.58,0.58,0.58}
\definecolor{gray59}{rgb}{0.59,0.59,0.59}
\definecolor{gray5}{rgb}{0.05,0.05,0.05}
\definecolor{gray60}{rgb}{0.60,0.60,0.60}
\definecolor{gray61}{rgb}{0.61,0.61,0.61}
\definecolor{gray62}{rgb}{0.62,0.62,0.62}
\definecolor{gray63}{rgb}{0.63,0.63,0.63}
\definecolor{gray64}{rgb}{0.64,0.64,0.64}
\definecolor{gray65}{rgb}{0.65,0.65,0.65}
\definecolor{gray66}{rgb}{0.66,0.66,0.66}
\definecolor{gray67}{rgb}{0.67,0.67,0.67}
\definecolor{gray68}{rgb}{0.68,0.68,0.68}
\definecolor{gray69}{rgb}{0.69,0.69,0.69}
\definecolor{gray6}{rgb}{0.06,0.06,0.06}
\definecolor{gray70}{rgb}{0.70,0.70,0.70}
\definecolor{gray71}{rgb}{0.71,0.71,0.71}
\definecolor{gray72}{rgb}{0.72,0.72,0.72}
\definecolor{gray73}{rgb}{0.73,0.73,0.73}
\definecolor{gray74}{rgb}{0.74,0.74,0.74}
\definecolor{gray75}{rgb}{0.75,0.75,0.75}
\definecolor{gray76}{rgb}{0.76,0.76,0.76}
\definecolor{gray77}{rgb}{0.77,0.77,0.77}
\definecolor{gray78}{rgb}{0.78,0.78,0.78}
\definecolor{gray79}{rgb}{0.79,0.79,0.79}
\definecolor{gray7}{rgb}{0.07,0.07,0.07}
\definecolor{gray80}{rgb}{0.80,0.80,0.80}
\definecolor{gray81}{rgb}{0.81,0.81,0.81}
\definecolor{gray82}{rgb}{0.82,0.82,0.82}
\definecolor{gray83}{rgb}{0.83,0.83,0.83}
\definecolor{gray84}{rgb}{0.84,0.84,0.84}
\definecolor{gray85}{rgb}{0.85,0.85,0.85}
\definecolor{gray86}{rgb}{0.86,0.86,0.86}
\definecolor{gray87}{rgb}{0.87,0.87,0.87}
\definecolor{gray88}{rgb}{0.88,0.88,0.88}
\definecolor{gray89}{rgb}{0.89,0.89,0.89}
\definecolor{gray8}{rgb}{0.08,0.08,0.08}
\definecolor{gray90}{rgb}{0.90,0.90,0.90}
\definecolor{gray91}{rgb}{0.91,0.91,0.91}
\definecolor{gray92}{rgb}{0.92,0.92,0.92}
\definecolor{gray93}{rgb}{0.93,0.93,0.93}
\definecolor{gray94}{rgb}{0.94,0.94,0.94}
\definecolor{gray95}{rgb}{0.95,0.95,0.95}
\definecolor{gray96}{rgb}{0.96,0.96,0.96}
\definecolor{gray97}{rgb}{0.97,0.97,0.97}
\definecolor{gray98}{rgb}{0.98,0.98,0.98}
\definecolor{gray99}{rgb}{0.99,0.99,0.99}
\definecolor{gray9}{rgb}{0.09,0.09,0.09}
\definecolor{gray}{rgb}{0.75,0.75,0.75}
\definecolor{green1}{rgb}{0.00,1.00,0.00}
\definecolor{green2}{rgb}{0.00,0.93,0.00}
\definecolor{green3}{rgb}{0.00,0.80,0.00}
\definecolor{green4}{rgb}{0.00,0.55,0.00}
\definecolor{greenyellow}{rgb}{0.68,1.00,0.18}
\definecolor{green}{rgb}{0.00,1.00,0.00}
\definecolor{grey0}{rgb}{0.00,0.00,0.00}
\definecolor{grey100}{rgb}{1.00,1.00,1.00}
\definecolor{grey10}{rgb}{0.10,0.10,0.10}
\definecolor{grey11}{rgb}{0.11,0.11,0.11}
\definecolor{grey12}{rgb}{0.12,0.12,0.12}
\definecolor{grey13}{rgb}{0.13,0.13,0.13}
\definecolor{grey14}{rgb}{0.14,0.14,0.14}
\definecolor{grey15}{rgb}{0.15,0.15,0.15}
\definecolor{grey16}{rgb}{0.16,0.16,0.16}
\definecolor{grey17}{rgb}{0.17,0.17,0.17}
\definecolor{grey18}{rgb}{0.18,0.18,0.18}
\definecolor{grey19}{rgb}{0.19,0.19,0.19}
\definecolor{grey1}{rgb}{0.01,0.01,0.01}
\definecolor{grey20}{rgb}{0.20,0.20,0.20}
\definecolor{grey21}{rgb}{0.21,0.21,0.21}
\definecolor{grey22}{rgb}{0.22,0.22,0.22}
\definecolor{grey23}{rgb}{0.23,0.23,0.23}
\definecolor{grey24}{rgb}{0.24,0.24,0.24}
\definecolor{grey25}{rgb}{0.25,0.25,0.25}
\definecolor{grey26}{rgb}{0.26,0.26,0.26}
\definecolor{grey27}{rgb}{0.27,0.27,0.27}
\definecolor{grey28}{rgb}{0.28,0.28,0.28}
\definecolor{grey29}{rgb}{0.29,0.29,0.29}
\definecolor{grey2}{rgb}{0.02,0.02,0.02}
\definecolor{grey30}{rgb}{0.30,0.30,0.30}
\definecolor{grey31}{rgb}{0.31,0.31,0.31}
\definecolor{grey32}{rgb}{0.32,0.32,0.32}
\definecolor{grey33}{rgb}{0.33,0.33,0.33}
\definecolor{grey34}{rgb}{0.34,0.34,0.34}
\definecolor{grey35}{rgb}{0.35,0.35,0.35}
\definecolor{grey36}{rgb}{0.36,0.36,0.36}
\definecolor{grey37}{rgb}{0.37,0.37,0.37}
\definecolor{grey38}{rgb}{0.38,0.38,0.38}
\definecolor{grey39}{rgb}{0.39,0.39,0.39}
\definecolor{grey3}{rgb}{0.03,0.03,0.03}
\definecolor{grey40}{rgb}{0.40,0.40,0.40}
\definecolor{grey41}{rgb}{0.41,0.41,0.41}
\definecolor{grey42}{rgb}{0.42,0.42,0.42}
\definecolor{grey43}{rgb}{0.43,0.43,0.43}
\definecolor{grey44}{rgb}{0.44,0.44,0.44}
\definecolor{grey45}{rgb}{0.45,0.45,0.45}
\definecolor{grey46}{rgb}{0.46,0.46,0.46}
\definecolor{grey47}{rgb}{0.47,0.47,0.47}
\definecolor{grey48}{rgb}{0.48,0.48,0.48}
\definecolor{grey49}{rgb}{0.49,0.49,0.49}
\definecolor{grey4}{rgb}{0.04,0.04,0.04}
\definecolor{grey50}{rgb}{0.50,0.50,0.50}
\definecolor{grey51}{rgb}{0.51,0.51,0.51}
\definecolor{grey52}{rgb}{0.52,0.52,0.52}
\definecolor{grey53}{rgb}{0.53,0.53,0.53}
\definecolor{grey54}{rgb}{0.54,0.54,0.54}
\definecolor{grey55}{rgb}{0.55,0.55,0.55}
\definecolor{grey56}{rgb}{0.56,0.56,0.56}
\definecolor{grey57}{rgb}{0.57,0.57,0.57}
\definecolor{grey58}{rgb}{0.58,0.58,0.58}
\definecolor{grey59}{rgb}{0.59,0.59,0.59}
\definecolor{grey5}{rgb}{0.05,0.05,0.05}
\definecolor{grey60}{rgb}{0.60,0.60,0.60}
\definecolor{grey61}{rgb}{0.61,0.61,0.61}
\definecolor{grey62}{rgb}{0.62,0.62,0.62}
\definecolor{grey63}{rgb}{0.63,0.63,0.63}
\definecolor{grey64}{rgb}{0.64,0.64,0.64}
\definecolor{grey65}{rgb}{0.65,0.65,0.65}
\definecolor{grey66}{rgb}{0.66,0.66,0.66}
\definecolor{grey67}{rgb}{0.67,0.67,0.67}
\definecolor{grey68}{rgb}{0.68,0.68,0.68}
\definecolor{grey69}{rgb}{0.69,0.69,0.69}
\definecolor{grey6}{rgb}{0.06,0.06,0.06}
\definecolor{grey70}{rgb}{0.70,0.70,0.70}
\definecolor{grey71}{rgb}{0.71,0.71,0.71}
\definecolor{grey72}{rgb}{0.72,0.72,0.72}
\definecolor{grey73}{rgb}{0.73,0.73,0.73}
\definecolor{grey74}{rgb}{0.74,0.74,0.74}
\definecolor{grey75}{rgb}{0.75,0.75,0.75}
\definecolor{grey76}{rgb}{0.76,0.76,0.76}
\definecolor{grey77}{rgb}{0.77,0.77,0.77}
\definecolor{grey78}{rgb}{0.78,0.78,0.78}
\definecolor{grey79}{rgb}{0.79,0.79,0.79}
\definecolor{grey7}{rgb}{0.07,0.07,0.07}
\definecolor{grey80}{rgb}{0.80,0.80,0.80}
\definecolor{grey81}{rgb}{0.81,0.81,0.81}
\definecolor{grey82}{rgb}{0.82,0.82,0.82}
\definecolor{grey83}{rgb}{0.83,0.83,0.83}
\definecolor{grey84}{rgb}{0.84,0.84,0.84}
\definecolor{grey85}{rgb}{0.85,0.85,0.85}
\definecolor{grey86}{rgb}{0.86,0.86,0.86}
\definecolor{grey87}{rgb}{0.87,0.87,0.87}
\definecolor{grey88}{rgb}{0.88,0.88,0.88}
\definecolor{grey89}{rgb}{0.89,0.89,0.89}
\definecolor{grey8}{rgb}{0.08,0.08,0.08}
\definecolor{grey90}{rgb}{0.90,0.90,0.90}
\definecolor{grey91}{rgb}{0.91,0.91,0.91}
\definecolor{grey92}{rgb}{0.92,0.92,0.92}
\definecolor{grey93}{rgb}{0.93,0.93,0.93}
\definecolor{grey94}{rgb}{0.94,0.94,0.94}
\definecolor{grey95}{rgb}{0.95,0.95,0.95}
\definecolor{grey96}{rgb}{0.96,0.96,0.96}
\definecolor{grey97}{rgb}{0.97,0.97,0.97}
\definecolor{grey98}{rgb}{0.98,0.98,0.98}
\definecolor{grey99}{rgb}{0.99,0.99,0.99}
\definecolor{grey9}{rgb}{0.09,0.09,0.09}
\definecolor{grey}{rgb}{0.75,0.75,0.75}
\definecolor{honeydew1}{rgb}{0.94,1.00,0.94}
\definecolor{honeydew2}{rgb}{0.88,0.93,0.88}
\definecolor{honeydew3}{rgb}{0.76,0.80,0.76}
\definecolor{honeydew4}{rgb}{0.51,0.55,0.51}
\definecolor{honeydew}{rgb}{0.94,1.00,0.94}
\definecolor{hotpink}{rgb}{1.00,0.41,0.71}
\definecolor{indianred}{rgb}{0.80,0.36,0.36}
\definecolor{ivory1}{rgb}{1.00,1.00,0.94}
\definecolor{ivory2}{rgb}{0.93,0.93,0.88}
\definecolor{ivory3}{rgb}{0.80,0.80,0.76}
\definecolor{ivory4}{rgb}{0.55,0.55,0.51}
\definecolor{ivory}{rgb}{1.00,1.00,0.94}
\definecolor{khaki1}{rgb}{1.00,0.96,0.56}
\definecolor{khaki2}{rgb}{0.93,0.90,0.52}
\definecolor{khaki3}{rgb}{0.80,0.78,0.45}
\definecolor{khaki4}{rgb}{0.55,0.53,0.31}
\definecolor{khaki}{rgb}{0.94,0.90,0.55}
\definecolor{lavenderblush}{rgb}{1.00,0.94,0.96}
\definecolor{lavender}{rgb}{0.90,0.90,0.98}
\definecolor{lawngreen}{rgb}{0.49,0.99,0.00}
\definecolor{lemonchiffon}{rgb}{1.00,0.98,0.80}
\definecolor{lightblue}{rgb}{0.68,0.85,0.90}
\definecolor{lightcoral}{rgb}{0.94,0.50,0.50}
\definecolor{lightcyan}{rgb}{0.88,1.00,1.00}
\definecolor{lightgoldenrod}{rgb}{0.93,0.87,0.51}
\definecolor{lightgoldenrod}{rgb}{0.98,0.98,0.82}
\definecolor{lightgray}{rgb}{0.83,0.83,0.83}
\definecolor{lightgreen}{rgb}{0.56,0.93,0.56}
\definecolor{lightgrey}{rgb}{0.83,0.83,0.83}
\definecolor{lightpink}{rgb}{1.00,0.71,0.76}
\definecolor{lightsalmon}{rgb}{1.00,0.63,0.48}
\definecolor{lightsea}{rgb}{0.13,0.70,0.67}
\definecolor{lightsky}{rgb}{0.53,0.81,0.98}
\definecolor{lightslate}{rgb}{0.47,0.53,0.60}
\definecolor{lightslate}{rgb}{0.47,0.53,0.60}
\definecolor{lightslate}{rgb}{0.52,0.44,1.00}
\definecolor{lightsteel}{rgb}{0.69,0.77,0.87}
\definecolor{lightyellow}{rgb}{1.00,1.00,0.88}
\definecolor{limegreen}{rgb}{0.20,0.80,0.20}
\definecolor{linen}{rgb}{0.98,0.94,0.90}
\definecolor{magenta1}{rgb}{1.00,0.00,1.00}
\definecolor{magenta2}{rgb}{0.93,0.00,0.93}
\definecolor{magenta3}{rgb}{0.80,0.00,0.80}
\definecolor{magenta4}{rgb}{0.55,0.00,0.55}
\definecolor{magenta}{rgb}{1.00,0.00,1.00}
\definecolor{maroon1}{rgb}{1.00,0.20,0.70}
\definecolor{maroon2}{rgb}{0.93,0.19,0.65}
\definecolor{maroon3}{rgb}{0.80,0.16,0.56}
\definecolor{maroon4}{rgb}{0.55,0.11,0.38}
\definecolor{maroon}{rgb}{0.69,0.19,0.38}
\definecolor{mediumaquamarine}{rgb}{0.40,0.80,0.67}
\definecolor{mediumblue}{rgb}{0.00,0.00,0.80}
\definecolor{mediumorchid}{rgb}{0.73,0.33,0.83}
\definecolor{mediumpurple}{rgb}{0.58,0.44,0.86}
\definecolor{mediumsea}{rgb}{0.24,0.70,0.44}
\definecolor{mediumslate}{rgb}{0.48,0.41,0.93}
\definecolor{mediumspring}{rgb}{0.00,0.98,0.60}
\definecolor{mediumturquoise}{rgb}{0.28,0.82,0.80}
\definecolor{mediumviolet}{rgb}{0.78,0.08,0.52}
\definecolor{midnightblue}{rgb}{0.10,0.10,0.44}
\definecolor{mintcream}{rgb}{0.96,1.00,0.98}
\definecolor{mistyrose}{rgb}{1.00,0.89,0.88}
\definecolor{moccasin}{rgb}{1.00,0.89,0.71}
\definecolor{navajowhite}{rgb}{1.00,0.87,0.68}
\definecolor{navyblue}{rgb}{0.00,0.00,0.50}
\definecolor{navy}{rgb}{0.00,0.00,0.50}
\definecolor{oldlace}{rgb}{0.99,0.96,0.90}
\definecolor{olivedrab}{rgb}{0.42,0.56,0.14}
\definecolor{orange1}{rgb}{1.00,0.65,0.00}
\definecolor{orange2}{rgb}{0.93,0.60,0.00}
\definecolor{orange3}{rgb}{0.80,0.52,0.00}
\definecolor{orange4}{rgb}{0.55,0.35,0.00}
\definecolor{orangered}{rgb}{1.00,0.27,0.00}
\definecolor{orange}{rgb}{1.00,0.65,0.00}
\definecolor{orchid1}{rgb}{1.00,0.51,0.98}
\definecolor{orchid2}{rgb}{0.93,0.48,0.91}
\definecolor{orchid3}{rgb}{0.80,0.41,0.79}
\definecolor{orchid4}{rgb}{0.55,0.28,0.54}
\definecolor{orchid}{rgb}{0.85,0.44,0.84}
\definecolor{palegoldenrod}{rgb}{0.93,0.91,0.67}
\definecolor{palegreen}{rgb}{0.60,0.98,0.60}
\definecolor{paleturquoise}{rgb}{0.69,0.93,0.93}
\definecolor{paleviolet}{rgb}{0.86,0.44,0.58}
\definecolor{papayawhip}{rgb}{1.00,0.94,0.84}
\definecolor{peachpuff}{rgb}{1.00,0.85,0.73}
\definecolor{peru}{rgb}{0.80,0.52,0.25}
\definecolor{pink1}{rgb}{1.00,0.71,0.77}
\definecolor{pink2}{rgb}{0.93,0.66,0.72}
\definecolor{pink3}{rgb}{0.80,0.57,0.62}
\definecolor{pink4}{rgb}{0.55,0.39,0.42}
\definecolor{pink}{rgb}{1.00,0.75,0.80}
\definecolor{plum1}{rgb}{1.00,0.73,1.00}
\definecolor{plum2}{rgb}{0.93,0.68,0.93}
\definecolor{plum3}{rgb}{0.80,0.59,0.80}
\definecolor{plum4}{rgb}{0.55,0.40,0.55}
\definecolor{plum}{rgb}{0.87,0.63,0.87}
\definecolor{powderblue}{rgb}{0.69,0.88,0.90}
\definecolor{purple1}{rgb}{0.61,0.19,1.00}
\definecolor{purple2}{rgb}{0.57,0.17,0.93}
\definecolor{purple3}{rgb}{0.49,0.15,0.80}
\definecolor{purple4}{rgb}{0.33,0.10,0.55}
\definecolor{purple}{rgb}{0.63,0.13,0.94}
\definecolor{red1}{rgb}{1.00,0.00,0.00}
\definecolor{red2}{rgb}{0.93,0.00,0.00}
\definecolor{red3}{rgb}{0.80,0.00,0.00}
\definecolor{red4}{rgb}{0.55,0.00,0.00}
\definecolor{red}{rgb}{1.00,0.00,0.00}
\definecolor{rosybrown}{rgb}{0.74,0.56,0.56}
\definecolor{royalblue}{rgb}{0.25,0.41,0.88}
\definecolor{saddlebrown}{rgb}{0.55,0.27,0.07}
\definecolor{salmon1}{rgb}{1.00,0.55,0.41}
\definecolor{salmon2}{rgb}{0.93,0.51,0.38}
\definecolor{salmon3}{rgb}{0.80,0.44,0.33}
\definecolor{salmon4}{rgb}{0.55,0.30,0.22}
\definecolor{salmon}{rgb}{0.98,0.50,0.45}
\definecolor{sandybrown}{rgb}{0.96,0.64,0.38}
\definecolor{seagreen}{rgb}{0.18,0.55,0.34}
\definecolor{seashell1}{rgb}{1.00,0.96,0.93}
\definecolor{seashell2}{rgb}{0.93,0.90,0.87}
\definecolor{seashell3}{rgb}{0.80,0.77,0.75}
\definecolor{seashell4}{rgb}{0.55,0.53,0.51}
\definecolor{seashell}{rgb}{1.00,0.96,0.93}
\definecolor{sienna1}{rgb}{1.00,0.51,0.28}
\definecolor{sienna2}{rgb}{0.93,0.47,0.26}
\definecolor{sienna3}{rgb}{0.80,0.41,0.22}
\definecolor{sienna4}{rgb}{0.55,0.28,0.15}
\definecolor{sienna}{rgb}{0.63,0.32,0.18}
\definecolor{skyblue}{rgb}{0.53,0.81,0.92}
\definecolor{slateblue}{rgb}{0.42,0.35,0.80}
\definecolor{slategray}{rgb}{0.44,0.50,0.56}
\definecolor{slategrey}{rgb}{0.44,0.50,0.56}
\definecolor{snow1}{rgb}{1.00,0.98,0.98}
\definecolor{snow2}{rgb}{0.93,0.91,0.91}
\definecolor{snow3}{rgb}{0.80,0.79,0.79}
\definecolor{snow4}{rgb}{0.55,0.54,0.54}
\definecolor{snow}{rgb}{1.00,0.98,0.98}
\definecolor{springgreen}{rgb}{0.00,1.00,0.50}
\definecolor{steelblue}{rgb}{0.27,0.51,0.71}
\definecolor{tan1}{rgb}{1.00,0.65,0.31}
\definecolor{tan2}{rgb}{0.93,0.60,0.29}
\definecolor{tan3}{rgb}{0.80,0.52,0.25}
\definecolor{tan4}{rgb}{0.55,0.35,0.17}
\definecolor{tan}{rgb}{0.82,0.71,0.55}
\definecolor{thistle1}{rgb}{1.00,0.88,1.00}
\definecolor{thistle2}{rgb}{0.93,0.82,0.93}
\definecolor{thistle3}{rgb}{0.80,0.71,0.80}
\definecolor{thistle4}{rgb}{0.55,0.48,0.55}
\definecolor{thistle}{rgb}{0.85,0.75,0.85}
\definecolor{tomato1}{rgb}{1.00,0.39,0.28}
\definecolor{tomato2}{rgb}{0.93,0.36,0.26}
\definecolor{tomato3}{rgb}{0.80,0.31,0.22}
\definecolor{tomato4}{rgb}{0.55,0.21,0.15}
\definecolor{tomato}{rgb}{1.00,0.39,0.28}
\definecolor{turquoise1}{rgb}{0.00,0.96,1.00}
\definecolor{turquoise2}{rgb}{0.00,0.90,0.93}
\definecolor{turquoise3}{rgb}{0.00,0.77,0.80}
\definecolor{turquoise4}{rgb}{0.00,0.53,0.55}
\definecolor{turquoise}{rgb}{0.25,0.88,0.82}
\definecolor{violetred}{rgb}{0.82,0.13,0.56}
\definecolor{violet}{rgb}{0.93,0.51,0.93}
\definecolor{wheat1}{rgb}{1.00,0.91,0.73}
\definecolor{wheat2}{rgb}{0.93,0.85,0.68}
\definecolor{wheat3}{rgb}{0.80,0.73,0.59}
\definecolor{wheat4}{rgb}{0.55,0.49,0.40}
\definecolor{wheat}{rgb}{0.96,0.87,0.70}
\definecolor{whitesmoke}{rgb}{0.96,0.96,0.96}
\definecolor{white}{rgb}{1.00,1.00,1.00}
\definecolor{yellow1}{rgb}{1.00,1.00,0.00}
\definecolor{yellow2}{rgb}{0.93,0.93,0.00}
\definecolor{yellow3}{rgb}{0.80,0.80,0.00}
\definecolor{yellow4}{rgb}{0.55,0.55,0.00}
\definecolor{yellowgreen}{rgb}{0.60,0.80,0.20}
\definecolor{yellow}{rgb}{1.00,1.00,0.00}

\definecolor{kit-green100}{rgb}{0,.59,.51}
\definecolor{kit-green70}{rgb}{.3,.71,.65}
\definecolor{kit-green50}{rgb}{.50,.79,.75}
\definecolor{kit-green30}{rgb}{.69,.87,.85}
\definecolor{kit-green15}{rgb}{.85,.93,.93}

\definecolor{kit-blue100}{rgb}{.27,.39,.67}
\definecolor{kit-blue70}{rgb}{.49,.57,.76}
\definecolor{kit-blue50}{rgb}{.64,.69,.83}
\definecolor{kit-blue30}{rgb}{.78,.82,.9}
\definecolor{kit-blue15}{rgb}{.89,.91,.95}

\definecolor{kit-yellow100}{cmyk}{0,.05,1,0}
\definecolor{kit-yellow70}{cmyk}{0,.035,.7,0}
\definecolor{kit-yellow50}{cmyk}{0,.025,.5,0}
\definecolor{kit-yellow30}{cmyk}{0,.015,.3,0}
\definecolor{kit-yellow15}{cmyk}{0,.0075,.15,0}

\definecolor{kit-orange100}{cmyk}{0,.45,1,0}
\definecolor{kit-orange70}{cmyk}{0,.315,.7,0}
\definecolor{kit-orange50}{cmyk}{0,.225,.5,0}
\definecolor{kit-orange30}{cmyk}{0,.135,.3,0}
\definecolor{kit-orange15}{cmyk}{0,.0675,.15,0}

\definecolor{kit-red100}{cmyk}{.25,1,1,0}
\definecolor{kit-red70}{cmyk}{.175,.7,.7,0}
\definecolor{kit-red50}{cmyk}{.125,.5,.5,0}
\definecolor{kit-red30}{cmyk}{.075,.3,.3,0}
\definecolor{kit-red15}{cmyk}{.0375,.15,.15,0} 

\usepackage{float}                  

\definecolor{lime}{HTML}{A6CE39}
\DeclareRobustCommand{\orcidicon}{\begin{tikzpicture}
	\draw[lime, fill=lime] (0,0) 
	circle [radius=0.16] 
	node[white] {{\fontfamily{qag}\selectfont \tiny ID}};
	\draw[white, fill=white] (-0.0625,0.095) 
	circle [radius=0.007];
	\end{tikzpicture}
	\hspace{-2mm}
}

\foreach \x in {A, ..., Z}{\expandafter\xdef\csname orcid\x\endcsname{\noexpand\href{https://orcid.org/\csname orcidauthor\x\endcsname}{\noexpand\orcidicon}}
}

\newcommand{\rflow}{\textsc{RFlow}}
\newcommand{\rvol}{\textsc{RVol}}
\newcommand{\eroa}{\textsc{EROa}}

\definecolor{KITgreen}{RGB}{140 182 60}
\definecolor{KITyellow}{RGB}{252 229 0}
\definecolor{KITorange}{RGB}{223 155 27}
\definecolor{KITred}{RGB}{162, 34, 35}
\definecolor{KITblue}{RGB}{70 100 170}
\definecolor{myred}{RGB}{192 0 0}
\definecolor{myblue}{RGB}{42 200 224}

\begin{document}

\title[Investigating the shortcomings of the Flow Convergence Method for quantification of Mitral Regurgitation in a pulsatile in-vitro environment and a virtual environment]{Investigating the shortcomings of the Flow Convergence Method for quantification of Mitral Regurgitation in a pulsatile in-vitro environment and with Computational Fluid Dynamics}

\author*[3]{\fnm{Robin} \sur{Leister \orcidB}} \email{robin.leister@kit.edu}
\equalcont{These authors contributed equally to this work.}

\author[1,2]{\fnm{Roger} \sur{Karl  \orcidA}}\nomail \equalcont{These authors contributed equally to this work.}

\author[4]{\fnm{Lubov} \sur{Stroh}}\nomail

\author[2]{\fnm{Derliz} \sur{Mereles}}\nomail

\author[2]{\fnm{Matthias} \sur{Eden}}\nomail

\author[3]{\fnm{Luis} \sur{Neff}}\nomail

\author[1]{\fnm{Raffaele} \sur{de Simone}}\nomail

\author[1]{\fnm{Gabriele} \sur{Romano}}\nomail

\author[3]{\fnm{Jochen} \sur{Kriegseis \orcidK}}\nomail

\author[1]{\fnm{Matthias} \sur{Karck}}\nomail

\author[4]{\fnm{Christoph} \sur{Lichtenstern}}\nomail

\author[2]{\fnm{Norbert} \sur{Frey}}\nomail

\author[3]{\fnm{Bettina} \sur{Frohnapfel \orcidF}}\nomail

\author[3]{\fnm{Alexander} \sur{Stroh \orcidC}}\nomail
\equalcont{These authors contributed equally to this work.}

\author[1,2]{\fnm{Sandy} \sur{Engelhardt \orcidE}}\nomail
\equalcont{These authors contributed equally to this work.}

\affil[1]{\orgdiv{Department of Cardiac Surgery}, \orgname{Heidelberg University Hospital, Heidelberg, Germany}}

\affil[2]{\orgdiv{Department of Internal Medicine III}, \orgname{Heidelberg University Hospital, Heidelberg, Germany}}

\affil*[3]{\orgdiv{Institute of Fluid Mechanics (ISTM)}, \orgname{Karlsruhe Institute of Technology (KIT)}}

\affil[4]{\orgdiv{Department of Anaesthesiology}, \orgname{Heidelberg University Hospital, Heidelberg, Germany}}

\abstract{
The flow convergence method includes calculation of the proximal isovelocity surface area (PISA) and is widely used to classify mitral regurgitation (MR) with echocardiography. It constitutes a primary decision factor for determination of treatment and should therefore be a robust quantification method. 
However, it is known for its tendency to underestimate MR and its dependence on user expertise. The present work systematically compares different pulsatile flow profiles arising from different regurgitation orifices using transesophageal echocardiographic (TEE) probe and particle image velocimetry (PIV) as a reference in an \textit{in-vitro} environment.  
It is found that the inter-observer variability using echocardiography is small compared to the systematic underestimation of the regurgitation volume for large orifice areas (up to 52\%) where a violation of the flow convergence 
 method assumptions occurs.
From a flow perspective, a starting vortex was found as a dominant flow pattern in the regurgant jet for all orifice shapes and sizes. 
A series of simplified computational fluid dynamics (CFD) simulations indicate that selecting a suboptimal aliasing velocity during echocardiography measurements might be a primary source of potential underestimation in MR characterization via the PISA-based method, reaching up to 40\%. 
In this study, it has been noted in clinical observations that physicians often select an aliasing velocity higher than necessary for optimal estimation in diagnostic procedures.
}

\maketitle

\section{Introduction}
\label{sec1}
Mitral regurgitation (MR) is one of the most common valvular heart conditions \cite{Nkomo.2006}. It is   caused by the retrograde flow of blood from the left ventricle into the left atrium through the mitral valve (MV). 
Echocardiography, particularly when conducted via transesophageal approach (TEE), is instrumental in diagnosing the pathology and assessing the severity of mitral regurgitation~\cite{EnriquezSarano.2005}. 
Mitral regurgitation can be addressed through various therapeutic strategies, with the choice of treatment contingent upon multiple factors including the patient's symptoms, the severity of the regurgitation, age, and overall health status.
Particularly in cases when the decision for specific treatment strongly depends on the assessment of TEE, its accuracy and reliability are of great importance. 

 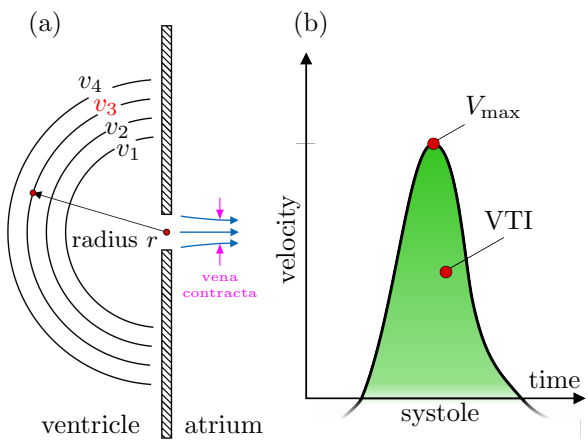
\begin{figure}[h]
        \centering
\begin{tikzpicture}
        \node[anchor=south west,inner sep=0] at (1,0)
	{ \includegraphics[width=0.4\linewidth]{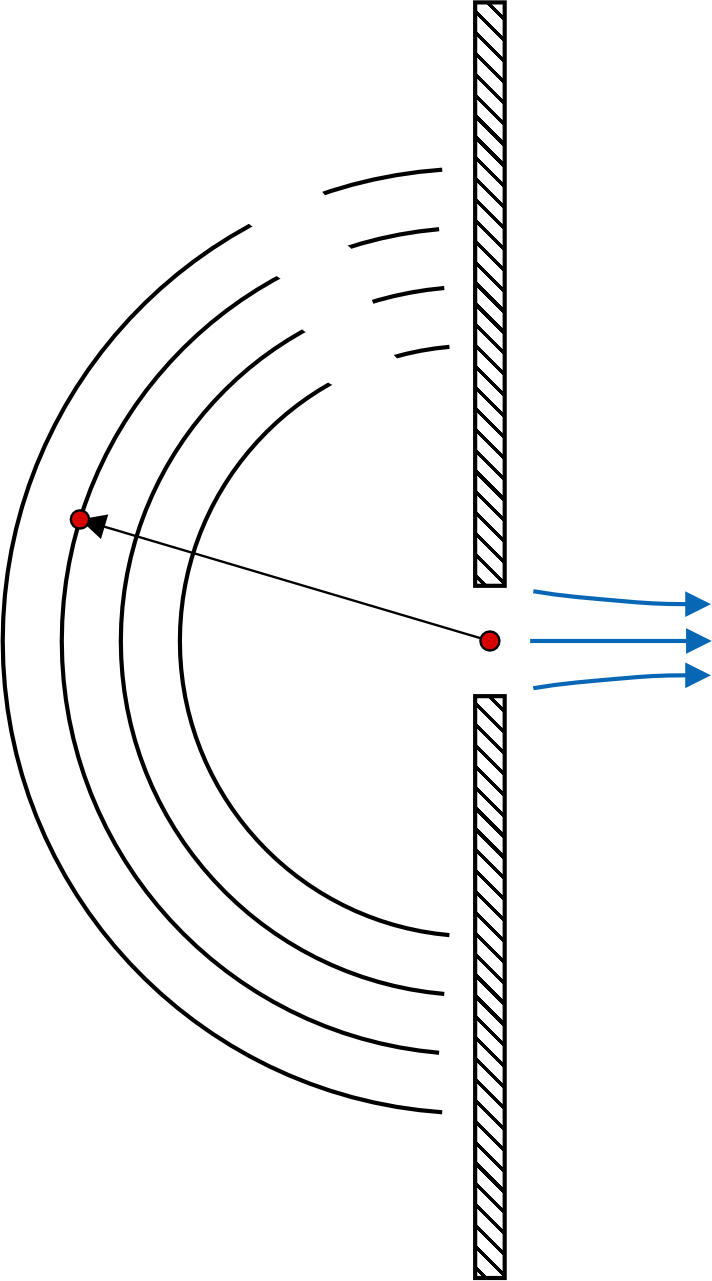}};
   \node[anchor=south west,inner sep=0] at (4.8,0.0)
	{ \includegraphics[width=0.5\linewidth]{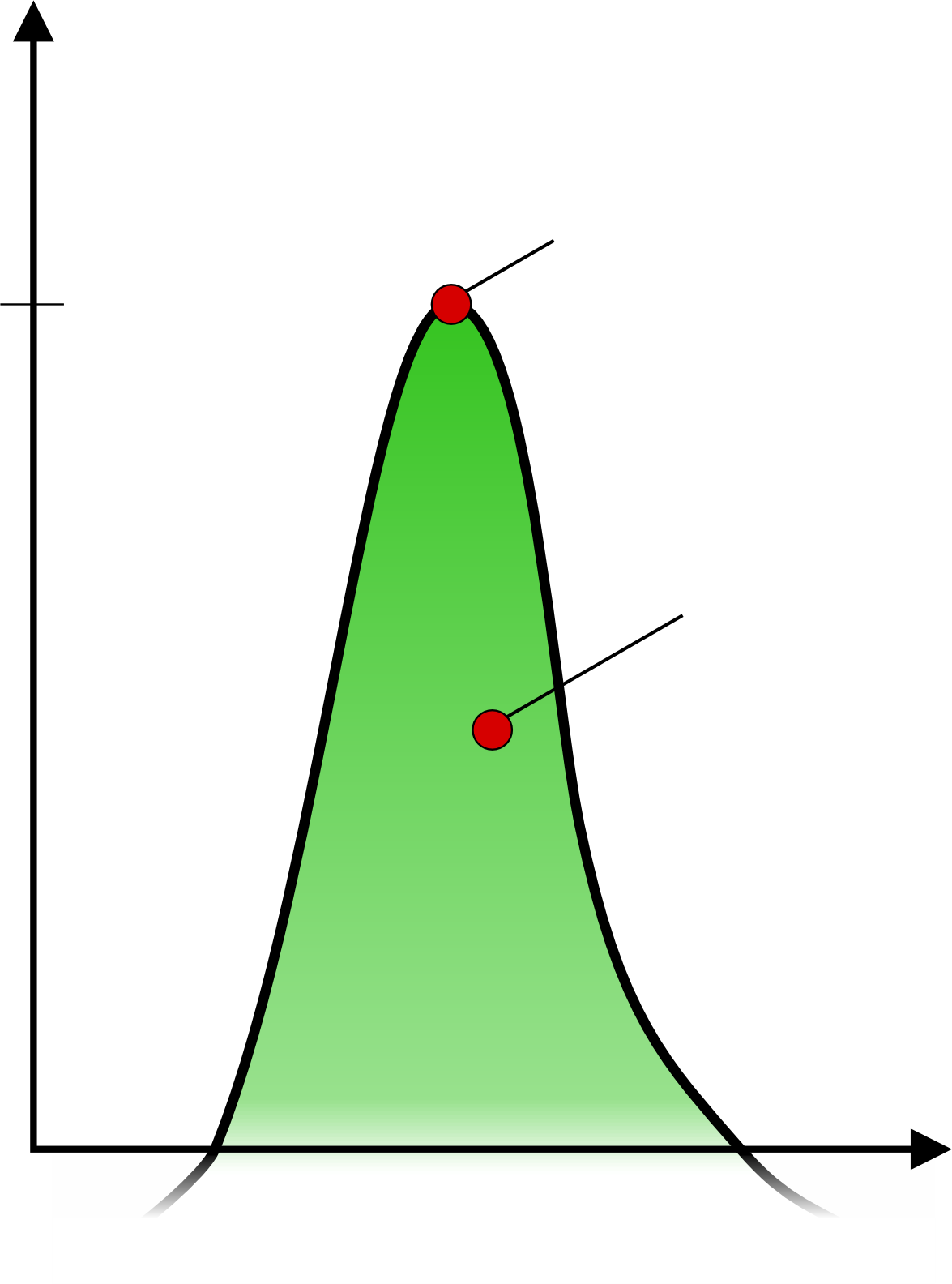}};
  \node at (1.5, 5.5) {(a)};
   \node at (5.0, 5.5) {(b)};
  \node at (2.6, 3.8) {$v_\mathrm{1}$};
  \node at (2.45, 4.1) {$v_\mathrm{2}$};
  \node[red] at (2.3, 4.4) {$v_\mathrm{3}$};
  \node at (2.1, 4.7) {$v_\mathrm{4}$};
  \node at (2.1, 0.2) {ventricle};
   \node at (3.85, 0.2) {atrium};
   \node[font=\small] at (2.42, 2.65) {radius $r$};

\draw[-latex,magenta] (3.8,3.2) -- (3.8,2.9);

\draw[latex-,magenta] (3.8,2.6) -- (3.8,2.3) node[below,font=\tiny,align=center]{vena \\ contracta};
    \node[rotate=90] at (4.7, 2.65) {velocity};
    \node at (6.7, 0.33) {systole};
     \node at (8.2, 0.8) {time};
     \node at (7.4, 4.4) {$V_\mathrm{max}$};
     \node at (7.6, 2.9) {VTI};

\end{tikzpicture}         \caption{(a) Principle of flow convergence: 
        The flow approaching the aperture is described as an idealized sink flow in which increasing flow velocity $v_\mathrm{4}<v_\mathrm{3}<v_\mathrm{2}<v_\mathrm{1}$ is found on concentric hemispheric shells with decreasing surface area. In the shown example $v_\mathrm{3}$ is chosen as the aliasing velocity $V_\mathrm{a}$. It is multiplied with the area of the corresponding hemispheric shell located at distance $r$ from the aperture  to obtain the regurgitation flow (\rflow). The shell area $2\pi r^2$ is referred to as proximal isovelocity surface area (PISA).\\ 
        (b) Schematic of the velocity time integral (VTI) and the maximum speed ($V_\mathrm{max}$) obtained during systole.}
        \label{fig:sketch_isovelocity}
    \end{figure}
Typical quantitative measures for the severity of mitral regurgitation via TEE are the effective regurgitant orifice area (\eroa), regurgitation volume (\rvol{}), the rejection fraction, the vena contracta width (VCW) or area~\cite{salcedo2009framework}. 
The flow convergence method is an established procedure and recommended for determining \eroa{} and \rvol{} whenever possible~\cite{bargiggia1991new,Lancellotti.2010, EnriquezSarano.2005}.
The method, however, is based on several simplifications due to limitations imposed by the restricted abilities of ultrasound technique to acquire the exact geometry of the regurgitant orifice and the abilities to measure flow velocity only in the direction aligned with the probe \cite{Coisne.2002}. 
In order to calculate {\rvol} and {\eroa{}} the so-called radius of a proximal isovelocity surface on the ventricular side of the leaflet has to be measured in color-Doppler mode (cf. Figure \ref{fig:sketch_isovelocity}a) to calculate the proximal isovelocity surface area (PISA), which is then multiplied with the aliasing velocity $V_\mathrm{a}$ to obtain the regurgitation flow:
\begin{equation}
\mathrm{RFlow} = 2 \pi r^2 \cdot V_\mathrm{a}.
\label{eq1}
\end{equation}
This PISA method, however, relies on several assumptions. 
The orifice is assumed to be circular and infinitesimally small such that the flow approaching the orifice can be treated as (half of) an idealized sink flow. 
For such a radially symmetric flow in which all velocity vectors point towards the orifice, a hemispherical shell corresponds to an isovelocity contour. 
A single velocity measurement on this hemisphere is thus sufficient to estimate the flow rate across the hemisphere and thus through the orifice. 
It is important to note that the evaluation of the flow rate from this velocity measurement requires the assumption that the measured velocity vector $v$ is perpendicular to the hemisphere surface.

In a next step, $\mathrm{RFlow}$ is used to calculate \eroa{} based on the maximum observed velocity $V_\text{max}$ within the regurgitation jet
\begin{equation}
 \mathrm{EROa} = \frac{\mathrm{RFlow}}{V_\mathrm{max}},
\label{eq2}
\end{equation}
which is then multiplied with the velocity time integral VTI measured in continious-wave-Doppler (CW-Doppler, Figure \ref{fig:sketch_isovelocity}b) mode during the systole to obtain the regurgitant volume \rvol{}
\begin{equation}
\ \mathrm{RVol} = \mathrm{EROa} \cdot \mathrm{VTI}.
\label{eq3}
\end{equation}

An accurate determination of the PISA-radius $r$ poses a nontrivial challenge in practical scenarios: since the TEE measurement delivers only the velocity component aligned with the ultrasound beam, the isovelocity contours observed during the measurement strongly differ from the assumed hemispherical shell. 
Hence, the ultrasound beam is ideally aligned perpendicular to the orifice plane and the measurement location is chosen such that the ultrasound beam points towards the orifice.
The combination of this factor with the selection of the aliasing velocity $V_\mathrm{a}$, at which the isovelocity contour is examined, introduces a  measurement uncertainty in the practical application of the method in clinical context.
Moreover, the geometry of the orifice can vary considerably, ranging from slit-like apertures to the presence of multiple openings.
Consequently, the assumption of an idealized sink flow is violated such that not all velocity vectors are perpendicular to the PISA~\cite{this2016proximal}. 
Calculating \rvol{} and \eroa{} based on potentially oversimplified assumptions, in turn, is known to lead to underestimation of the severity of mitral regurgitation~\cite{Lancellotti.2010, Iwakura.2006}. 
Coisne~\textit{et al.} reports underestimations of up to 44.2\% even for circular orifices~\cite{Coisne.2002}. 
 However, it has been demonstrated that the estimated flow rates might  also be overestimated since the measured velocity is not strictly perpendicular to the surface~\cite{Barclay.1993, Coisne.2002}.

First approaches to investigate the limitations of the  flow convergence method were carried out comparing it to cardiac catheterization in patients~\cite{Dujardin.1997, LesniakSobelga.2002}. 
These \textit{in-vivo} experiments, however, do not allow to assess different orifice shapes in a reproducible environment. 
In awareness of this problem Coisne~\textit{et al.} \cite{Coisne.2002} showed a promising approach to quantitatively assess the regurgitation flow in an \textit{in-vitro} environment. 
They conducted some advanced 3D echocardiographic reconstructions to quantify the error for the conventional PISA technique. 
Yet, limitations on spatial and temporal  resolution persisted. 
Sonntag~\textit{et al.}~\cite{Sonntag.2014} overcame this limitation by comparing the flow convergence method to highly resolved measurements from particle image velocimetry (PIV) and computational fluid dynamics (CFD). 
The authors used a continuous flow set-up  and Mitral Regurgitation Orifice Phantoms (MROP) with three generic orifice shapes. 
While this approach successfully showed correlation between flows observed through echocardiography and PIV, no quantitative comparison between the flow convergence method and PIV-based estimation was obtained.

In this work, we have build an \textit{in-vitro} set-up with \textit{pulsatile} physiological flow and pressures across specifically designed mitral valve orifice templates to produce regurgitation jets. In this set-up, we aim to quantitatively investigate the relation between orifice and regurgitation volume and show the short-comings of the clinically established PISA estimation, which is based on the flow convergence method. 
The contribution of the paper in comparison to existing work is three-fold:
\begin{itemize}
    \item We capture the  pulsatile regurgitation flow patterns arising from different orifice geometries and sizes with PIV under high spatio-temporal resolution and identify a novel \textit{temporal} regurgitation flow phenomenon (cf. Section \ref{sec:PIV_results}).

\item We employ the PIV measurement results to evaluate  clinical parameters (e.g., \eroa{}, \rvol{}) which are used to grade the severity of chronic mitral regurgitation. This ground truth is compared to echocardiography results obtained by three different physicians, such that an (estimated) inter-observer variability can be distinguished from possible systematic uncertainties in the echocardiography data due to the simplifying assumptions in the flow convergence method (cf. Section \ref{sec:US_results}).
    \item Last but not least, we perform a sensitivity analysis using CFD simulations to estimate influencing factors of the flow convergence method. 
    Hereby we focus on the choice of aliasing velocity $V_\mathrm{a}$, which can introduce essential errors into the PISA estimation (cf. Section~\ref{sec:cfd}). 
\end{itemize}

\section{Materials and Methods}\label{sec2}

\subsection{Mitral Regurgitation Orifice Phantoms}

\begin{figure}[h]\centering
\begin{tikzpicture}[]

\draw[fill=none,ultra thick](0,0) circle (0.66);
\node at (0,-1) {circle};

\draw[fill=none,ultra thick]  (3.5,0) arc(60:120:2.25);
\draw[fill=none,ultra thick]  (3.5,0) arc(-60:-120:2.25);
\node at (1.225+2.29/2,-1) {slit};

\draw[fill=none,ultra thick]  (4.4,0.45) arc(90:270:0.45);
\draw[ultra thick] (4.4,0.45)--(4.4+1.53,0);
\draw[ultra thick] (4.4,-0.45)--(4.4+1.53,0);
\node at (4.9,-1) {drop};

\draw[latex-latex] (0,-0.66) -- (0,0.66) node[left,midway]{$h$};

\draw[latex-latex] (1.225+2.29/2,-0.32) -- (1.225+2.29/2,0.32) node[right,midway]{$h$};

\draw[latex-latex] (4.4,-0.45) -- (4.4,0.45) node[right,midway]{$h$};

\draw[latex-latex] (1.225,0.6) -- (3.515,0.6) node[above,midway]{$w$};
\draw[latex-latex] (3.95,0.6) -- (3.95+1.98,0.6) node[above,midway]{$w$};

\end{tikzpicture} \caption{Different Mitral Regurgitation Orifice Phantoms (MROP) shapes: circle, slit (pointed oval), and drop; $h~=$~height (corresponds to the diameter for circle), $w~=$~width.}\label{shapes}
\end{figure}
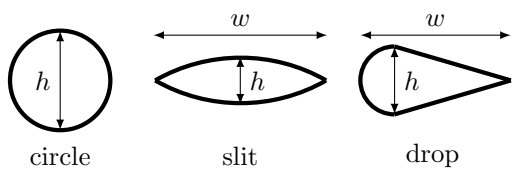

Similar to \cite{Sonntag.2014}, three different MROP-shapes are chosen for the study: circle, slit, and drop (Figure \ref{shapes}). 
The first chosen orifice geometry is a circle, which is selected due to the fact that the flow convergence method is based on the hemispherical PISA assumption, which typically appears upstream of a small circular opening \cite{salcedo2009framework}. 
However, it has been reported that patients often exhibit elongated \cite{Matsumura.2008} or asymmetrical opening geometries \cite{Ashikhmina.2015}. 
To reflect this, we incorporated additional asymmetries into the geometries under consideration -- progressing from the point-symmetrical shape of the circular orifice, to axial symmetries along two axes for the slit, and finally to a single axial symmetry in the drop shape.
The PISA shape follows the opening geometry in the vicinity of the orifice~\cite{Yosefy.2007}. 
Given that the deviation of the actual PISA shape from the ideal hemispherical form contributes to the uncertainty of the flow convergence method, this approach enables the exploration of how specific geometrical properties of the MROP may influence the confidence in the estimation.

The size of the MROPs is varied in three steps (small \textbf{S}, medium \textbf{M}, large \textbf{L}) as denoted in Table \ref{tab:shapelist} in order to produce different grades of mitral regurgitation jets according to established recommendations (Table \ref{Guidlines}). 
The MROPs are manufactured out of a $0.5$\;mm polyvinyl chloride (PVC) film (\textit{Outside Living Industries Deutschland GmbH, Bocholt, Germany}) with a laser cutter and geometrically characterized using a flatbed scanner.

\begin{table}[h]
\caption{Grading the severity of chronic mitral regurgitation by echocardiography as proposed by the American Society of Echocardiography \cite{ZOGHBI2017303}.}\label{Guidlines}\begin{tabular}{l c c c c }
\toprule
~ & Mild  & \multicolumn{2}{c}{Moderate} & Severe\\
\midrule
VCW [cm]   & \textless0.3  & \multicolumn{2}{c}{0.3-0.7} & \textgreater0.7   \\
\rvol [ml]    & \textless30   & 30-44  & 45-59 & \textgreater60 \\
\eroa{} [cm$^2$]  & \textless0.20 & 0.20-0.29   & 0.30-0.39  & \textgreater0.39  \\
\botrule
\end{tabular}                           
\end{table}

\begin{table}[ht!]
    \centering
   
      \caption[Sizes of mitral regurgitation  orifice phantoms (MROPs).]{Sizes of mitral regurgitation  orifice phantoms (MROPs).}
  
    \begin{tabular}{l c c c}
    \hline
       shape&height $h$&width $w$&area\\
       &[mm] &[mm]& [mm$^2$]\\
       \hline
        circle-S  & 4.7 &- & 17.1\\
        circle-M & 8.7 & -& 58.8\\
        circle-L & 12.2 & -& 116.7\\
        
        slit-S & 3.3 & 11.1 &27.0\\
        slit-M & 4.5 & 14.0 & 44.8\\
        slit-L & 7.3 & 22.9 & 115.1\\

       drop-S & 4.3 & 9.7 & 27.1 \\ drop-M & 6.5 & 13.7 & 52.1\\
       drop-L & 9.0 & 19.8 & 108.4 \\
        \hline
    \end{tabular}
 \footnotetext{S - small, M - medium, L - large}
    \label{tab:shapelist}
\end{table}

\subsection{Hemodynamic Simulator}
The test rig is based on the hemodynamic left-heart simulator introduced by Karl~\textit{et al.}~\cite{karl2023ex}, which has been augmented with an atrial window at the opposite side of the septum to include the optical access for the PIV set-up.
The PIV measurement is executed at the atrium side of the MROP, aiming not only to assess the regurgitant flow rate and volume but also to capture the dynamic behavior of the regurgitant jet.
The test rig was manufactured by 3D printing (\textit{Form 3B, Formlabs GmbH, Germany}; layer thickness $50$\,\textmu m). 
MROPs are installed at the position of the mitral valve between atrium and ventricle. 
Since the MROP cannot open and close, but has a fixed-size orifice, it replicates the orifice during mid-systole and not during mid-diastole. 
To enable the filling of the left ventricle, the aortic valve is removed so the aortic flow is enabled to be bidirectional. 
The frequency of the pump (\textit{ViVitro SuperPump, ViVitro Labs, Inc., Victoria, Canada}) is adjusted to 80~bpm and the stroke volume of the pump is set to reach a left ventricular pressure of approximately 120~mmHg. 
A mixture of 70\% water and 30\% glycerol is used as blood mimicking fluid. 1\% of corn starch is added to enhance the ultrasound back-scatter properties of the fluid for the ultrasound experiments. 
Likewise, polyamid particles with a mean diameter of $d_\mathrm{p}=20$ \textmu m were added as seeding for the PIV experiments, which corresponds to a characteristic particle response time of $\tau_\mathrm{p}=33\;$ \textmu s \cite{Raffel2018}.

Essentially, in the first place, the limitations of the simulator include a simplified geometry of the left ventricle and left atrium. This is considered to be acceptable for the current study, since only the flow in the immediate vicinity of the mitral valve opening is of interest. Additionally, the omitted aortic valve allows filling of the left ventricle through the aorta, as our mitral valve phantoms cannot open during diastole. 
Assuming that temporal variation in flow development linked to the aortic valve closure during diastole is negligible, the absence of the aortic valve can be also considered acceptable, as all measurements for determining regurgitation volume are conducted only during systole. 
Another limitation of the simulator is the pulsatile flow pump profile, which replicates a healthy rather than a pathological heartbeat. Combined with reduced peripheral resistance in the simulator, this leads to a higher flow rate through the aorta, which in turn results in a lower flow rate through the mitral valve and a presence of decreased regurgitation volume compared to a real heart. 
Although the regurgitation volumes are reduced, they are still reproducible and quantifiable using both the flow convergence method and PIV, which in turn allows for a comparison of these two methods.

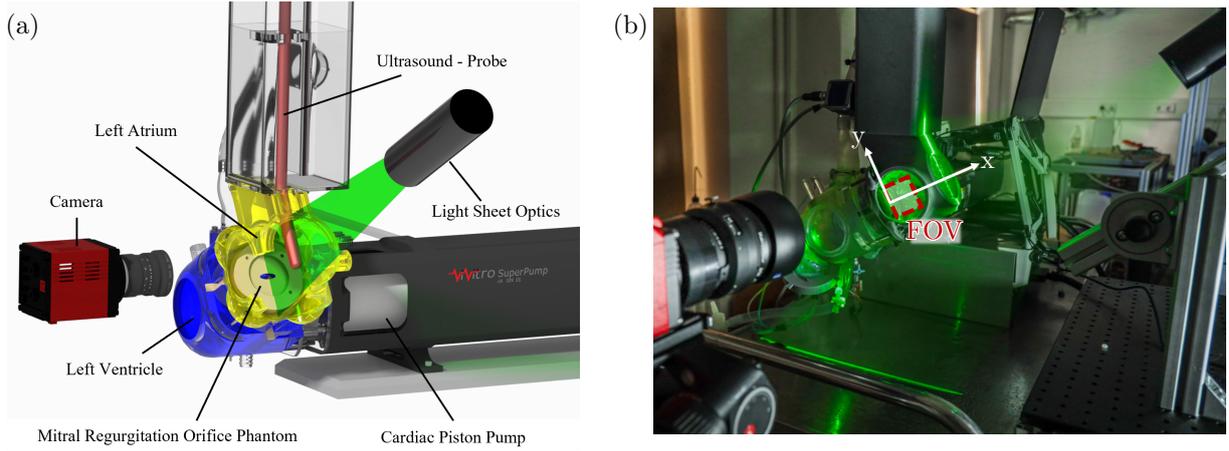
\begin{figure*}[ht]\centering
\begin{tikzpicture}
        \node[anchor=south west,inner sep=0] at (0,0)
	{ \includegraphics[width=0.47\linewidth]{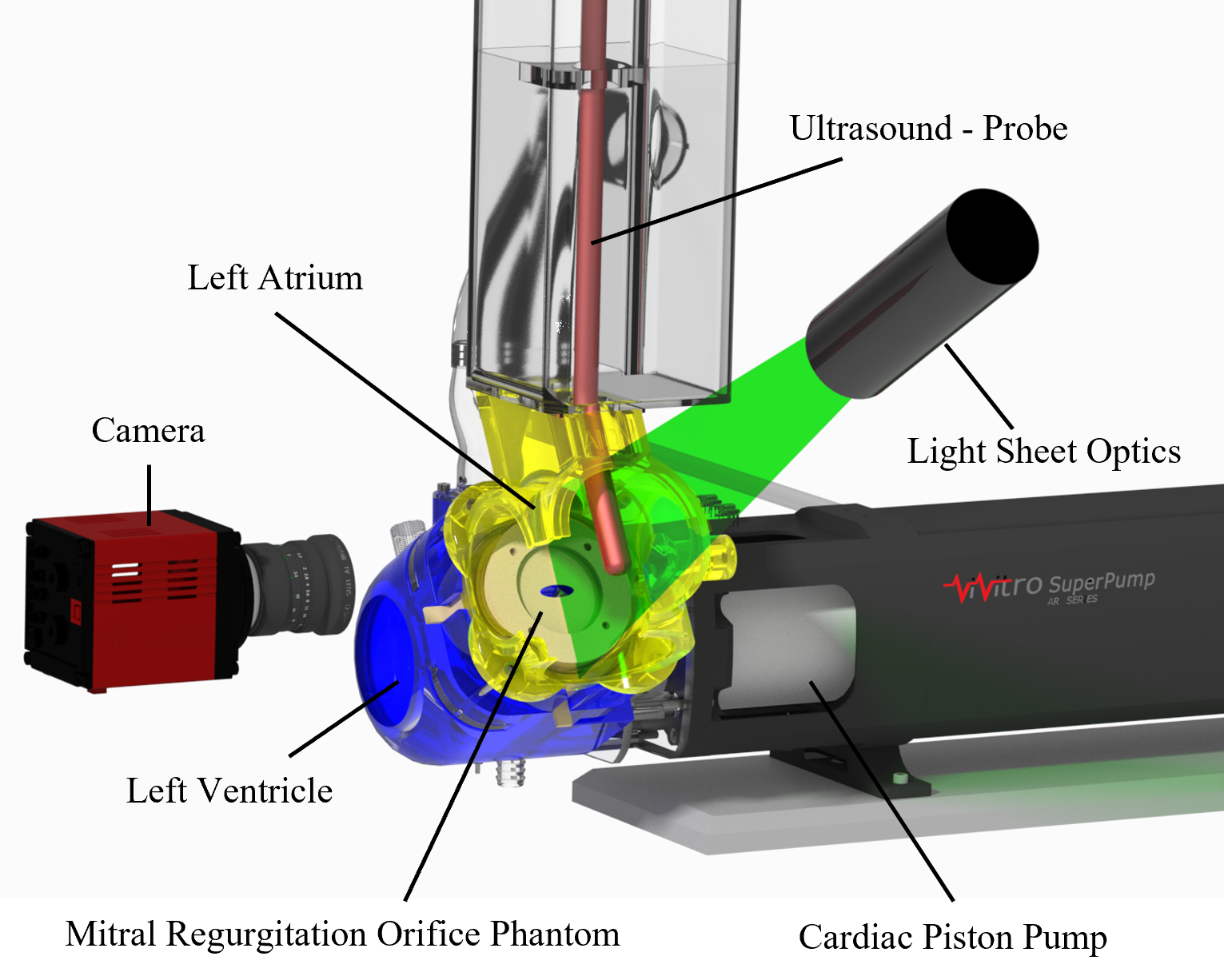}};
   \node[anchor=south west,inner sep=0] at (8.5,0.3)
	{ \includegraphics[width=0.47\linewidth]{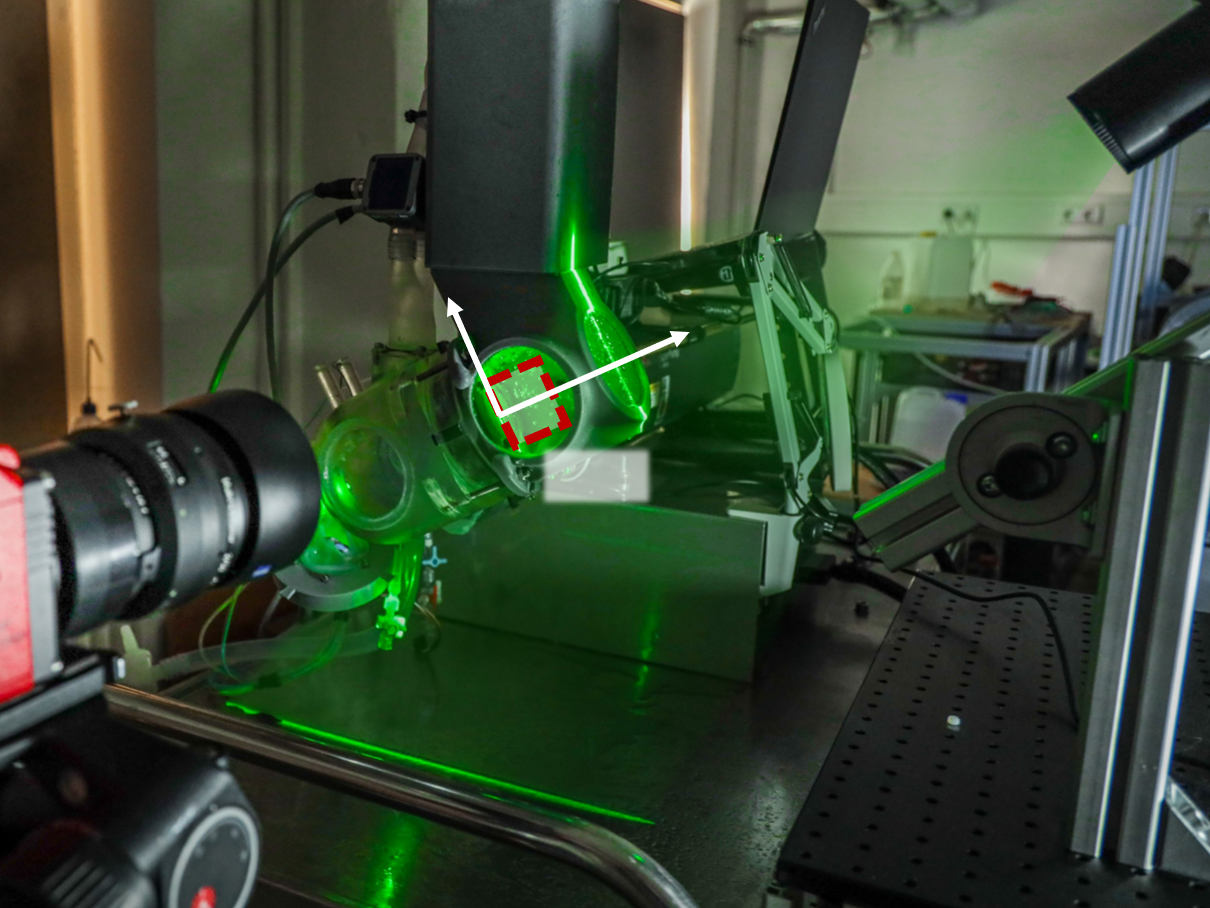}};
    \node at (0.2, 5.7) {(a)};
   \node at (8.2, 5.7) {(b)};

   \node[color=white] at (12.9, 3.9) {x};
   \node[color=white] at (11.2, 4.2) {y};
   \node[color=myred] at (12.2, 3.0) {\contour{white}{FOV}};
\end{tikzpicture} \caption{(a) Schematic of the test rig. Please note, that the ultrasound probe and the PIV set-up (camera and light sheet optics) were not used simultaneously. (b) Photograph of PIV set-up with field of view (FOV) and notation of the axes.}\label{fig:test_rig}
\end{figure*}

\section{Experimental Procedure}

A series of experiments based on TEE and PIV measurements has been conducted throughout the study. 
First, the regurgitation volume for each of the nine different MROPs  was estimated based on velocity information captured through PIV (Section \ref{sec:PIV_method}). The PIV-technique enables a direct measurement of the regurgitation jet and thus does not depend on auxiliary models like the flow convergence method. 
In general, PIV serves as a robust and well-assessed imaging technique for particle-based flow measurements (see e.g. \cite{Raffel2018, Adrian2011}).  
Afterwards, the regurgitation volume was measured by three different physicians using echocardiography (Section \ref{subsec:Ultrasound}).
The results of PIV and TEE experiments are then finally compared  to show possible differences in the resulting MR gradings.

\subsection{Particle Image Velocimetry}
\label{sec:PIV_method}

Figure \ref{fig:test_rig} presents the experimental set-up for the conducted PIV experiments, where two velocity components for two spatial directions (2D2C) are defined. The $x$-coordinate alignes with the regurgitant jet direction (streamwise direction), while the $y$-coordinate indicates the direction orthogonal to the main flow, which is parallel to the installed MROPs.

Two atrial windows positioned in $x$-$y$- and $y$-$z$-planes serve as optical access for camera and the laser sheet, respectively. 
To investigate the flow an \textit{ILA.PIV.sCMOS} camera (16 bit dynamic range, $6.5$\,\textmu m pixel size) was equipped with a $50$\,mm \textit{Zeiss Makro Planar} lens. The resulting magnification can be stated as $M=0.2$ (reproduction scale $s^{xy}=33.3\;$\textmu m/px; FOV size: $85 \times 72\; \mathrm{mm}^2$). As illumination source a double-pulsed \textit{Quantel Evergreen} Nd:YAG laser (210 mJ, $\lambda=532\;$nm, max. repetition rate 15 Hz) is used.
The recorded raw images are converted to velocity vectors with  \textit{PIVview} software leading to $1.88$ velocity vectors per mm, or one vector each $0.5323$\,mm.

Two experimental campaigns are conducted.
The first one deals with the cardiac phase-resolved analysis, where images are acquired for each of the 42 phase positions within a cardiac cycle and was only conducted for \textbf{circle-L} as exemplary case.
This campaign aims to shed light on the formation of the jet.
The second one is aimed to extract the regurgitation volume \rvol{} for all nine geometries from the extracted velocity information, where the streamwise velocity $u$ at the outlet is integrated along the $y$- and $z$-direction to receive the volume flow $\dot{V}$. This quantity is then integrated during the systole time to obtain \rvol{} via PIV.
An explanation of the differences of these evaluation strategies and a step-by-step guidance of the \rvol{} determination via PIV can be found in the Appendix \ref{sec:App_PIV} and \ref{sec:App_Rvol_PIV}.
For each plane 1,000 double-frame images were taken. 
For both axis-symmetric orifice shapes (i.e. ellipse and drop) the laser sheet spanned in $x$-$y$-plane was traversed along the $z$-coordinate with a separation distance of 1-2 mm depending on the orifice size and shape to incorporate 5-7 sampling layers in the calculations.

\subsection{Echocardiography}
\label{subsec:Ultrasound}
\begin{figure*}[ht]\centering
\begin{tikzpicture}
        \node[anchor=south west,inner sep=0] at (0,0)
	{ \includegraphics[height=0.3\linewidth]{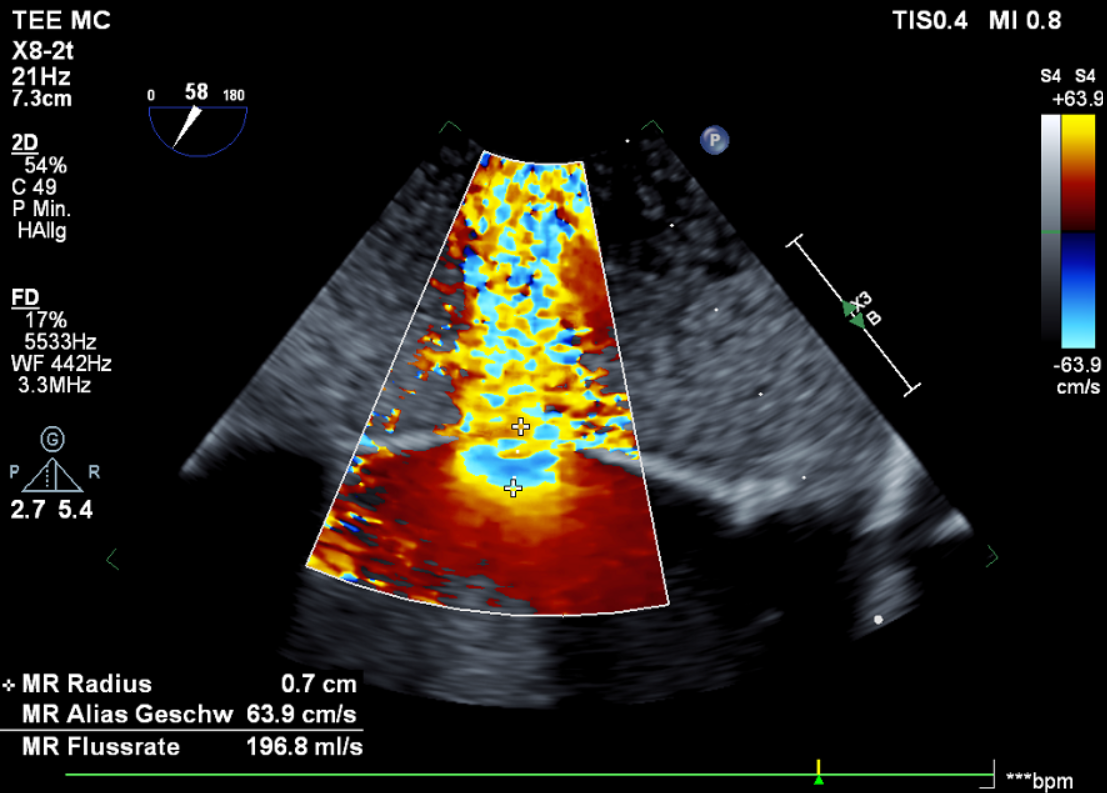}};
   \node[anchor=south west,inner sep=0] at (8.5,0)
	{ \includegraphics[height=0.3\linewidth]{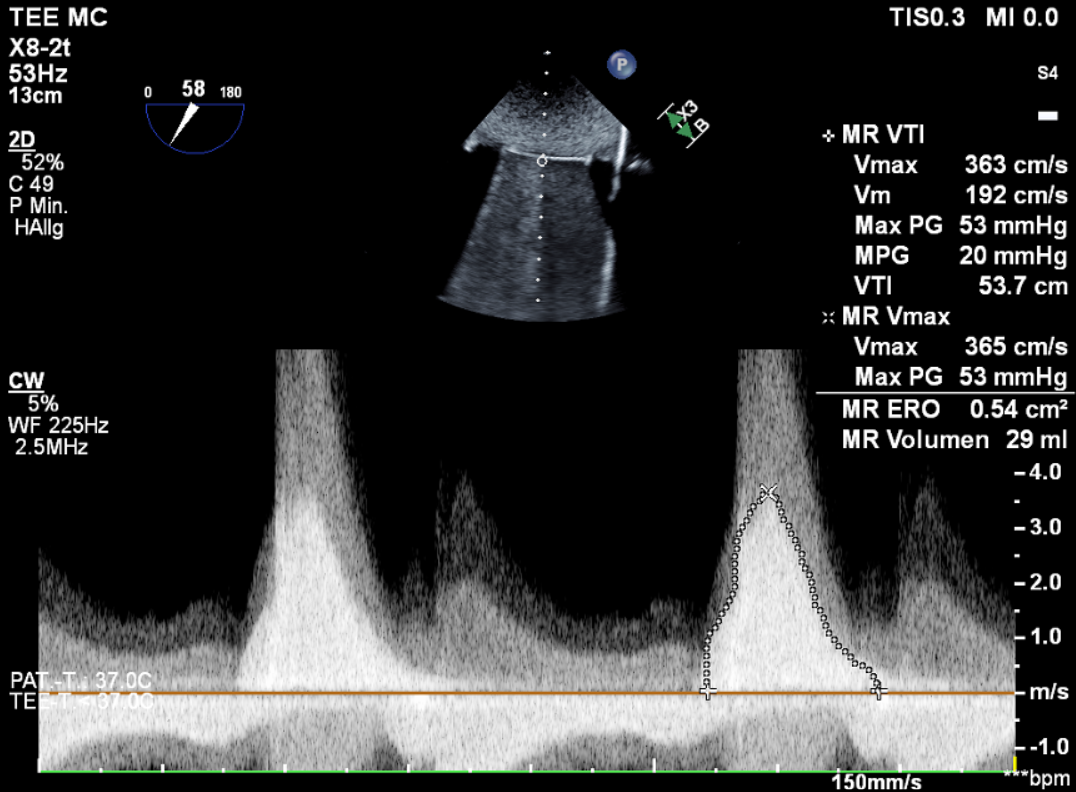}};
    \node[] at (-0.25, 4.5) {(a)};
   \node[] at (8.2, 4.5) {(b)};

\end{tikzpicture} \caption{Flow convergence method to determine the regurgitation volume by a physician. (a) In color-Doppler mode the aliasing velocity is adjusted and the PISA radius is measured. Subsequently the US machine calculates the flow rate. (b) In CW-Doppler mode the VTI area has to be marked, which translates in combination with the maximal velocity into the \eroa{} and \rvol{}.
(\textit{MR Radius}: PISA radius $r$; 
\textit{MR Alias Geschw}: aliasing velocity $V_\mathrm{a}$; 
\textit{MR Flussrate}: \rflow{}; 
\textit{MR ERO}: \eroa{};
\textit{MR Volumen}: \rvol{}). 
}\label{fig:US}
\end{figure*}

The ultrasound experiments were conducted by three experienced physicians. Each physician used a different echocardiographic system to display a realistic variety, which could occur in daily routine. 
\textit{Epiq Cvxi} and \textit{Epiq 7c} ultrasound system with a X8-2T TEE probe, and an \textit{IE33} ultrasound machine with a X7-2t TEE probe manufactured by \textit{Koninklijke Philips N.V., Amsterdam, Netherlands} have been utilized for the measurements. 
The physicians were asked to determine the \rvol{} the way they are used to do it in a clinical routine, which included the following steps:

\begin{itemize}
\item selecting aliasing speed $V_\mathrm{a}$, \item measuring PISA-radius $r$,
\item calculation of \rflow{} (Eq.~\ref{eq1}),
\item measuring VTI and $V_\mathrm{max}$, \item calculation of \eroa{} (Eq.~\ref{eq2}),
\item calculation of \rvol{} (Eq.~\ref{eq3}).
\end{itemize}
Figure~\ref{fig:US} presents two exemplary screenshots from the ultrasound systems used for the estimation of the \rvol{} in the present study.

\begin{figure*}[h]\centering
\begin{tikzpicture}
        \node[anchor=south west,inner sep=0] at (0,0)
	{\includegraphics[width=0.95\linewidth]{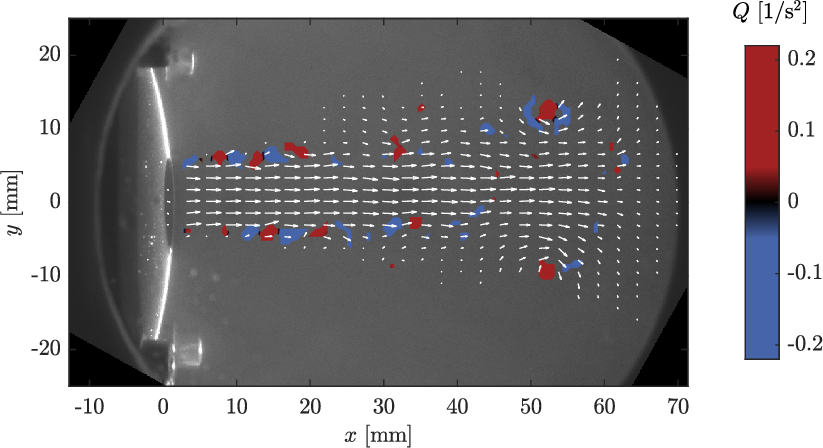}};
  \node[white] at (9.9,7.4) {starting};
\node[white] at (9.9,7.1) {vortex};
\draw [pen colour={white}, decorate, decoration = {calligraphic brace},thick](9.3,6.8) --  (10.5,6.8);
\node[rotate=90,white] at (3.0,4.7) {orifice};
\node[white] at (4.6,6.4) {Kelvin-Helmholtz};
\node[white] at (4.6,6.1) {instabilities};
\draw [pen colour={white}, decorate, decoration = {calligraphic brace},thick](3.45,5.85) --  (5.8,5.85);
 \end{tikzpicture}
\caption{Instantaneous flow field of the regurgitation jet at $t=245\;$ms with underlying vortex Q criterion~\cite{Hunt1988} to indicate vortical structures for \textbf{circle-L}.  Every third/fifth vector in y/x-direction is shown for clarity. In the background the raw data image with the PVC film and the orifice is visible.} \label{fig:pinhole13_phase_020}
\end{figure*}
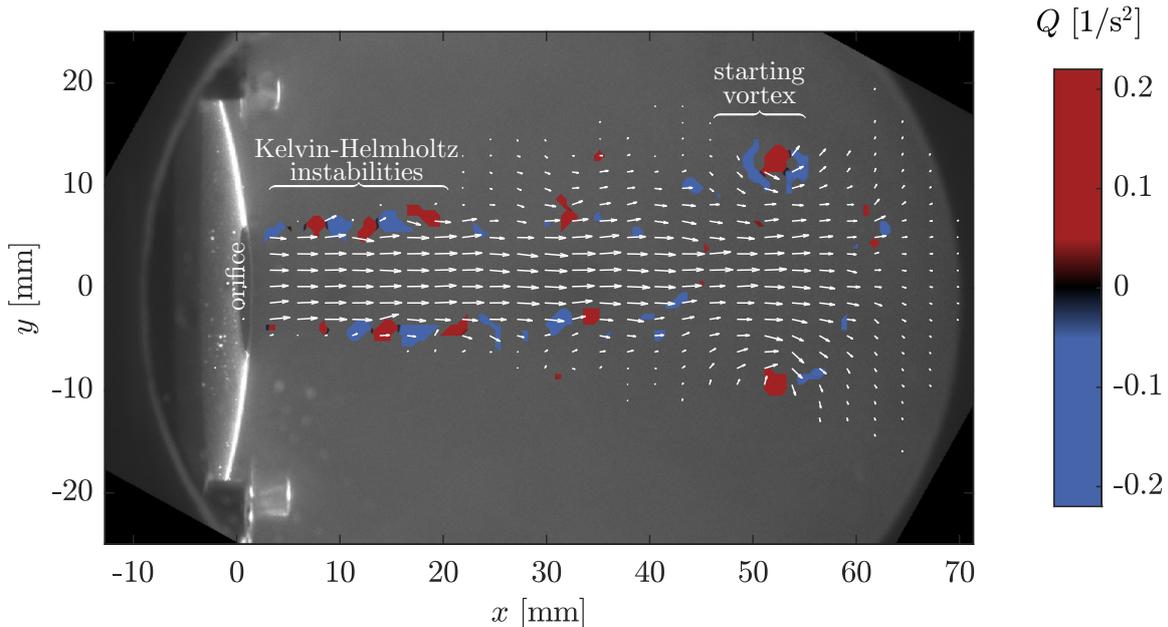

\section{Results}\label{sec:results}
The results gained by means of PIV experiments are twofold. 
First the cardiac phase-resolved analysis offers a deeper insight into the flow dynamics of the fluid system and second, the gained velocity fields are used to calculate the regurgitation volume, which can be compared to the ultrasound experiments. 
The cardiac phase-resolved measurements were conducted for selected MROP shapes. Exemplary results for \textbf{circle-L} are shown in the following.
\subsection{PIV-based Fluid Flow Analysis}
\label{sec:PIV_results}

\begin{figure*}[t]
\centering

\begin{tikzpicture}[baseline]
        \begin{axis}[ 
                cycle list name=color list,
                cycle list shift=4,
width=0.45\linewidth,
                height=0.45\linewidth,
                ylabel={streamwise velocity [m/s]},
                xlabel={$y/h$},
                xmin=-2, xmax=2,
                ymin=-0.5, ymax=4.5,
tick style={color=black},
legend style={at={(0.8,1)},fill=none,anchor=north,font=\tiny,draw=none},
                legend cell align={left},
                legend columns=1, axis on top
                ]

        \addplot +[thick] table [y=u, col sep=comma] {data/u075_.csv};
        \addlegendentry{$t=75$~ms}
        \addplot +[thick] table [y=u, col sep=comma] {data/u175.csv};
        \addlegendentry{$t=175$~ms}
        \addplot +[thick] table [y=u, col sep=comma] {data/u205.csv};
        \addlegendentry{$t=205$~ms}
        \addplot +[thick] table [y=u, col sep=comma] {data/u245.csv};
        \addlegendentry{$t=245$~ms}
        \addplot +[thick] table [y=u, col sep=comma] {data/u275.csv};
        \addlegendentry{$t=275$~ms}
        \addplot +[thick] table [y=u, col sep=comma] {data/u375.csv};
        \addlegendentry{$t=375$~ms}
        \addplot +[thick, dashed] table [y=u, col sep=comma] {data/u525.csv};
        \addlegendentry{$t=525$~ms}
        \addplot +[thick, dashed] table [y=u, col sep=comma] {data/u725.csv};
        \addlegendentry{$t=725$~ms}
        \node[below right] at (-2,4.5) {(a)};
    \end{axis}
    
    \end{tikzpicture}\hspace{1cm}
        \begin{tikzpicture}[baseline]
        \begin{axis}[
                  cycle list name=color list,
                cycle list shift=4,
                width=0.45\linewidth,
                height=0.450\linewidth,
                tick style={color=black},
                ylabel={max. velocity magnitude [m/s]},
                xlabel={$t$ [ms]},
                xmin=0, xmax=750,
                ymin=0, ymax=4.5,
legend style={at={(0.5,1.025)},anchor=south,font=\footnotesize,draw=none},
legend columns=3, axis on top
                ]
                
\addplot +[only marks] coordinates {
(75, 0.006702531)
};
\addplot +[only marks] coordinates {
(175, 0.07658)
};
\addplot +[only marks] coordinates {
(205, 1.7844579)
};
\addplot +[only marks] coordinates {
(245, 3.9749234)
};
\addplot +[only marks] coordinates {
(275, 3.5487075)
};
\addplot +[only marks] coordinates {
(375, 1.0624841)
};
\addplot +[only marks] coordinates {
(525, 0.011928687)
};
\addplot +[only marks] coordinates {
(725, 0.024339924)
};
\addplot[thick] table [y=u, col sep=comma] {data/u_t_new.csv};
\addplot[thick, dashed] table [y=u, col sep=comma] {data/u_t_dashed_new.csv};
\node[below right] at (-2,4.5) {(b)};
    \end{axis}
\end{tikzpicture}
 \caption{(a) Temporal evolution of velocity profiles for the regurgitation jet close to the orifice at $x/h=0.2$ (b) Temporal evolution of maximum velocity during one cardiac cycle; both  for \textbf{circle-L}. Dashed lines show points in time during diastole, which should not be compared to a realistic cardiac cycle.
\label{fig:velocityprofiles}}
\end{figure*}
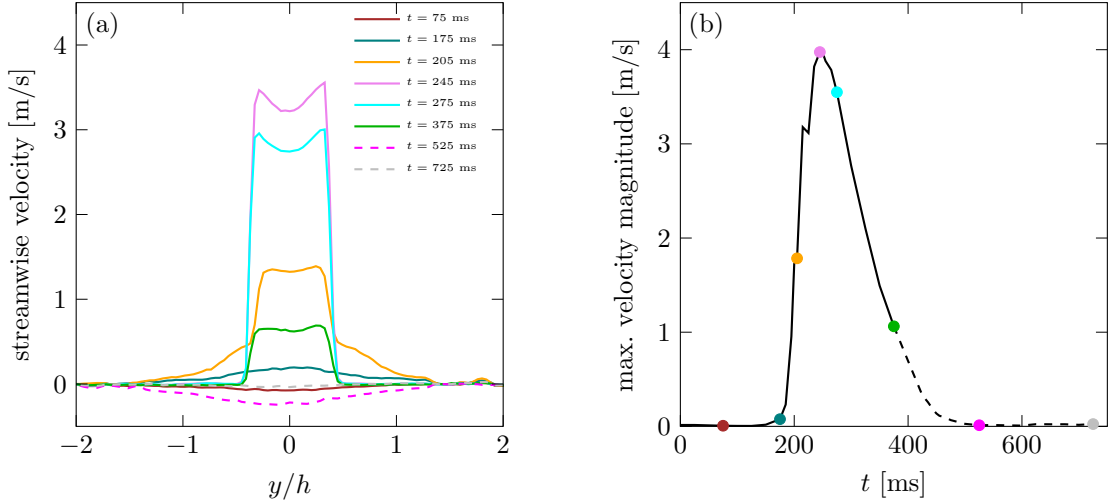

Figure \ref{fig:pinhole13_phase_020} shows the instantaneous velocity field for \textbf{circle-L} orifice as vector plot for a time instance, where the jet almost covers the entire atrium. 
The underlying color depicts the occurring vortical structures by means of the Q-criterion~\cite{Hunt1988}, where positive values represent regions with high rotation, while negative values show regions with high strain. 
The orifice of the jet is located on the left at $x$=0.
It can be seen that the appearing Kelvin-Helmholtz instabilities on the edge of the jet are weakly-pronounced as coherent structures and break up at around $20$\,mm distance downstream of the orifice. 
This represents a fluid instability phenomenon typically observed at the interface where two fluid layers are in motion at various distinct velocities.
To increase the statistical significance the subsequent analyses were conducted using an average of 100 images per phase.

Figure \ref{fig:velocityprofiles}(a) shows the velocity profiles near the  outlet \textbf{circle-L} for eight time instances, while Figure~\ref{fig:velocityprofiles}(b) shows the maximum velocity close to the outlet (up-to $20$\,mm) of all recorded phasings with marks of the same time instances. The evolution of the regurgitation jet in terms of maximum velocity is shown in Figure \ref{fig:velocityprofiles}(b), which is an analogon to the extraction of the VTI area in CW-Doppler ultrasound mode (Fig. \ref{fig:US}(a)). 
The overall behaviour corresponds well to the ultrasound data. The jet velocity experiences a non-monotonous rapid increase in time, with a local maximum at $210$ ms caused by the formation of the starting vortex and a global maximum at $245$ ms, followed by a continuous, exponential-like decrease.
For the time instances where the jet is most pronounced the formation of saddle-backed velocity profiles can be ascertained caused by the thin orifice. 
The shape of velocity profile utilized for estimation of \eroa{} from Eq.~\ref{eq2} in the flow convergence method is assumed to be uniform across the orifice exhibiting $V_\text{max}$ at any location (Fig.~\ref{fig:velocityprofiles}(a), $t=245$~ms).
This assumption is not fulfilled in the middle of the velocity profile, which might lead to underestimation of \eroa{} since a slightly larger velocity is assumed to be present everywhere across the orifice opening. As comparison, a block profile with uniform velocity $V_\text{max}$ would lead to a volume flow which is 8.6\% higher.

\subsection{Comparison Between PIV and Echocardiographic Findings}
\label{sec:US_results}

\begin{figure}[ht]
\centering
\begin{tikzpicture}[baseline]
    \begin{axis}[
    width=1\linewidth,
    height=0.82\linewidth,
                ybar,
                bar width = 4pt,
legend style={at={(0.825,0.995)},fill=white,draw=none,
                anchor=north,legend columns=1,font=\tiny},
                ylabel={$V_\text{max}$ [m/s]},
xtick=data,
nodes near coords align={vertical},
                x tick label style={rotate=45,anchor=east},
                xmin=0,
                               tick style={color=black},
                xmax=10,
                ymin=0,
                ymax=7.5,
                xticklabels={{circle-S},
{circle-M},
{circle-L},
{slit-S},
{slit-M},
{slit-L},
{drop-S},
{drop-M},
{drop-L},
},
                ]
        \addplot+[area legend] coordinates {
        (1,4.44) (2,4.05) (3,4.53) (4,3.66) (5,3.26) (6,3.64) (7, 3.31) (8, 3.52) (9, 3.49)};
         \addplot+[area legend,error bars/.cd, y dir=both, y explicit,error bar style=red]
         coordinates {
         (1,4.8284) +=(0,0.532) -=(0,0.532)
         (2,4.5677) +=(0,0.1838) -=(0,0.1838)
         (3,3.9535) +=(0,0.2858) -=(0,0.2858)
         
         (4,4.0123) +=(0,0.1242) -=(0,0.1242)
         (5,3.915) +=(0,0.1072) -=(0,0.1072)
         (6,3.5771) +=(0,0.162) -=(0,0.162)
         
         (7, 3.9768) +=(0,0.0916) -=(0,0.0916)
         (8, 3.8956) +=(0,0.1588) -=(0,0.1588)
         (9, 3.6836) +=(0,0.1806) -=(0,0.1806)
         };

         \addplot[only marks,mark=+,line legend]
         coordinates{
         (0.825,5)
         (1.825,3.6)
         (2.825,3.34)
         (3.825,3.99)
         (4.825,3.40)
         (5.825,3.62)         
         (6.825,3.69)
         (7.825,3.66)
         (8.825,3.65)
         }; 
         \addplot[only marks,mark=x,line legend]
         coordinates{
         (0.825,4.47)
         (1.825,4.61)
         (2.825,6.89)
         (3.825,4.13)
         (4.825,3.43)
         (5.825,4.51)         
         (6.825,3.54)
         (7.825,3.72)
         (8.825,3.81)
         };
         \addplot[only marks,mark=Mercedes star flipped,line legend]
         coordinates{
         (0.825,3.84)
         (1.825,3.94)
         (2.825,3.36)
         (3.825,2.86)
         (4.825,2.94)
         (5.825,2.79)         
         (6.825,2.70)
         (7.825,3.17)
         (8.825,3.0)         
         };              
         \legend{US,PIV,physician 1,physician 2,physician 3};
    \end{axis}

\end{tikzpicture} \caption{Maximum captured velocity for PIV and US. The ultrasound measurements are displayed individually (symbols) and as mean values (bars), while the PIV measurements show the $2 \sigma$-confidence interval superimposed to the mean values (see Appendix~\ref{sec:App_Uncertainty} for more details).}\label{fig:umax_piv_us}
\end{figure}
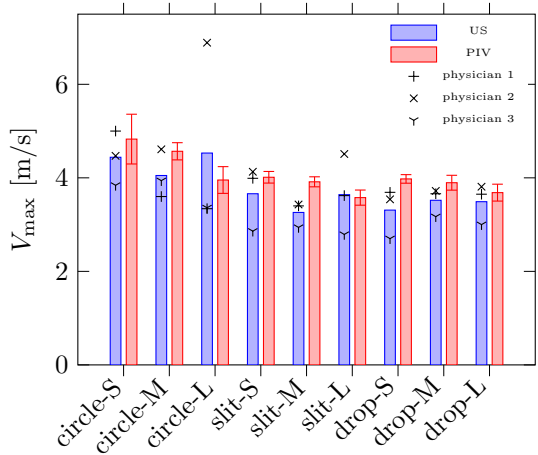

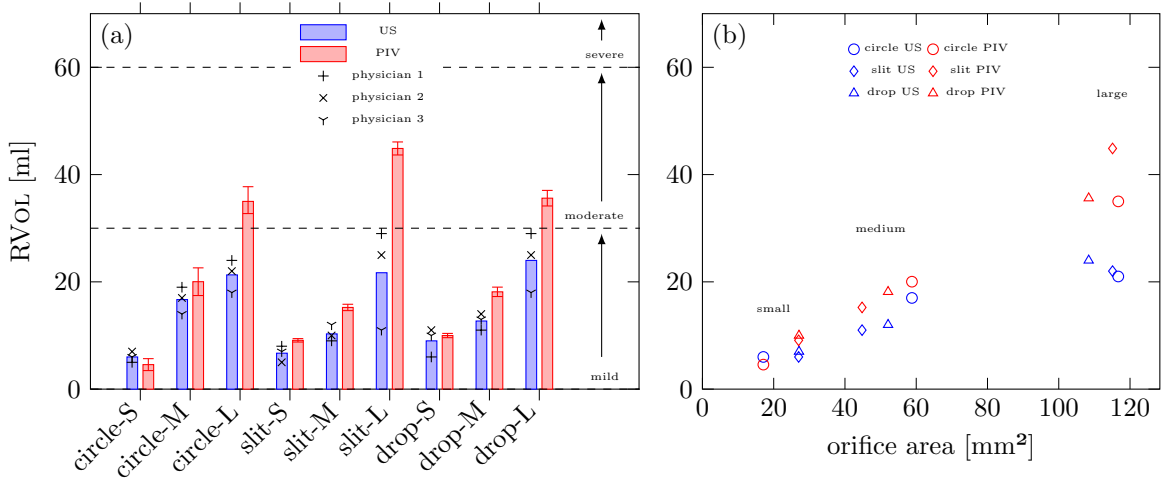
\begin{figure*}[ht]
\centering
\begin{tikzpicture}[baseline]
    \begin{axis}[
    width=0.55\linewidth,
    height=0.41\linewidth,
                ybar,
                bar width = 4pt,
                legend style={at={(0.5,0.995)},fill=white,draw=none,
                anchor=north,legend columns=1,font=\tiny},
                ylabel={\rvol{} [ml]},
                xtick=data,
                nodes near coords align={vertical},
                x tick label style={rotate=45,anchor=east},
                xmin=0,
                               tick style={color=black},
                xmax=11,
                ymin=0,
                ymax=70,
                xticklabels={%
{circle-S},
{circle-M},
{circle-L},
{slit-S},
{slit-M},
{slit-L},
{drop-S},
{drop-M},
{drop-L},
},
                ]
                         \draw[-latex] (10.25,6) -- (10.25,29);
         \draw[-latex] (10.25,35) -- (10.25,59);
         \draw[-latex] (10.25,65) -- (10.25,69);
        \addplot+[area legend] coordinates {
        (1,6) (2,16.7) (3,21.3) (4,6.7) (5,10.3) (6,21.7) (7, 9) (8, 12.7) (9, 24)};
                    \addplot+[area legend,error bars/.cd, y dir=both, y explicit,error bar style=red]
         coordinates {
         (1,4.57)  += (0,1.112) -= (0,1.112)
         (2,20.03)  += (0,2.58) -= (0,2.58)
         (3,35)  += (0,2.74) -= (0,2.27)
         
         (4,9.1)  += (0,0.32) -= (0,0.32)
         (5,15.23)  += (0,0.592) -= (0,0.592)
         (6,44.87)  += (0,1.214) -= (0,1.214)
         
         (7, 9.99)  += (0,0.398) -= (0,0.398)
         (8, 18.14)  += (0,0.8774) -= (0,0.8774)
         (9, 35.60)  += (0,1.452) -= (0,1.452)
         };

         \addplot[only marks,mark=+,line legend]
         coordinates{
         (0.835,5)
         (1.835,19)
         (2.835,24)
         (3.835,8)
         (4.835,9)
         (5.835,29)         
         (6.835,6)
         (7.835,11)
         (8.835,29)
         }; 
         \addplot[only marks,mark=x,line legend]
         coordinates{
         (0.835,7)
         (1.835,17)
         (2.835,22)
         (3.835,5)
         (4.835,10)
         (5.835,25)         
         (6.835,11)
         (7.835,14)
         (8.835,25)
         };
         \addplot[only marks,mark=Mercedes star flipped,line legend]
         coordinates{
         (0.835,6)
         (1.835,14)
         (2.835,18)
         (3.835,7)
         (4.835,12)
         (5.835,11)         
         (6.835,10)
         (7.835,13)
         (8.835,18)         
         };      
         \addplot[black,sharp plot,update limits=false,dashed] coordinates { (0,30) (12,30) }node [above,xshift=0pt,font=\tiny] at (10.1,30) {moderate};
         \addplot[black,sharp plot,update limits=false,dashed] coordinates { (0,0) (12,0)} node [above,xshift=4pt,font=\tiny] at (10.1,0) {mild};
         \addplot[black,sharp plot,update limits=false,dashed] coordinates { (0,60) (12,60) } node [above,font=\tiny,xshift=4pt] at (10.1,60) {severe};         
         \legend{US,PIV,physician 1,physician 2,physician 3};
         \node[below right] at (0,70) {(a)};
    \end{axis}

\end{tikzpicture} \begin{tikzpicture}[baseline]
    \begin{axis}[
    width=0.475\linewidth,
    height=0.41\linewidth,
        legend style={fill=none,draw=none,legend columns=2,legend cell align={center},font=\tiny},
legend style={at={(0.5,0.95)},anchor=north},
ymax=70,ymin=0,
        xmin=0,
        tick style={color=black},
        xlabel={orifice area [mm²]},
]
        \addplot [
        scatter,
        only marks,
        point meta=explicit symbolic,
        scatter/classes={
        Circle_US={mark=o,draw=blue},
        Circle_PIV={mark=o,draw=red},
        slit_US={mark=diamond,blue},
        slit_PIV={mark=diamond,red},
        Drop_US={mark=triangle,blue},
        Drop_PIV={mark=triangle,red}},
        ] 
        table [meta=label] {
        x y label
        17.1 6 Circle_US
        17.1 4.57 Circle_PIV
        27.1 7 Drop_US
        27.1 9.99 Drop_PIV
        27 6 slit_US
        27 9.1 slit_PIV
        44.8 11 slit_US
        44.8 15.23 slit_PIV
        52.1 12 Drop_US
        52.1 18.14 Drop_PIV
        58.8 17 Circle_US
        58.8 20.03 Circle_PIV
        108.4 24 Drop_US
        108.4 35.6 Drop_PIV
        115.1 22 slit_US
        115.1 44.87 slit_PIV
        116.7 21 Circle_US
        116.7 35 Circle_PIV
        };
        \legend{circle US, circle PIV, slit US, slit PIV, drop US, drop PIV},
        \node[font=\tiny] at (20,15) {small};
        \node[font=\tiny] at (50,30) {medium};
        \node[font=\tiny] at (115,55) {large};
        \node[below right] at (0,70) {(b)};
        \end{axis}
\end{tikzpicture} \caption{(a) The regurgitation volume of each MROP measured by ultrasound (US) with the flow convergence method and particle image velocimetry (PIV). The ultrasound measurements are displayed individually (symbols) and as mean values (bars), while the PIV measurements show the $2 \sigma$-confidence interval (see Appendix~\ref{sec:App_Uncertainty} for more details). (b) The mean regurgitation volume over orifice area for each MROP, where a systematic underestimation of US for larger areas becomes visible. }\label{fig:rvol-bars}
\end{figure*}

First, we examine the comparison of maximum captured velocity between the US and PIV measurements, as shown in Figure~\ref{fig:umax_piv_us}. 
The uncertainties in PIV velocity are represented by a $2\sigma$ confidence interval, ranging from 2\% to 11\% of the captured $V_\text{max}$, depending on the geometry.
It is important to note that $V_\text{max}$ in both PIV and US is estimated along the jet's centerline; however, for PIV, we extract $V_\text{max}$ at the position closest to the orifice.
We observe a reasonable agreement between the US and PIV values, though in 7 out of 9 MROPs, US measurements exhibit an underestimation of $V_\text{max}$ in the range of 5-10\%. 
This underestimation may be attributed to a slight misalignment of the US probe with the flow direction -- a known limitation of the ultrasound measurement technique~\cite{qin2021computational}, which inherently leads to an underestimation of the local velocity magnitude due to the one-dimensional nature of the US measurement. 
The comparison confirms that local velocity information can be consistently captured by both PIV and US, with a noticeable trend of velocity underestimation in the US measurements. 
This trend may also contribute to an underestimation of \rvol{} due to the underestimation of $V_a$ in Eqs.~\ref{eq1}-\ref{eq3}.

The evaluation of \rvol{} is carried out for all nine geometries based on the PIV images as explained in Section~\ref{sec:PIV_method} and Appendix~\ref{sec:App_PIV}. 
The results of the PIV and echocardiographic measurements are displayed in Figure~\ref{fig:rvol-bars} in terms of \rvol{}. The values for all intermediate steps of the data evaluation are provided in the Appendix in Table~\ref{tab:ultrasound}.
The mean \rvol{} measured by the flow convergence method is smaller for eight out of nine MROPs compared to the PIV measurements. 
Only one MROP, the \textbf{circle-S}, shows an overestimation in \rvol{} for echocardiographic measurements than for the PIV-based estimation.
For each shape the \rvol{} increased in correspondence to the size for both PIV and TEE measurements. 
Furthermore, mild regurgitation volumes measured by PIV (\textbf{S}- \& \textbf{M}-size) differed by an absolute error of 1--5~ml to the mean TEE \rvol{} and would be also classified as mild \rvol{}. 

Converted to a relative measure with a normalisation in regard to the PIV volumes this absolute error corresponds to a relative error of 10--30\%.
The regurgitation induced with \textbf{L}-sized geometries would be classified as moderate mitral regurgitation by PIV and as mild mitral regurgitation after applying the flow convergence method. 
The difference between the measurement methods ranges here from 12 to 23~ml (33 -- 52\%) with the highest deviation observed for the \textbf{slit-L}.
The circular and drop-shaped orifices show very similar results not only in the estimated \rvol{} (35--36~ml) but also in the deviation between the measurement techniques (12--14~ml, around 33\%).
This observation can be attributed to the greater degree of morphological similarity between the two structures, characterized by their rounded shape, as opposed to the slender angled aperture of \textbf{slit-L}.
The inter-observer variability is small in absolute values for \textbf{S} and \textbf{M} MROPs (Maximum of 5 ml, which corresponds to 30\%, when compared to the mean). However, the variation in the results among physicians with \textbf{slit-L} and \textbf{drop-L} is significantly larger than for the other MROPs. Here a maximum difference of 18 ml occurs, which corresponds to 83\% when compared to the mean.\\
Figure~\ref{fig:rvol-bars}(b) highlights the fact, that the large orifice area appears to be positively correlated with the higher estimation deviation for \rvol{} observed between PIV and TEE measurements. 

\subsection{CFD-based Analysis of Flow Convergence Method}
\label{sec:cfd}

We utilize a simplified simulation set-up described in Appendix~\ref{sec:App_CFD} to conduct a sensitivity analysis of the flow convergence method on CFD velocity information. 
We apply PISA approximation to the CFD data and investigate the effect of choosing a different aliasing velocity $V_\mathrm{a}$ and the effect of the orifice shape on the PISA calculation.
Based on the study we derive recommendations for PISA estimations. 
Please note that due to the assumption of a stationary flow in the simulations, we are not able to analyze the dynamics of the process (e.g. estimation of VTI), but can precisely quantify the effects of orifice geometry and the choice of aliasing velocity $V_\mathrm{a}$ on the estimation of PISA radius, which directly affects \rflow{} and \eroa{}.

\subsubsection{Estimation of PISA Radius}
\begin{figure*}[h]
  \centering
    \begin{tikzpicture}[baseline]
\begin{pgfonlayer}{background}
\includegraphics[width=0.3\linewidth]{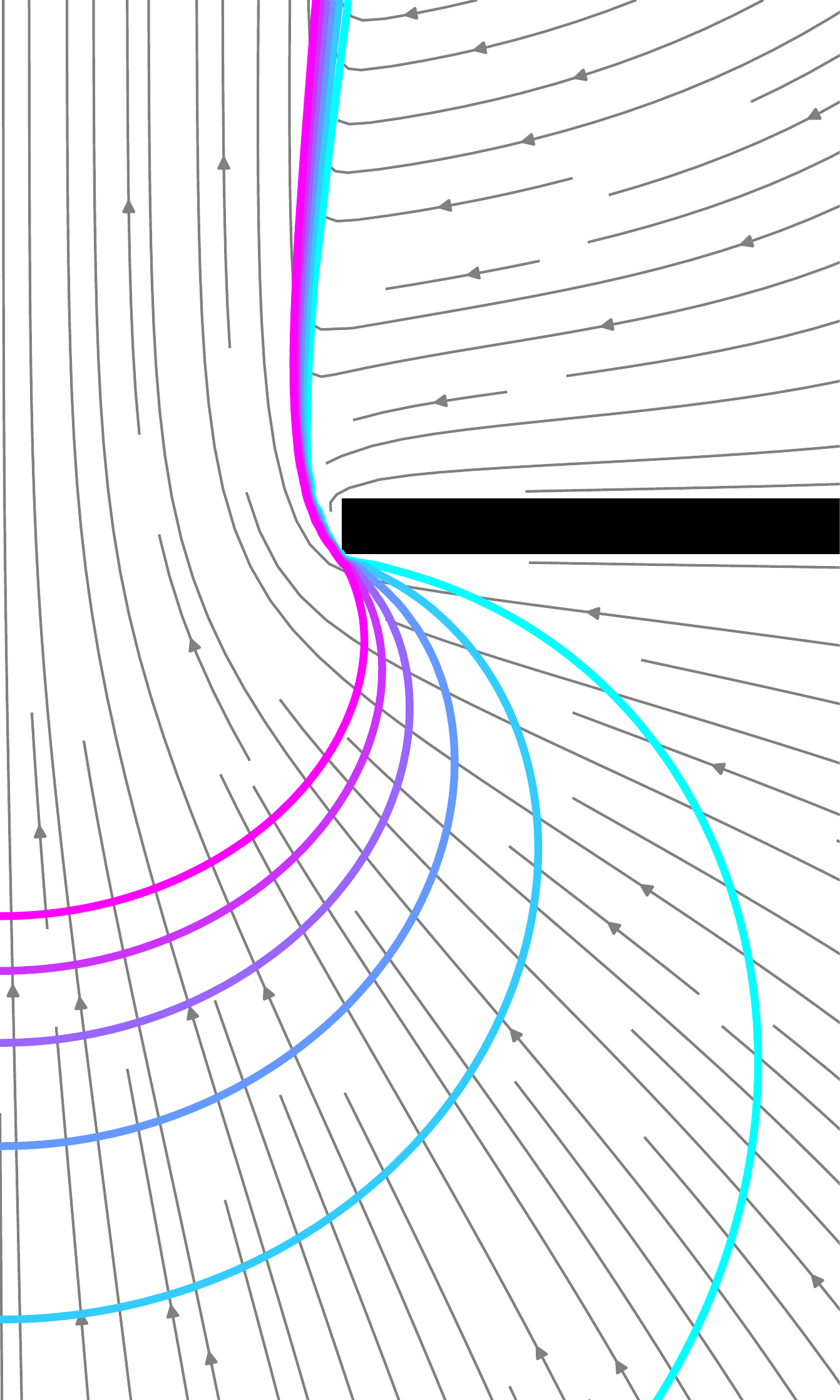};
\end{pgfonlayer}   
\begin{axis}[
                        height=0.5\linewidth,
                        clip=false,
                        colorbar,
                        colormap={cool}{rgb255(0cm)=(0,255,255); rgb255(1cm)=(128,128,255);
                        rgb255(2cm)=(255,0,255);},
point meta max=60, 
                        point meta min=10,  
                        colorbar style={
            at={(0.675,0.65)},               anchor=below south west,    width=0.05*\pgfkeysvalueof{/pgfplots/parent axis width},
            height=0.4*\pgfkeysvalueof{/pgfplots/parent axis width},
            title={\small{$V_{a}$ [cm/s]}},
            title style={yshift=-5pt},
        ytick={10,20,30,40,50,60},
        yticklabel style={
        font=\tiny,
align=right,
            /pgf/number format/.cd,
                fixed,
            fixed zerofill,
            precision=0,
        ,tick style={color=black}}
        },
                        ytick align=outside,
                        xtick align=outside,
                        axis on top,
                        scale only axis,
                        width=1\linewidth,
                        axis equal image,           
                        xlabel={$y$ [mm]},
                        ylabel={$x$ [mm]},
                        xmin=0, xmax=15,
                        ymin=-15, ymax=10,
                        tick style={color=black},
]
\draw[thick,-latex,DarkGreen] (0,0) -- (9.7,-9.7) node[midway,above,rotate=-45,fill=white,draw=none,font=\tiny,opacity=1,text opacity=1]{PISA radius $r$};

\node[DarkGreen,below right,font=\footnotesize,fill=white] at (0,-13.6) {PISA for $V_a$=20cm/s};

\filldraw[DarkGreen] (0,0) circle (2pt);

\filldraw[DarkGreen] (0,-13.6) circle (2pt);

\draw[DarkGreen,ultra thick,dashed] (13.6,0) arc (0:-90:13.6) ;

\draw[draw=black,fill=white] (7.5,1.5) rectangle ++(6.6,-6.7);   
\node[below right,fill=white] at (0,10) {(a)};
    \end{axis}
\end{tikzpicture}     \begin{tikzpicture}[baseline,trim axis right,]
                \begin{axis}[yshift=0.3\linewidth,
                        tick style={color=black},
                        colormap={cool}{rgb255(0cm)=(0,255,255); rgb255(1cm)=(128,128,255);
                        rgb255(2cm)=(255,0,255);},
                cycle list={[samples of colormap={5 of cool}]},
                    width=0.45\linewidth,
                    height=0.3\linewidth,
                    xlabel={aliasing velocity $V_a$ [cm/s]},
                    ylabel={deviation in RFlow [\%]},
                    xmin=10, xmax=60,
                    ymin=-100, ymax=30,
                    legend style={at={(0,0)},anchor=south west,font=\tiny,fill=none,draw=none},
legend entries={4.7 mm,8.7 mm,12.2 mm,16.0 mm,20.0 mm},
legend image code/.code={
    \fill [#1] (0cm,-0.1cm) rectangle (0.6cm,0.1cm);
},
legend cell align={right}
                ]
\addplot[fill=grey!50!white,draw=none,forget plot]
coordinates {
(31.2,30) (31.2,-100) (47.5,-100) (47.5,30)
};
\addplot +[thick]
coordinates {     
(10.000000000,22.616776343)(12.500000000,15.552988379)(15.000000000,11.457929659)(17.500000000,8.788673402)
(25.000000000,4.186497945)(27.500000000,3.232789228)(30.000000000,2.427053835)(32.500000000,1.723518008)(35.000000000,1.101123256)(37.500000000,0.538061981)(40.000000000,0.015026595)(42.500000000,-0.466600002)(45.000000000,-0.914521246)(47.500000000,-1.344719795)(50.000000000,-1.761510977)(52.500000000,-2.147206665)(55.000000000,-2.534203375)(57.500000000,-2.902108752)(60.000000000,-3.241767440)
};

\addplot +[thick]
coordinates {
(10.000000000,19.964024773)(12.500000000,12.252108137)(15.000000000,7.528548725)(17.500000000,4.417797397)(20.000000000,1.538696110)(22.500000000,-0.417433492)(25.000000000,-2.202847728)(27.500000000,-3.775808082)(30.000000000,-5.201944287)(32.500000000,-6.522956531)(35.000000000,-7.767231651)(37.500000000,-8.945593112)(40.000000000,-10.086960549)(42.500000000,-11.177822381)(45.000000000,-12.249753975)(47.500000000,-13.283355424)(50.000000000,-14.310787265)(52.500000000,-15.294107756)(55.000000000,-16.338731277)(57.500000000,-17.260809821)(60.000000000,-18.078893307)
};

\addplot +[thick]
coordinates {
(10.000000000,16.215325812)(12.500000000,7.740979932)(15.000000000,2.128087588)(17.500000000,-2.495386831)(20.000000000,-5.356389863)(22.500000000,-8.260381397)(25.000000000,-10.851250346)(27.500000000,-13.221297488)(30.000000000,-15.450316299)(32.500000000,-17.587238966)(35.000000000,-19.615088233)(37.500000000,-21.595881007)(40.000000000,-23.492348062)(42.500000000,-25.359651309)(45.000000000,-27.196286980)(47.500000000,-28.999964614)(50.000000000,-30.804707001)(52.500000000,-32.903692454)(55.000000000,-34.706571620)(57.500000000,-35.955016855)(60.000000000,-37.649045138)
};

\addplot +[thick]
coordinates { 
(10.000000000,10.726908927)(12.500000000,1.014894541)(15.000000000,-5.753330475)(17.500000000,-11.109978351)(20.000000000,-15.519561604)(22.500000000,-19.548584784)(25.000000000,-23.254155607)(27.500000000,-26.738423056)(30.000000000,-30.054557082)(32.500000000,-33.243018060)(35.000000000,-36.322970094)(37.500000000,-39.311617013)(40.000000000,-42.238797504)(42.500000000,-45.450460110)(45.000000000,-47.881820122)(47.500000000,-50.584940881)(50.000000000,-53.230779601)(52.500000000,-55.828281655)(55.000000000,-58.371980563)(57.500000000,-60.861357939)(60.000000000,-63.295866283)
};

\addplot +[thick]
coordinates {  
(10.000000000,3.363102924)(12.500000000,-7.885186583)(15.000000000,-16.964546143)(17.500000000,-22.867790830)(20.000000000,-28.814126607)(22.500000000,-34.230897171)(25.000000000,-39.283301702)(27.500000000,-44.062161689)(30.000000000,-48.619576797)(32.500000000,-52.986273925)(35.000000000,-57.647893424)(37.500000000,-61.208245417)(40.000000000,-65.066349059)(42.500000000,-68.776720182)(45.000000000,-72.333879489)(47.500000000,-75.733832284)(50.000000000,-78.969381480)(52.500000000,-82.032802030)(55.000000000,-84.914820753)(57.500000000,-87.603252597)(60.000000000,-90.084366446)
};

\addplot[dashed,grey]
coordinates {
(0,0) (60,0)
};

\addplot[dashed,grey]
coordinates {
(0,20) (60,20)
};
\addplot[dashed,grey]
coordinates {
(0,-20) (60,-20)
};

\node[font=\tiny,anchor=east,left,rotate=-1] at (60,3) {circle-S};
\node[font=\tiny,anchor=east,left,rotate=-3] at (60,-11) {circle-M};
\node[font=\tiny,anchor=east,left,rotate=-7] at (60,-31) {circle-L};

\draw[stealth-stealth] (22,-20) -- (22,20) node [midway,right,font=\tiny]{\contour{white}{$\pm 20 \%$}};

\draw[stealth-stealth] (31.2,17) -- (47.5,17) node [midway,above,font=\tiny]{physicians choice};

\draw [thick,-latex] plot [smooth] coordinates {(44,10)  (39,-75)} node [below,font=\tiny] {larger orifice};
\node[below left,fill=white] at (60,30) {(b)};
\end{axis}

                \begin{axis}[
                        tick style={color=black},
width=0.45\linewidth,
                    height=0.3\linewidth,
                    xlabel={aliasing velocity $V_a$ [cm/s]},
                    ylabel={deviation in RFlow [\%]},
                    xmin=10, xmax=60,
                    ymin=-50, ymax=30,
                    legend style={at={(0,0)},anchor=south west,font=\tiny,fill=none,draw=none},
legend entries={circle-L,slit-L,drop-L},
legend image code/.code={
    \fill [#1] (0cm,-0.1cm) rectangle (0.6cm,0.1cm);
},
legend cell align={right}
                ]
\addplot[fill=grey!50!white,draw=none,forget plot]
coordinates {
(31.2,30) (31.2,-100) (47.5,-100) (47.5,30)
};
\addplot [red,thick]
coordinates {     
(10.000000000,16.215325812)(12.500000000,7.740979932)(15.000000000,2.128087588)(17.500000000,-2.495386831)(20.000000000,-5.356389863)(22.500000000,-8.260381397)(25.000000000,-10.851250346)(27.500000000,-13.221297488)(30.000000000,-15.450316299)(32.500000000,-17.587238966)(35.000000000,-19.615088233)(37.500000000,-21.595881007)(40.000000000,-23.492348062)(42.500000000,-25.359651309)(45.000000000,-27.196286980)(47.500000000,-28.999964614)(50.000000000,-30.804707001)(52.500000000,-32.903692454)(55.000000000,-34.706571620)(57.500000000,-35.955016855)(60.000000000,-37.649045138)
};

\addplot [DarkGreen,thick]
coordinates {
(10.000000000,11.986285817)(12.500000000,2.699991814)(15.000000000,-3.568282228)(17.500000000,-9.001282149)(20.000000000,-12.383887748)(22.500000000,-15.807764949)(25.000000000,-18.909875697)(27.500000000,-21.707578004)(30.000000000,-24.334690148)(32.500000000,-26.830691576)(35.000000000,-29.141108058)(37.500000000,-31.413121449)(40.000000000,-33.539601025)(42.500000000,-35.592198174)(45.000000000,-37.582582252)(47.500000000,-39.510492999)(50.000000000,-41.382980654)(52.500000000,-43.211406626)(55.000000000,-44.982571597)(57.500000000,-46.676017999)(60.000000000,-48.335821746)
};

\addplot [blue,thick]
coordinates {
(10.000000000,12.512326104)(12.500000000,3.446989900)(15.000000000,-2.527006361)(17.500000000,-7.969588091)(20.000000000,-10.754674979)(22.500000000,-13.883722070)(25.000000000,-16.628109431)(27.500000000,-19.143363023)(30.000000000,-21.454708445)(32.500000000,-23.620885452)(35.000000000,-25.690018456)(37.500000000,-27.628765427)(40.000000000,-29.479462892)(42.500000000,-31.274452575)(45.000000000,-33.008135510)(47.500000000,-34.684772755)(50.000000000,-36.315616557)(52.500000000,-37.916861334)(55.000000000,-39.499516862)(57.500000000,-40.966283001)(60.000000000,-42.433651913)
};

\addplot[dashed,grey]
coordinates {
(0,0) (60,0)
};

\addplot[dashed,grey]
coordinates {
(0,20) (60,20)
};
\addplot[dashed,grey]
coordinates {
(0,-20) (60,-20)
};

\draw[stealth-stealth] (22,-20) -- (22,20) node [midway,right,font=\tiny,fill=white]{\contour{white}{$\pm 20 \%$}};

\draw[stealth-stealth] (31.2,7) -- (47.5,7) node [midway,above,font=\tiny]{physicians choice};
\node[below left,fill=white] at (60,30) {(c)};
\end{axis}
\end{tikzpicture}     \caption{Left panel (a): streamlines with isolines extracted from $V_x$ at different aliasing velocities $V_\mathrm{a}$ overlayed with the PISA estimation for the circular large orifice (\textbf{circle-L}) at $V_\mathrm{a}=20$ cm/s.
    Right panel: deviation introduced by the choice of aliasing velocity threshold into the PISA estimation at the constant flow rate of \rflow$_\text{CFD}=244.9$~ml/s for (b) circular orifice with various size and (c) three large orifices with different shape.
    \label{fig:cfd3}}
\end{figure*}
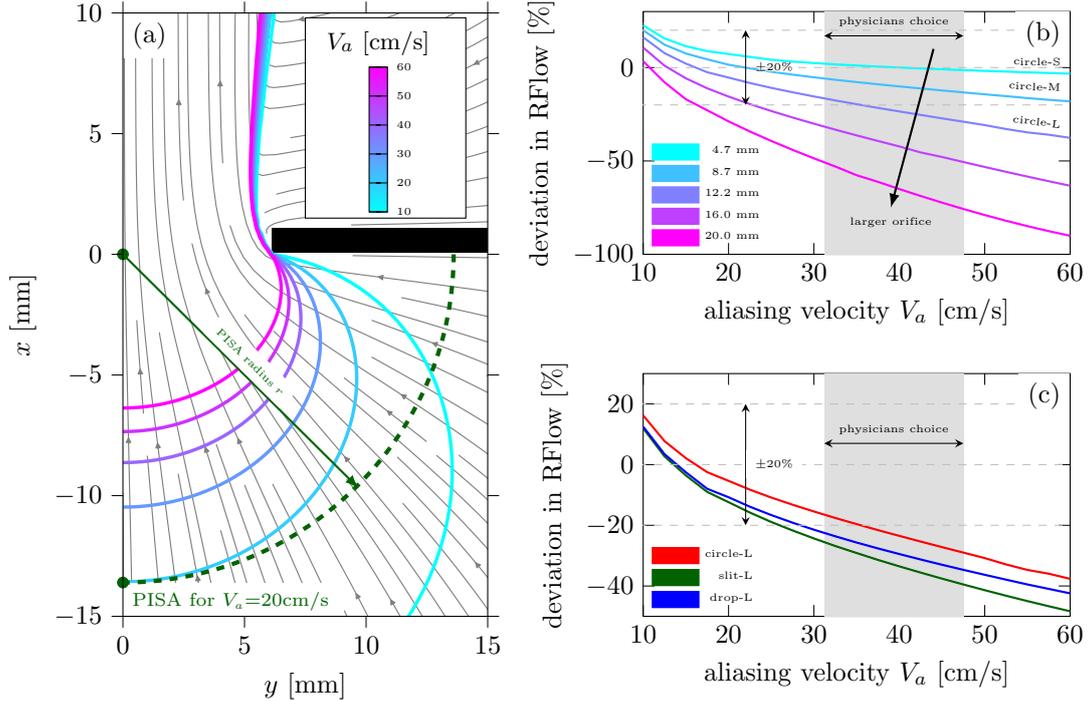

Figure~\ref{fig:cfd3}(a) exemplarly presents the velocity field from the CFD simulation for large circular orifice (\textbf{circle-L}) overlayed with several isovelocity contours.
Those contours mark the velocity envelopes observed during the TEE measurement (see Fig.~\ref{fig:US}) extracted as the isosurface only for the streamwise velocity component $V_x$ at various thresholds $V_\mathrm{a}$ in the range from 10 to 20 cm/s. 
This corresponds to the configuration when the ultrasound beam is aligned with the main flow direction $x$.
The PISA radius $r$ extracted from the velocity field spans the distance from the orifice center to the position along the $x$-axis where the local velocity $V_x$ is equal to the chosen aliasing velocity $V_\mathrm{a}$.
The hemispherical approximation of PISA is marked with the green dashed line.
Please note that all presented isovelocity contours neither coincide with the hemispherical PISA, nor exhibit a surface featuring velocity vectors oriented perpendicular to the surface.
This discrepancy arises from the inherent assumptions embedded within the PISA estimation method and is well documented~\cite{qin2021computational,this2016proximal,mao2020comparative}.
In reality, however, a misalignment of the ultrasonic beam can also introduce a significant error into the estimation of PISA radius especially for non-axisymmetric flows and flows through complex orifice geometries. 
For further information on the misalignment error the reader is reffered to the publication by Qin et. al~\cite{qin2021computational}.

\subsubsection{Effect of Aliasing Velocity and Orifice Shape}

The importance of appropriate selection for aliasing velocity $V_\mathrm{a}$ is well recognized in the literature~\cite{this2020pipeline,mao2020comparative,lee2023impact,jamil2017feasibility}.
At the same time, however, the recommended range for the choice of the aliasing velocity is very broad, ranging from 15 to 60~cm/s~\cite{chandra2011three,bargiggia1991new,biner2010reproducibility,lambert2007proximal}.
The quantification of the \rflow{} error introduced into PISA-based estimation by the choice of aliasing velocity for circular geometries of various sizes is shown in Figure~\ref{fig:cfd3}(b).
Here we add two additional simulations for circular orifices with the diameter of 16 and 20~mm in order to better visualize the trends.
The calculation based on higher aliasing velocities tends to immensely underestimate the \rflow{} for large orifices, while an overestimation is observed at lower aliasing velocities, especially for smaller orifices.
The underestimation is, however, much more severe for the larger orifices. 
In order to maintain a reasonable estimation error within $\pm 20\%$, the aliasing velocity should 
not exceed 35, 22 and 16~cm/s for the three largest circular geometries; for the smallest two, any chosen $V_\mathrm{a}$ remains within the estimation error of $\pm 20\%$.
Based on the procedure presented in Appendix~\ref{sec:app_opt_va}, the optimal aliasing velocities for the considered circular orifices are estimated to be 39, 22 and 16 cm/s for \textbf{circle-S, -M, -L}, respectively.
It should be noted that the aliasing velocity was adjusted for every specific geometry size only by one out of three physicians, while the other two kept the velocity constant throughout the measurement.
At the same time the chosen velocities are mostly higher than the optimal ones, inherently leading to an underestimation of \rflow{}.

These findings might also shed light on the single MROP in our study (\textbf{circle-S}), where a higher \rvol{} for echocardiography measurement (6~ml) has been observed than for the PIV estimation (4.6~ml).
We hypothesize that this might originate from the overestimation in \rflow{}, which is observed when the chosen aliasing velocity is lower than the optimal one, especially for the small considered geometries.
In the case of \textbf{circle-S} $V_\mathrm{a}<39$ cm/s an overestimation in \rflow{} is observed, which increases for lower $V_\mathrm{a}$ and can exceed 20\% at $V_\mathrm{a}=10$ cm/s (Figure~\ref{fig:cfd3}b).

Figure~\ref{fig:cfd3}(c) presents the introduced error at different aliasing velocities for three considered shapes with the largest cross-sections (\textbf{circle-L}, \textbf{slit-L}, \textbf{drop-L}).
The error increases for non-circular shapes by 5-10\% depending on the chosen $V_\mathrm{a}$.
It is also evident that the error for the drop-shaped orifice falls between that of circular and slit-shaped orifices, indicating a flow behavior more akin to the circular orifice than the slit-shaped MROP.
The optimal aliasing velocity has been estimated to be 16, 13 and 14 cm/s for \textbf{circle-}, \textbf{slit-} and \textbf{drop-L}, correspondingly.

\section{Discussion}
We first discuss important findings from a fluid dynamic perspective and will subsequently put them in comparison to other measurement and simulation techniques. 

\subsection{Fluid dynamic findings}
The evaluation of the instantaneous velocity field in Figure \ref{fig:pinhole13_phase_020} gives evidence that the flow develops 
Kelvin-Helmholtz instabilities directly behind the outlet. 
The most dominant vortical structure is the main starting vortex traveling downstream to the opposite side of the atrium and eventually interacting with the outer wall. 
Overall, the starting vortex of the jet plays a dominant role in the velocity distribution and its manipulation caused by geometrical variances is worth further investigation.
The saddle-backed exit velocity profiles shown in Figure \ref{fig:velocityprofiles}~(a) are a known characteristics for special orifice geometries like the orifices used in the present study \cite{Tsuchiya1986, Deo2007} as well as for pulsating pipe flows in general \cite{Womersley1955}.
For sharp-corner orifice geometries this is attributable to the formation of a vena contracta \cite{Quinn1989}. A clear indication for this cause is the time-resolved evolution of the velocity profiles shown in Figure \ref{fig:velocityprofiles}(a). The saddle-backed profiles are not present for low velocities or the reverse flow, but develop predominately for the highest outlet velocities.

\subsection{Comparative findings}
In this study the accuracy of the flow convergence method was assessed in a hemodynamic reproducible \textit{in-vitro} environment. Quantitative assessment was done by juxtaposing the flow convergence method and PIV to classify mitral regurgitation by measuring the \rvol{} produced by different MV phantoms.

The most obvious finding is that the \rvol{} determined by physicians with the flow convergence methods underestimates the \rvol{} for eight out of nine MROPs. 
This finding aligns well with the previous findings~\cite{Lancellotti.2010, Iwakura.2006, Coisne.2002}, which report an underestimation of up to 44\%. 
In the present study 
the underestimation reached 52\% for large geometries such as \textbf{slit-L}.

For all large orifices this underestimation would ascertain mild severity instead of moderate severity for mitral regurgitation, which might potentially affect therapy decisions. This is clinically significant because patients with moderate mitral valve regurgitation have a higher 1-year mortality rate (15-45\% vs. 7\%)~\cite{Cioffi.2005}. These patients should be monitored annually according to ESC/EACTS guidelines as opposed to every 3-4 years~\cite{Vahanian.2022}, and may require mitral valve repair earlier -- even in the absence of symptoms~\cite{EnriquezSarano.1994}. 
For mild mitral regurgitation present at mid- and small-sized apertures, the difference between the two methods was small in absolute values.
In cases of moderate to severe mitral regurgitation, the observed deviation assumes significant proportions and may pose potential challenges.
Although a profound statistical evaluation wasn't possible for the given sample size of physicians, the second finding shows some valuable indications concerning the inter-observer variability. 
This quantity predominantly depends on shape and size.
For all small and medium MROPs the variability remains small in terms of relative error and for the circular orifice shapes the variability might be considered minor to negligible for all sizes. 
For the large slit and drop, however, the variability was significantly higher. 
This raises the question about the robustness of the flow convergence method in the context of elliptical or more complex orifice geometries, particularly given the prevalence of such shapes in patient populations~\cite{Lancellotti.2010}. Moreover, the inter-observer variability in ultrasound estimations further complicates the precise quantification of deviations compared to PIV estimations. This, in turn, makes it difficult to identify a specific geometric property of the orifice that could be correlated with these deviations.

The limitations of PISA-based methods have been also confirmed through CFD simulations.
CFD allows for an accurate extraction of isovelocity contours within the simulation domain and hence can deliver reference PISA estimations for various orifice geometries under different flow conditions.

The trends identified in CFD simulations validate the assertions made by This et al.~\cite{this2020pipeline}, highlighting that elevated aliasing velocities are more likely linked to challenges in accurately assessing \rvol{}.
Considering the results of \rvol{} estimation presented in the previous section in Figure~\ref{fig:rvol-bars}, the trend of underestimation at higher aliasing velocities might partially explain the strong underestimation of \rvol{} for larger orifices (\textbf{L}-size) compared to smaller ones (\textbf{S}- \& \textbf{M}-size) since \rvol{} is proportional to \rflow{} and the values for $V_\mathrm{a}$ chosen by the three physicians are rather high mostly located in the range between 31.2 and 47.5~cm/s, which might translate into the \rflow{} estimation error of 17-30\% for \textbf{circle-L} according to our CFD simulation results.
Furthermore, please note that according to our data the physicians tend to choose a rather high aliasing velocity, which is associated with larger estimation deviations especially for large orifices.
The opposite trend of an overestimation in \rvol{} observed for a single case of \textbf{circle-S} might be similarly linked to the overestimation of \rflow{} even at higher velocities, since the optimal $V_\mathrm{a}$ for these cases is higher than for the larger geometries. Finally, it is demonstrated that the choice of aliasing velocity may have a greater impact on the accuracy of PISA-based \rvol{} estimation than variations in orifice geometry.
This finding, along with the observed inter-observer variability in the selected US input parameters, may account for the underestimation of \rvol{} observed in the US measurements.

In, summary, the present work compares the measurement of regurgitation jet by ultrasound accomplished by three different physicians with three different systems. 
Complementary, PIV extracts the flow field information, which can be used for a ground truth comparison, since a direct velocity measurement of the jet is possible. 
Overall, nine different artificial valve geometries, consisting of three different shapes and three sizes, are taken under consideration. 
A systematic underestimation of the regurgitation volume for large orifice areas has been found, where a violation of the flow convergence assumptions is significant. 
The inter-observer variability becomes larger for these large orifice areas. 
The extraction of PISA for these shapes presents a challenge especially at high aliasing velocity, which are often used by the physicians.
From the flow perspective, an occurring starting vortex was found as a long-lasting dominant flow pattern, which might interact with the left atrial wall.
The present set-up can be used as high-fidelity experimental tool, where also patient-specific heart valves can be compared with ultrasound and PIV, opening the possibility for an in-depth flow analysis with short feedback-loop to actual practicing physicians.

\backmatter

\section*{Acknowledgements}
We gratefully acknowledge the financial support for AS by the Young Investigator Network (YIN) at the Karlsruhe Institute of Technology (KIT) within the YIN start-up grant 2023. The work was partly supported by Informatics for life funded by the Klaus Tschira Foundation.
\section*{Declarations}
The authors declare that they have no conflict of interest.

\section*{Author contribution statements}
RL: conceptualization, data acquisition (PIV), data evaluation (PIV \& CFD), writing--original draft and visualization;
RK: conceptualization, data acquisition (PIV \& US), data evaluation (US), writing--original draft and visualization;
LS: data acquisition (US);
DM: data acquisition (US);
ME: data acquisition (US);
LN: data acquisition (PIV), data evaluation (PIV);
RdS: conceptualization;
GR: conceptualization;
JK: writing-review and editing;
MK: project administration;
CL: project administration; data interpretation (US);
NF: project administration;
BF: supervision, project administration, writing--review and editing;
AS: conceptualization, data acquisition (CFD), data evaluation, writing--original draft, visualization, funding acquisition;
SE: project administration, conceptualization, writing--original draft, Funding acquisition

\bibliography{PIV_literature,PISA_lit,CFD_lit}

\clearpage

\FloatBarrier

\appendix
\section{Appendix}

\subsection{Further information on PIV measurements}
\label{sec:App_PIV}
\begin{figure}
    \centering
    \begin{tikzpicture}
     \node[anchor=south west,inner sep=0] at (0,0){ \includegraphics[width=\linewidth]{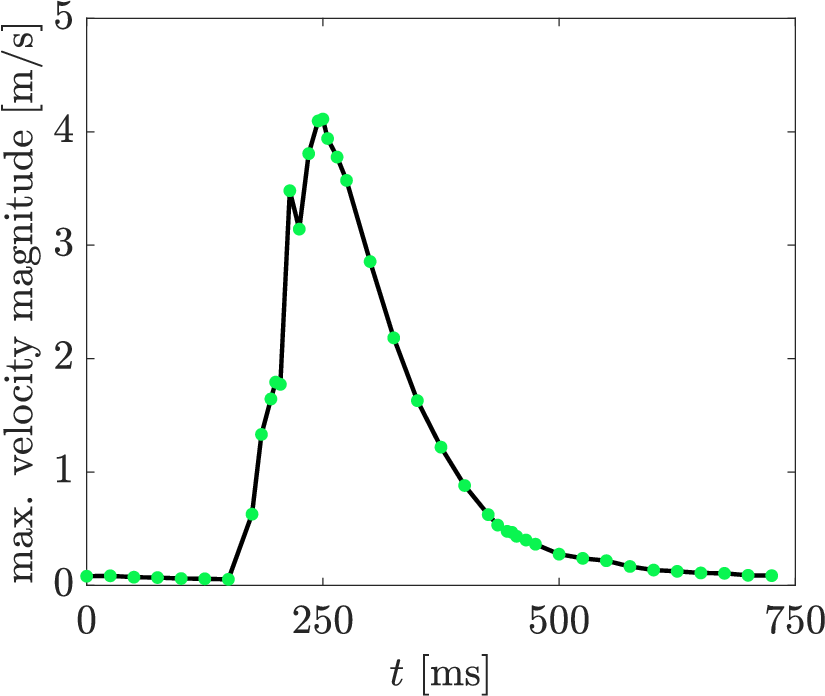}};
      \draw (3.4,4.0) -- (4.0,5.7);
       \draw (3.4,4.0) -- (4.2,4.2);
      \node at (5.5, 5.9) {100 images averaged};
      \node at (5.5, 5.6) {for cardiac-phase};
      \node at (5.5, 5.3) {resolved results};

       \node at (5.5, 4.2) {1 image used for};
      \node at (5.5, 3.9) {instantaneous};
      \node at (5.5, 3.6) {flow field evaluation};

      \node at (4.0,7.45) {1000 images along entire};
      \node at (4.0,7.1) {cycle for regurgitation volume};
                \draw [decorate,
               decoration = {calligraphic brace},thick] (1.0,6.8) --  (7.0,6.8);
     \end{tikzpicture}
    \caption{For each of the 42 phase positions within the cardiac cycle, 100 double-images were recorded and averaged for the \emph{cardiac-phase resolved} evaluation, while only one image was used for an \emph{instantaneous flow field evaluation}. For the regurgitation volume extraction another 1000 images were recorded along the entire cycle.}
    \label{fig:Appendix:PIV_explanation}
\end{figure}
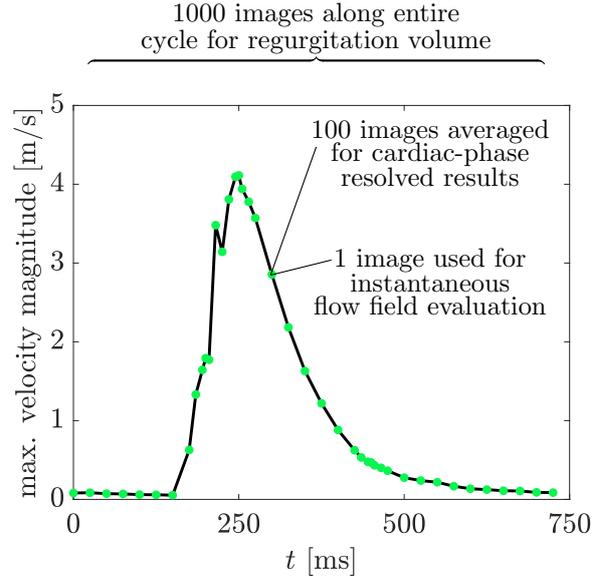

The PIV measurements were carried out as noted in Section \ref{sec:PIV_method}. The following subsection clarifies the difference between the evaluation techniques: 
\begin{itemize}
    \item Extraction of the regurgitation volume with 1,000 images for each geometry and plane respectively. These images were recorded with a frequency of 15 Hz along various phase positions of the cardiac cycle. For the calculation of the regurgitation volume only positive velocity values at the outlet were considered.
    \item \emph{Cardiac-phase resolved} evaluation: For this purpose Figure \ref{fig:Appendix:PIV_explanation} is modified from Figure \ref{fig:velocityprofiles} to visually explain the different evaluation techniques.
The figure shows each point were \emph{cardiac-phase resolved} PIV measurements were conducted. This is a number of 42 different phase positions within the cardiac cycle. For each point 100 double-images were recorded and averaged to increase the statistical significance. This evaluation method is used in Figure~\ref{fig:velocityprofiles}.
\item For the extraction of an \emph{instantaneous flow field} one random sample of these 100 double-images was chosen to also account for a sufficient contribution of the instabilities, which aren't visible in phase-resolved results. This evaluation method is used in Figure~\ref{fig:pinhole13_phase_020}.
\end{itemize}

\subsection{\rvol{} determination via PIV}
\label{sec:App_Rvol_PIV}
\begin{figure}
    \centering
    \begin{tikzpicture}
     \draw[loosely dashed,very thin] (0.125,0.5) -- (0.125,5.5) ;
     \node[anchor=south west,inner sep=0] at (0,0){ \includegraphics[width=0.2\linewidth]{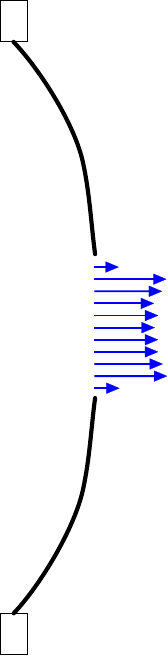}};
      \node at (1.4, 5.7) {side view};
      \node at (5, 5.7) {front view};

    \draw[-latex,thin] (0.125,3) -- (3,3) node[below]  {$x$};
        \draw[-latex,thin] (0.125,3) -- (0.125,5) node[left] {$y$};
\draw[*-,thin,DarkGreen] (0.45,1) -- (1.25, 0.75) node[align=center,right] {bent \\ MROP};
        \node[rotate=90,font=\small,DarkGreen,fill=white] at (0.6,3) {orifice};
        \node at (2.6, 3.5) {\textcolor{blue}{$u(x_0,y,z,t)$}};

      \draw[-latex,thin] (5,3) -- (6,3) node[below]  {$z$};
        \draw[-latex,thin] (5,3) -- (5,5) node[left] {$y$};

 \draw[latex-latex,thin,DarkGreen] (0.15,4.7) -- (0.85,4.7) node[right] {$x_0$};        

\draw[very thin,blue!30!white] (0.875,1.25) -- (0.875,4.75);
\draw[dotted,very thick,blue] (0.875,1.5) -- (0.875,4.5);

\draw[very thin,blue!30!white] (5,1.25) -- (5,4.75);
\draw[dotted,very thick,blue] (5,1.5) -- (5,4.5);
\draw[very thin,blue!30!white](5.3,1.25) -- (5.3,4.75);
\draw[dotted,very thick,blue] (5.3,1.5) -- (5.3,4.5);
\draw[very thin,blue!30!white] (4.7,1.25) -- (4.7,4.75);
\draw[dotted,very thick,blue] (4.7,1.5) -- (4.7,4.5);
\draw[very thin,blue!30!white](5.15,1.25) -- (5.15,4.75);
\draw[dotted,very thick,blue] (5.15,1.5) -- (5.15,4.5);
\draw[very thin,blue!30!white](4.85,1.25) -- (4.85,4.75);
\draw[dotted,very thick,blue] (4.85,1.5) -- (4.85,4.5);

\begin{scope}[rotate=90,xshift=-0.5cm,yshift=-5cm,scale=0.7]
     \draw[fill=none,ultra thick]  (4.4,0.45) arc(90:270:0.45);
 \draw[ultra thick] (4.4,0.45)--(4.4+1.53,0);
 \draw[ultra thick] (4.4,-0.45)--(4.4+1.53,0);
 \end{scope}

\draw [decorate,               decoration = {calligraphic brace,mirror},thick] (4.5,1.0) --  (5.5,1.0) node[below,midway,align=center,yshift=-6pt]  {measurement \\ planes};

     \end{tikzpicture}     \caption{Schematic of the extracted PIV data for the calculation of \rvol{} for \textbf{drop-M} MROP: $u(x_0,y,z,t)$ during systole is used for further processing.}
    \label{fig:Appendix:PIV_explanation2}
\end{figure}
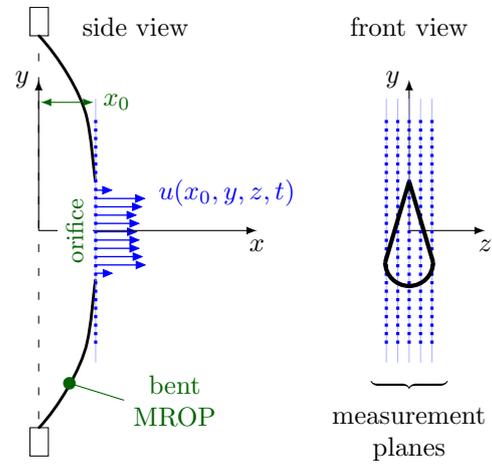

In the following, a step-by-step guidance from the extracted velocity data to the calculation of the regurgitation volume is provided. As exemplary case the MROP \textbf{drop-M} is chosen. As indicated in Figure \ref{fig:Appendix:PIV_explanation2} (front view), five $z$-planes are measured for this geometry, resulting in 5,000 raw-data double images.
Each double image is evaluated with the mentioned PIV software (\textit{PIVview}) and the extracted velocity data set is loaded into \textit{MATLAB}, where further post-processing takes place. The vector fields are rotated by $29^\circ$, such that the main flow direction is aligned with the $x$-axis. 
For the determination of the regurgitation volume the velocity in stream-wise direction as close to the outlet as possible, is used. For the present case of \textbf{drop-M} a distance of $x_0=1.1\,$mm from the outlet is used, which corresponds to two interrogation area distances. Since an overlap of 50\% is used, this is the first interrogation area without contact to the solid. 
The selected location represents an optimal balance between two competing factors: reflections at the MROP and flow entrainment. To minimize the inclusion of entrained fluid in the measurements, positioning the control volume boundary as far upstream as possible is ideal, though this introduces the risk of erroneous local velocity vectors caused by laser-light reflections.

\begin{figure}[h]
    \centering
    \includegraphics[width=\linewidth]{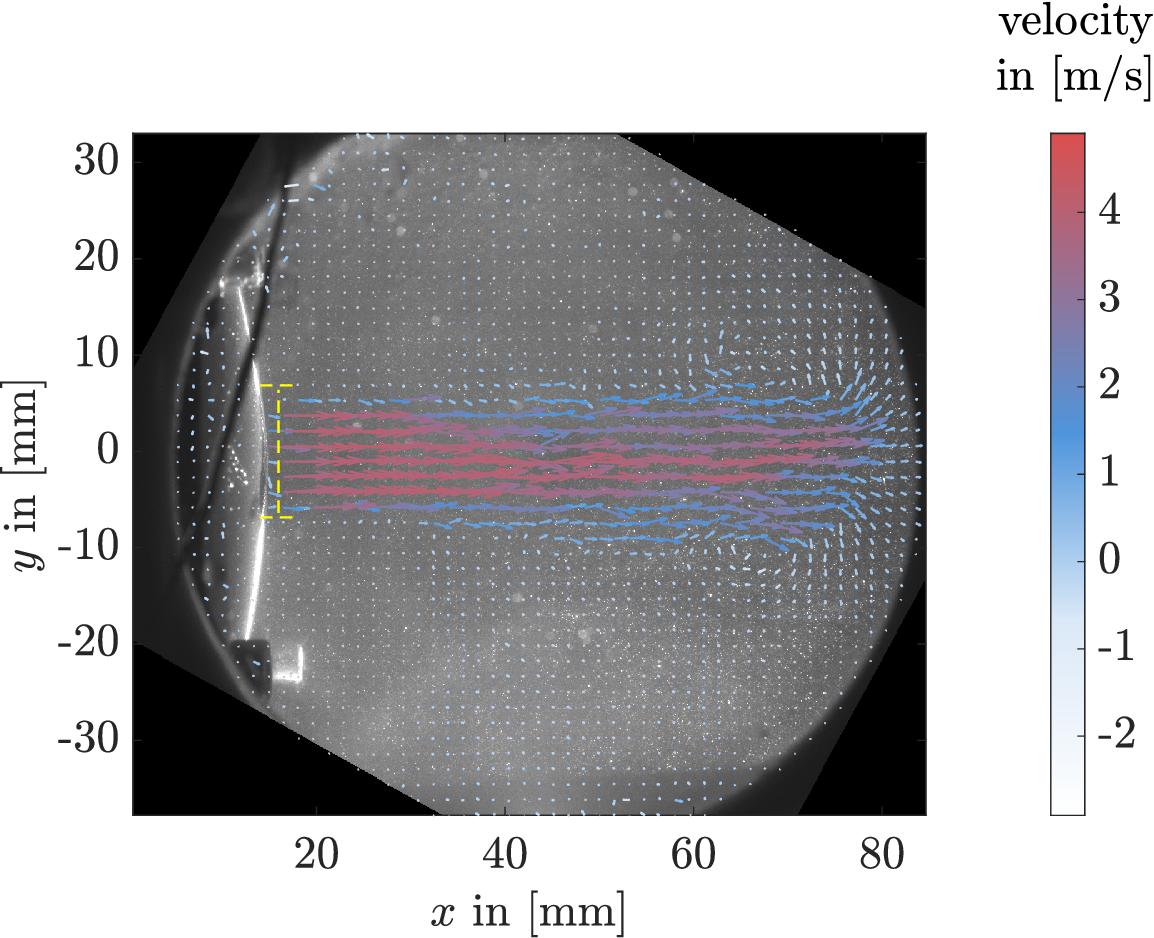}
    \caption{Instantaneous flow field for \textbf{drop-M} as colored vector plot overlayed with the raw data image. The integration limits and the $x$-position of the \rvol{} extraction are indicated with yellow lines.}
    \label{fig:app:drop_M_single}
\end{figure}

Figure \ref{fig:Appendix:PIV_explanation2} shows the PIV data handling with the streamwise velocity profile $u(x_0,y,z,t)$, which is used for spatio-temporal integration. 
This velocity is extracted for each $z$-plane and the volumetric flow rate $\dot{V}(t)$ is calculated as follows:
\begin{equation}
   \dot{V}(t)=\iint u(x_0,y,z,t)\mathrm{d}y \mathrm{d}z.
\end{equation}
 For a precise \rvol{} determination only positive velocity values, i.e. flow in the positive $x$-direction are considered, which represent the values during systole. The volume flow for each time step is then integrated in time to calculate \rvol{}:
 \begin{equation}
  \rvol{}=\int \dot{V}(t) \mathrm{d}t.
\end{equation}
For the exemplary case of the MROP \textbf{drop-M} Figure~\ref{fig:app:drop_M_single} shows an instantaneous flow field with the limits of the integration in $y$-direction and the $x$-position of the velocity extraction.

\subsection{PIV uncertainty estimation}
\label{sec:App_Uncertainty}
Figure \ref{fig:umax_piv_us} and \ref{fig:rvol-bars} indicate a $2 \sigma$-confidence interval for the maximum velocity at the outlet $V_\mathrm{max}$ and the regurgitation volume \rvol{}. For every MROP 1,000 snapshots were evaluated. These snapshots have different cardiac-cycle positions, as explained in \ref{sec:App_PIV}, which renders a straight-forward implementation of statistical moments from this temporally undersampled data set difficult. 
Instead, the standard deviation was calculated by dividing the 1,000 individual sets into 10 blocks with a number of 100 individual snapshots. For each of these 10 blocks the value of $V_\mathrm{max}$ and value of \rvol{} was calculated. In order to ensure a statistically significant number of recorded jet events across the entire recorded sample number of 1,000 snapshots, the standard deviation  $\sigma$ across the 10 determined statistical mean values was subsequently determined. This statistically more meaningful uncertainty estimation is represented by the error bars in the respective figures, which include a $2 \sigma$ region or $95.4 \%$ of Gaussian distributed data points.

\subsection{Computational Fluid Dynamics}
\label{sec:App_CFD}

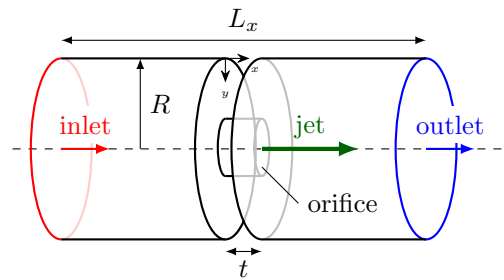
\begin{figure}[h]\centering
\begin{tikzpicture}[scale=0.8,thick]

\draw[thin,dashed] (-0.3,-0.5) -- (7.8,-0.5);

\draw[thin,latex-latex] (0.5,1.3) -- (6.5,1.3) node [above,midway]{$L_x$};

\draw[thin,latex-latex] (3.2,-2.2) -- (3.8,-2.2) node [below,midway]{$t$};

\draw[thin,-latex] (1.8,-0.5) -- (1.8,1) node [right,midway]{$R$};

\draw[gray] (3.8,0) arc (90:270:0.125 and 0.475);
\draw[gray] (3.8,0) arc (90:-90:0.125 and 0.475);

\draw[gray] (3.31,0) -- (3.8,0);
\draw[gray] (3.31,-0.95) -- (3.8,-0.95);

\draw[red!20!white] (0.5,1) arc (90:-90:0.5 and 1.5);
\draw[gray] (3.8,1) arc (90:-90:0.5 and 1.5);
\draw[gray] (3.2,1) arc (90:-90:0.5 and 1.5);

\draw[blue] (6.5,1) arc (90:270:0.5 and 1.5);
\draw[blue] (6.5,1) arc (90:-90:0.5 and 1.5);
\draw (3.8,1) arc (90:270:0.5 and 1.5);

\draw (3.2,1) arc (90:270:0.5 and 1.5);

\draw (3.2,1) arc (90:52:0.5 and 1.5);
\draw (3.2,-2) arc (-90:-52:0.5 and 1.5);

\draw[red] (0.5,1) arc (90:270:0.5 and 1.5);

\draw (3.2,0) arc (90:270:0.125 and 0.475);

\draw (3.2,0) -- (3.31,0);
\draw (3.2,-0.95) -- (3.31,-0.95);

\draw (0.5,1) -- (3.2,1);
\draw (0.5,-2) -- (3.2,-2);

\draw (3.8,1) -- (6.5,1);
\draw (3.8,-2) -- (6.5,-2);

\draw[-latex,red] (0.5,-0.5) -- (1.3,-0.5) node [above,midway,fill=white,yshift=2pt]{inlet};
\draw[-latex,blue] (6.5,-0.5) -- (7.3,-0.5) node [above,midway,fill=white,yshift=2pt]{outlet};
\draw[thin] (3.8,-0.8) -- (4.4,-1.4) node[right] {orifice};

\draw[-latex,DarkGreen,ultra thick] (3.8,-0.5) -- (5.4,-0.5) node [above,midway,yshift=0pt]{jet};

\draw[thin,-Stealth] (3.2,1) -- (3.6,1) node[font=\tiny,below,xshift=2pt]{\contour{white}{$x$}};
\draw[thin,-Stealth] (3.2,1) -- (3.2,0.6) node[font=\tiny,below]{$y$};

\end{tikzpicture} \caption{Sketch of the considered fluid mechanical environment in CFD simulations.\label{fig:simconf}}
\end{figure}
In order to assess the effects of assumptions introduced in the flow convergence method and especially the estimation of \rflow{} through PISA we consider a simplified numerical simulation of a flow through orifice as depicted in Figure~\ref{fig:simconf}.
The simulation domain ($L_x=40$~cm) consists of two cylindrical chambers with $R=5$~cm connected in the middle of the domain by the orifice of variable geometry (thickness $t=1$mm, geometry description see Table~\ref{tab:shapelist}).
We consider an incompressible steady state flow with a constant flow rate of \rflow$_\text{CFD}=244.9$~ml/s applied at the inlet boundary.
The flow rate represents the maximal velocity $V_\mathrm{max}$ in the peak of systole (Figure~\ref{fig:sketch_isovelocity}) and is chosen based on the results of the PIV measurement for the MROP \textbf{circle-L}, so the same magnitude of the velocity is achieved in the regurgitation jet ($V\approx 3.4$~m/s).
Please note, that this chosen reference flow rate might change for different configurations, since it depends on the choice of boundary conditions and the resistance introduced by the investigated orifice. 
At the outlet the zero-gradient velocity boundary condition is applied, while the no-slip boundary condition is applied for the side walls.
For the pressure field the zero-gradient boundary condition is applied at the inlet and the side walls of the domain, while a constant fixed pressure is prescribed at the outlet.
To accelerate the calculations the simulation domain is considered to be axisymmetric for the circular orifice; for the non-axisymmetric orifice geometry of the slit- and the drop-shape symmetrical quarter- and half-domain configurations are utilized, respectively.
The simulations are carried out with the open-source solver \texttt{simpleFoam} from the OpenFOAM v2306 simulation framework~\cite{weller1998tensorial}.
The solver utilizes RANS equations based on the $k$-$\omega$-SST turbulence model~\cite{menter2003ten} with the wall-resolved approach.
The solution algorithm is terminated after the normalized residuals reach $1\cdot~10^{-6}$ and $1\cdot 10^{-5}$ for the velocity and the pressure field, respectively.
The fluid properties are chosen in correspondence to that of the fluid utilized in the experiment ($\nu=3.3\cdot 10^{-6}$\,m\textsuperscript{2}/s).
The mesh generation is performed utilizing the OpenFOAM tool \texttt{snappyHexMesh}.
The domain-size and resolution independence are ensured through a prior grid-independence and domain-size study.

\subsection{CFD Estimation of the Optimal Aliasing Velocity}
\label{sec:app_opt_va}
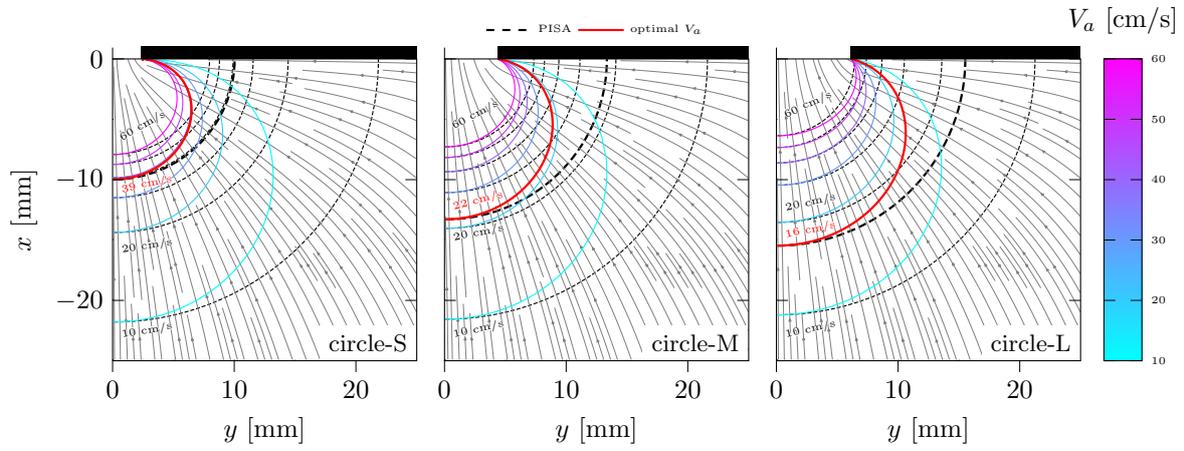
\begin{figure*}[h]
\begin{tikzpicture}   
\begin{pgfonlayer}{background}
\includegraphics[width=0.25\textwidth]{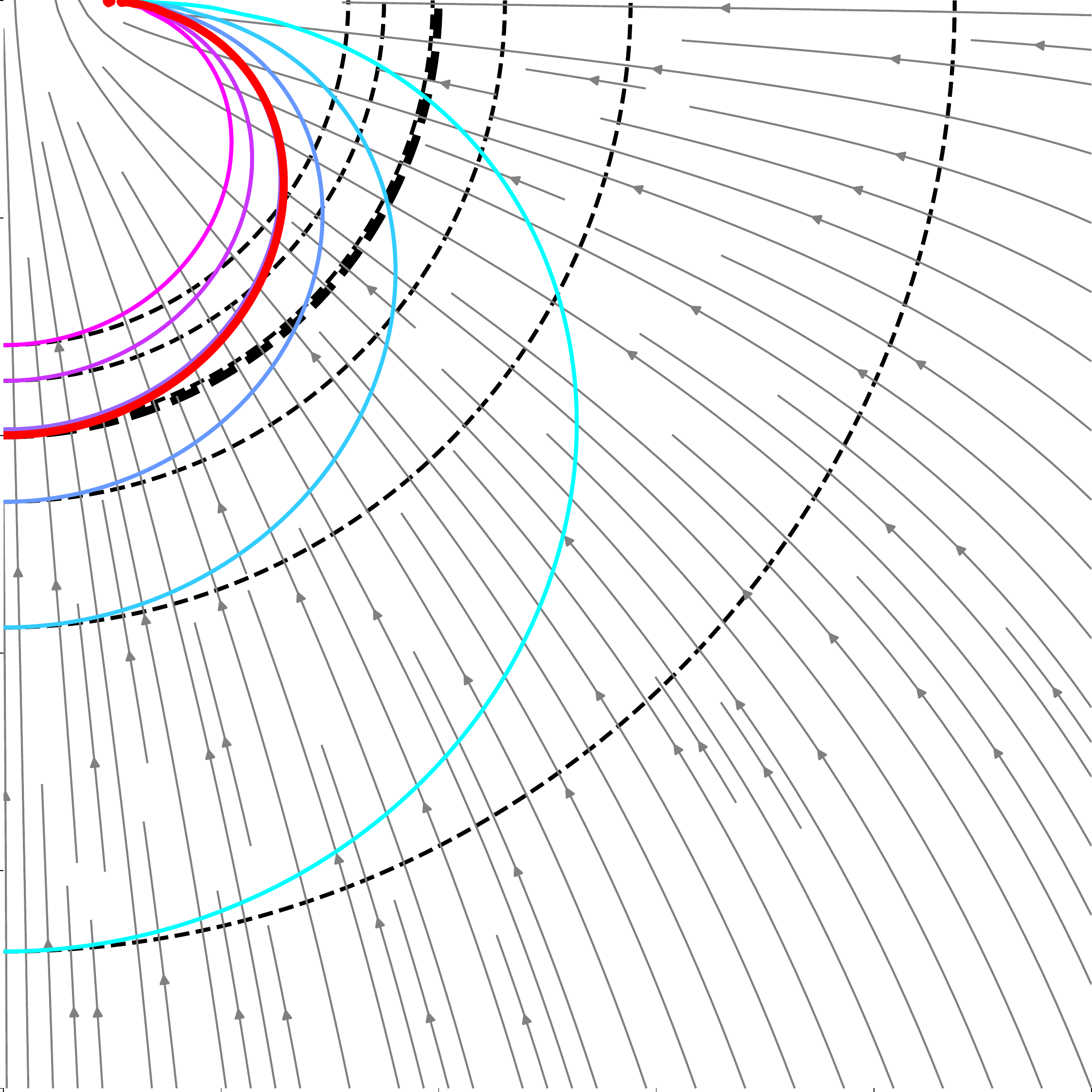};
\end{pgfonlayer}   
\begin{axis}[
xtick align=outside,
ylabel near ticks,
                        xlabel near ticks,
                        clip=false,
                        scale only axis,
                        width=0.29\textwidth,
                        axis equal image,
xlabel={$y$ [mm]},
                        ylabel={$x$ [mm]},
                        xmin=0, xmax=25,
                        ymin=-25, ymax=0,
                        tick style={color=black},
                        axis on top,
legend style={fill=white,draw=none,legend columns=1,legend cell align={center},font=\tiny},
colorbar style={
        title={$V/V_\text{max}$},
        yticklabel style={
        font=\tiny,
align=right,
            /pgf/number format/.cd,
                fixed,
            fixed zerofill,
            precision=1,
        ,tick style={color=black}}
        }
                    ]
 \addlegendimage{red, solid, thick}                 
 \addlegendimage{blue, solid, thick}                 
 \addlegendimage{DarkGreen, solid, thick}       
 \addlegendimage{black, dashed, thick}                 
\draw[draw=none, fill=black] (2.35,0) rectangle (25,1);               \node[above left,font=\small,fill=white]  at (current axis.south east) {{circle-S}};        
\node[above, right,font=\tiny,rotate=10] at (0,-23) {\contour{white}{$10$ cm/s}};
\node[above, right,font=\tiny,rotate=10] at (0,-16) {\contour{white}{$20$ cm/s}};
\node[above, right,font=\tiny,red,rotate=12] at (0,-11) {\contour{white}{$39$ cm/s}};
\node[above, right,font=\tiny,rotate=30] at (0,-7) {\contour{white}{$60$ cm/s}};

\end{axis}
\end{tikzpicture}     \begin{tikzpicture}   
\begin{pgfonlayer}{background}
\includegraphics[width=0.25\textwidth]{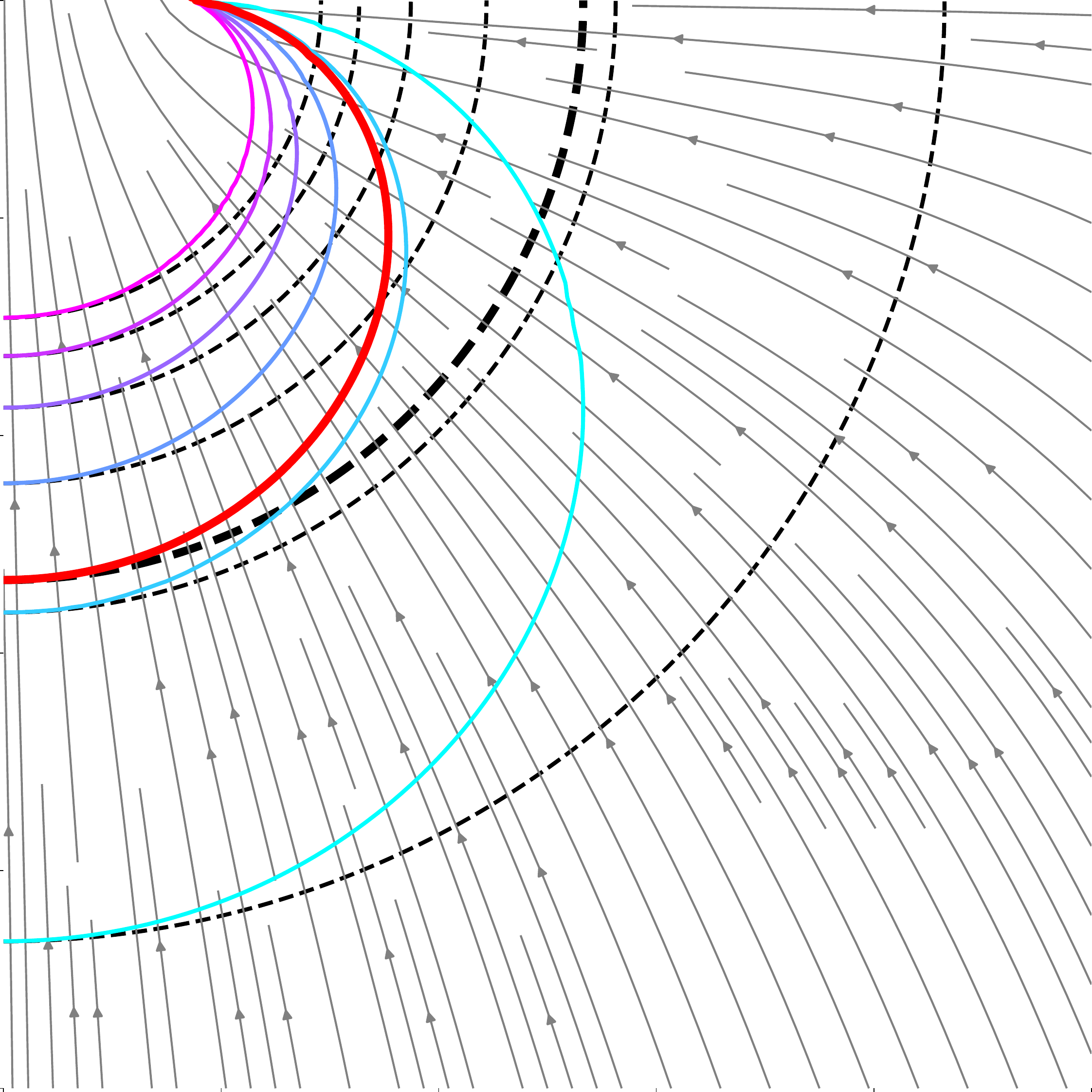};
\end{pgfonlayer}   
\begin{axis}[
xtick align=outside,
                        ylabel near ticks,
                        xlabel near ticks,
                        clip=false,
                        scale only axis,
                        width=0.29\textwidth,
                        axis equal image,
xlabel={$y$ [mm]},
                        yticklabel={\empty},
xmin=0, xmax=25,
                        ymin=-25, ymax=0,
                        tick style={color=black},
                        axis on top,
                        legend entries={PISA, optimal $V_a$},
                        legend style={fill=none,draw=none,legend columns=2,legend cell align={left},font=\tiny,
                        at={(axis cs:12.5,1)},
                        anchor=south},
colorbar style={
        title={$V/V_\text{max}$},
        yticklabel style={
        font=\tiny,
align=right,
            /pgf/number format/.cd,
                fixed,
            fixed zerofill,
            precision=1,
        ,tick style={color=black}}
        }
                    ]
 \addlegendimage{black, dashed, thick}                 
 \addlegendimage{red, solid, thick}                 
        
\draw[draw=none, fill=black] (4.35,0) rectangle (25,1);                 \node[above left,font=\small,fill=white]  at (current axis.south east) {{circle-M}};

\node[above, right,font=\tiny,rotate=12] at (0,-23) {\contour{white}{$10$ cm/s}};
\node[above, right,font=\tiny,rotate=12] at (0,-15) {\contour{white}{$20$ cm/s}};
\node[above, right,font=\tiny,red,rotate=15] at (0,-12.5) {\contour{white}{$22$ cm/s}};
\node[above, right,font=\tiny,rotate=28] at (0,-6) {\contour{white}{$60$ cm/s}};
    \end{axis}
\end{tikzpicture}     \begin{tikzpicture}   
\begin{pgfonlayer}{background}
\includegraphics[width=0.25\textwidth]{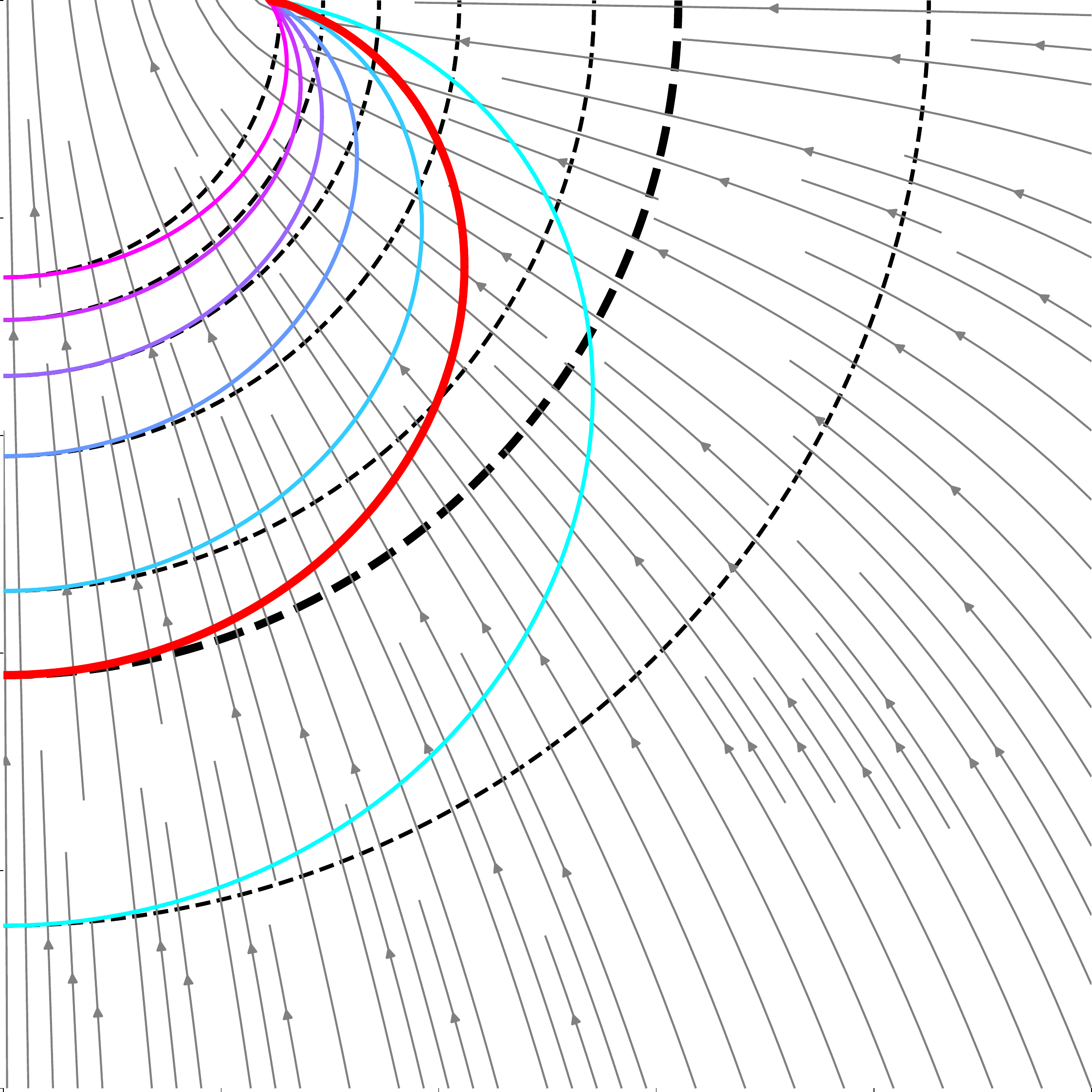};
\end{pgfonlayer}   
\begin{axis}[
xtick align=outside,
                        ylabel near ticks,
                        xlabel near ticks,
                        clip=false,
                        scale only axis,
                        width=0.29\textwidth,
                        axis equal image,
                        colorbar,
                        colormap={cool}{rgb255(0cm)=(0,255,255); rgb255(1cm)=(128,128,255);
                        rgb255(2cm)=(255,0,255);},
point meta max=60, 
                        point meta min=10,                
                        xlabel={$y$ [mm]},
                        yticklabel={\empty},
xmin=0, xmax=25,
                        ymin=-25, ymax=0,
                        tick style={color=black},
                        axis on top,
legend style={fill=white,draw=none,legend columns=1,legend cell align={center},font=\tiny},
colorbar style={
        title={$V_{a}$ [cm/s]},
        ytick={10,20,30,40,50,60},
        yticklabel style={
        font=\tiny,
align=right,
            /pgf/number format/.cd,
                fixed,
            fixed zerofill,
            precision=0,
        ,tick style={color=black}}
        }
                    ]
 \addlegendimage{red, solid, thick}                 
 \addlegendimage{blue, solid, thick}                 
 \addlegendimage{DarkGreen, solid, thick}       
 \addlegendimage{black, dashed, thick}                 
\draw[draw=none, fill=black] (6.1,0) rectangle (25,1);     \node[above left,font=\small,fill=white]  at (current axis.south east) {{circle-L}};

\node[above, right,font=\tiny,rotate=12] at (0,-23) {\contour{white}{$10$ cm/s}};
\node[above, right,font=\tiny,rotate=12] at (0,-12.5) {\contour{white}{$20$ cm/s}};
\node[above, right,font=\tiny,red,rotate=12] at (0,-14.75) {\contour{white}{$16$ cm/s}};
\node[above, right,font=\tiny,rotate=25] at (0,-5) {\contour{white}{$60$ cm/s}};
    \end{axis}
\end{tikzpicture}     \caption{Comparison of $V_x$-based isovelocity contours with PISA estimation for circular orifice at various aliasing velocities.
    Streamlines are plotted in the background.\label{fig:vacomp}}
\end{figure*}
\begin{figure*}[h]
\begin{tikzpicture}[baseline]
\begin{pgfonlayer}{background}
\includegraphics[width=0.25\textwidth]{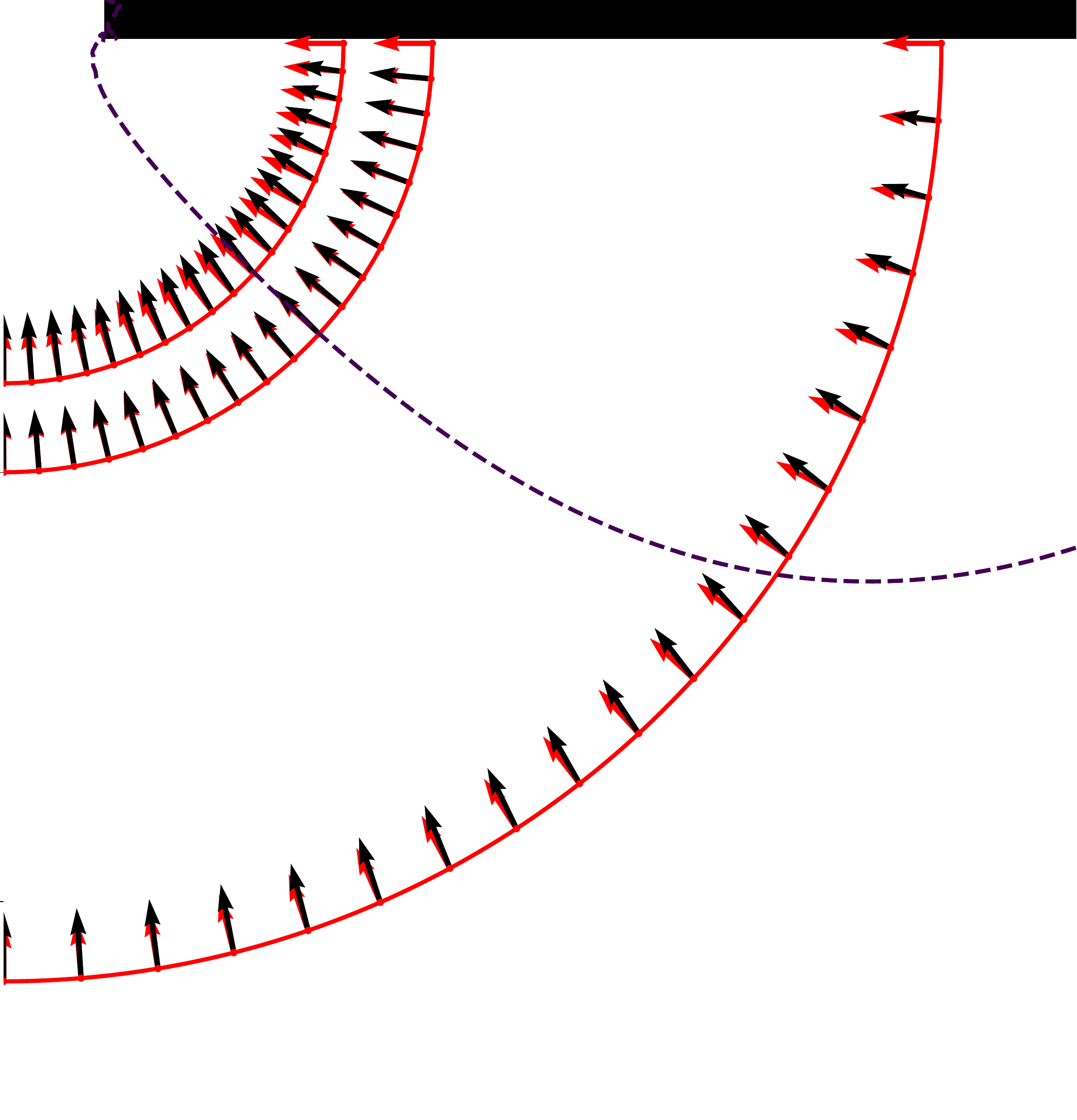};
\end{pgfonlayer}   
\begin{axis}[
xtick align=outside,
ylabel near ticks,
                        xlabel near ticks,
                        clip=false,
                        scale only axis,
                        width=0.3\textwidth,
                        axis equal image,
xlabel={$y$ [mm]},
                        ylabel={$x$ [mm]},
                        xmin=0, xmax=25,
                        ymin=-25, ymax=1,
                        tick style={color=black},
                        axis on top,
legend entries={{$V_x=V_y$ CFD}},
                        legend style={fill=none,draw=none,legend columns=2,legend cell align={left},font=\small,
                        at={(axis cs:12.5,1)},
                        anchor=south},
colorbar style={
        title={$V/V_\text{max}$},
        yticklabel style={
        font=\tiny,
align=right,
            /pgf/number format/.cd,
                fixed,
            fixed zerofill,
            precision=1,
        ,tick style={color=black}}
        }
                    ]
 \addlegendimage{black, dashed, thick}                 
\draw[blue,dashed] (0,0) -- (25,-25);
\node[above left,font=\small,fill=white]  at (current axis.south east) {{circle-S}};        
\node[above, right,font=\tiny,rotate=10] at (0,-23) {\contour{white}{$10$ cm/s}};
\node[above, right,font=\tiny,red,rotate=12] at (0,-12) {\contour{white}{$39$ cm/s}};
\node[above, right,font=\tiny,rotate=25] at (0,-5) {\contour{white}{$60$ cm/s}};

\node[left,font=\tiny] at (25,-11) {\contour{white}{$V_y$>$V_x$}};

\node[left,font=\tiny] at (25,-14) {\contour{white}{$V_y$<$V_x$}};

\end{axis}
\end{tikzpicture}     \begin{tikzpicture}[baseline]
\begin{pgfonlayer}{background}
\includegraphics[width=0.25\textwidth]{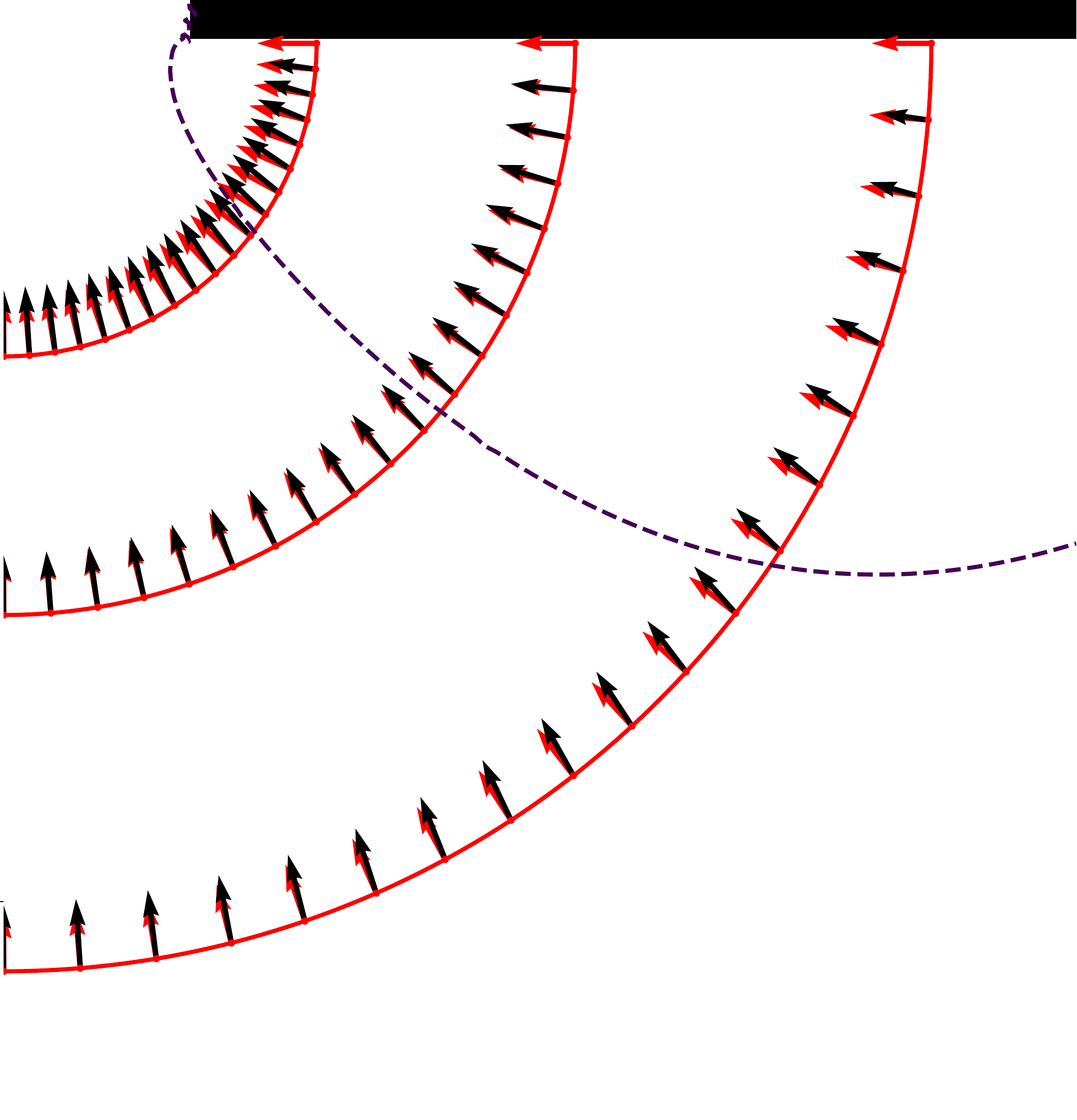};
\end{pgfonlayer}   
\begin{axis}[
xtick align=outside,
                        ylabel near ticks,
                        xlabel near ticks,
                        clip=false,
                        scale only axis,
                        width=0.3\textwidth,
                        axis equal image,
xlabel={$y$ [mm]},
                        yticklabel={\empty},
xmin=0, xmax=25,
                        ymin=-25, ymax=1,
                        tick style={color=black},
                        axis on top,
                        legend entries={CFD, ideal},
                        legend style={fill=none,draw=none,legend columns=2,legend cell align={left},font=\small,
                        at={(axis cs:12.5,1)},
                        anchor=south},
colorbar style={
        title={$V/V_\text{max}$},
        yticklabel style={
        font=\tiny,
align=right,
            /pgf/number format/.cd,
                fixed,
            fixed zerofill,
            precision=1,
        ,tick style={color=black}}
        }
                    ]
 \addlegendimage{black, solid, thick}                 
 \addlegendimage{red, solid, thick}                 
       \draw[blue,dashed] (0,0) -- (25,-25); 
\node[above left,font=\small,fill=white]  at (current axis.south east) {{circle-M}};

\node[above, right,font=\tiny,rotate=12] at (0,-23) {\contour{white}{$10$ cm/s}};
\node[above, right,font=\tiny,red,rotate=15] at (0,-15) {\contour{white}{$22$ cm/s}};
\node[above, right,font=\tiny,rotate=15] at (0,-9) {\contour{white}{$60$ cm/s}};
\node[left,font=\tiny] at (25,-11) {\contour{white}{$V_y$>$V_x$}};

\node[left,font=\tiny] at (25,-14) {\contour{white}{$V_y$<$V_x$}};

\end{axis}
\end{tikzpicture}     \begin{tikzpicture}[baseline]
\begin{pgfonlayer}{background}
\includegraphics[width=0.25\textwidth]{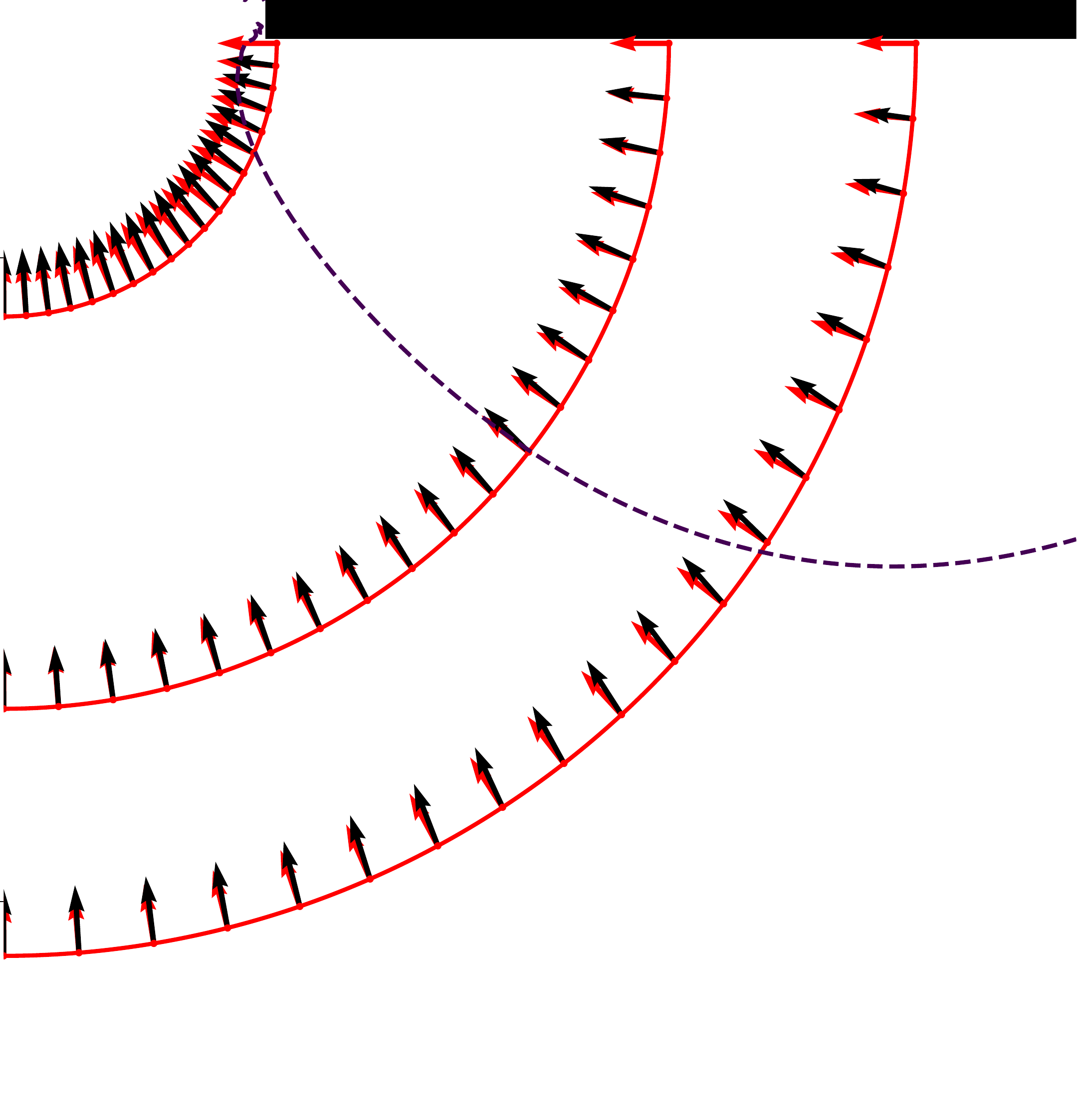};
\end{pgfonlayer}   
\begin{axis}[
xtick align=outside,
                        ylabel near ticks,
                        xlabel near ticks,
                        clip=false,
                        scale only axis,
                        width=0.3\textwidth,
                        axis equal image,
colormap={cool}{rgb255(0cm)=(0,255,255); rgb255(1cm)=(128,128,255);
                        rgb255(2cm)=(255,0,255);},
point meta max=60, 
                        point meta min=10,                
                        xlabel={$y$ [mm]},
                        yticklabel={\empty},
xmin=0, xmax=25,
                        ymin=-25, ymax=1,
                        tick style={color=black},
                        axis on top,
legend entries={{$V_x=V_y$ ideal}},
                        legend style={fill=none,draw=none,legend columns=2,legend cell align={left},font=\small,
                        at={(axis cs:12.5,1)},
                        anchor=south},
colorbar style={
        title={$V_{a}$ [cm/s]},
        ytick={10,20,30,40,50,60},
        yticklabel style={
        font=\tiny,
align=right,
            /pgf/number format/.cd,
                fixed,
            fixed zerofill,
            precision=0,
        ,tick style={color=black}}
        }
                    ]
 \addlegendimage{blue, dashed, thick}                 
 \addlegendimage{, solid, thick}                 
 \addlegendimage{DarkGreen, solid, thick}       
 \addlegendimage{black, dashed, thick}             
\draw[blue,dashed] (0,0) -- (25,-25); 
\node[above left,font=\small,fill=white]  at (current axis.south east) {{circle-L}};

\node[above, right,font=\tiny,rotate=12] at (0,-23) {\contour{white}{$10$ cm/s}};
\node[above, right,font=\tiny,red,rotate=12] at (0,-17) {\contour{white}{$16$ cm/s}};
\node[above, right,font=\tiny,rotate=15] at (0,-8) {\contour{white}{$60$ cm/s}};
\node[left,font=\tiny] at (25,-11) {\contour{white}{$V_y$>$V_x$}};

\node[left,font=\tiny] at (25,-14) {\contour{white}{$V_y$<$V_x$}};

    \end{axis}
\end{tikzpicture}     \caption{Comparison of the velocity vectors of ideal PISA estimation with the velocity vectors from CFD simulation for circular orifice at various aliasing velocities. The aliasing velocity marked with red label represents the optimal one. Dashed line marks the separation between the $V_x$- and $V_y$-dominant regions.\label{fig:vacompvec}}
\end{figure*}
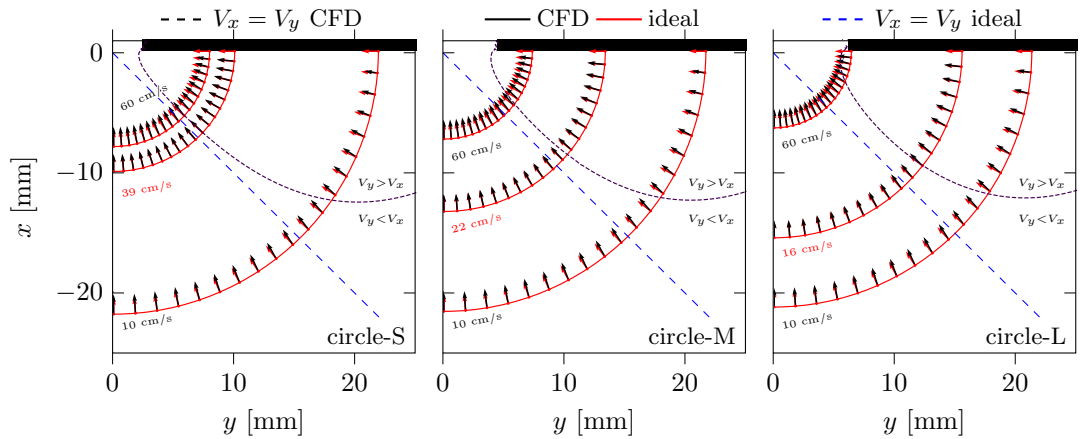

To shed light on the effect of the aliasing-velocity-($V_\mathrm{a}$)-choice on the estimation of PISA radius we carry out CFD-simulations with the circular orifice in three sizes corresponding to the experimentally investigated ones (cf. Figure \ref{fig:rvol-bars}) and extract the PISA at different $V_\mathrm{a}$ ranging from 10 to 60~cm/s.
The results for the evaluated PISA for each size are shown in Figure~\ref{fig:vacomp}.
Since the \rflow{} in the CFD-simulation is prescribed and known (\rflow$_\text{CFD}=244.9$~ml/s), it is possible to compute the optimal value for $V_\mathrm{a}$, which would give an ideal prediction for PISA-based \rflow{}.
The optimal $V_\mathrm{a}$ are estimated to be 39, 22 and 16~cm/s for small, medium and large circular orifices, respectively; the values and their corresponding isovelocity contours are marked in Figure~\ref{fig:vacomp} with red color.
For larger orifices with high $V_\mathrm{a}$ we observe the dominance of the streamwise velocity close to the orifice entrance -- this leads to the fact that the velocity vectors along PISA at higher $V_\mathrm{a}$ are not perpendicular to the PISA surface but rather aligned with the jet direction.
This violates the main PISA-assumption.
A similar effect can be seen at lower aliasing velocities for smaller orifice, where again the streamwise velocity component is shown to be dominant.
This effect is emphasized in Figure~\ref{fig:vacompvec}, where the deviation of real CFD-based velocity vectors (black) along the hemispherical PISA are compared with the idealized vectors sitting perpendicularly to the PISA surface (red).
The best agreement between the PISA assumption and the velocity field can be observed at the optimal $V_\mathrm{a}$ -- for this aliasing velocity the CFD vectors almost entirely coincide with the ideal ones.
In the figure the dashed line splits the observed area into the streamwise velocity dominant part ($V_y<V_x$) and the radial velocity dominant region ($V_y>V_y$).
For the ideal PISA assumption the line should be marking the bisector at $45^\circ$ as shown with the blue dashed line in the figures.
The PISA obtained for the optimal aliasing velocity crosses the demarkation line $V_x=V_y$ from CFD at the closest position to the $V_x=V_y$ for the ideal case.
This reaffirms that the calculated optimal aliasing velocity aligns most closely with the assumptions of the PISA method.

\begin{table*}[ht!]
\centering
\caption{Ultrasound Measurements and corresponding PIV measurement}
\tiny
\begin{tabular}{ccccccccccccc}
\toprule
Physician & US-system & TEE-Probe & MROP & Freq. & $V_\mathrm{a}$ & Radius & \rflow{} & $V_\mathrm{max}$ & \eroa{} & VTI & \rvol{} & \rvol{} (PIV) \\
& & & & [Hz] & [cm/s] & [cm] & [ml/s] & [cm/s] & [cm$^2$] &[cm] &[ml] & [ml]\\
\midrule
 &  &  & Circle-S & 23 & 38.5 & 0.4 & 38.9 & 500 & 0.08 & 66.2 & 5 & 5.59 \\
1 & EPIQ Cvxi & X8-2T & Circle-M & 22 & 38.5 & 0.8 & 154.8 & 360 & 0.43 & 44.9 & 19 & 22.43 \\
 &  &  & Circle-L & 19 & 38.5 & 0.9 & 196.0 & 334 & 0.59 & 40.0 & 24 & 35.1 \\
\hline
 & &  & Slit-S & 19 & 38.5 & 0.5 & 60.4 & 399 & 0.15 & 50.7 & 8 & 9.1\\
1 & EPIQ Cvxi & X8-2T & Slit-M & 22 & 38.5 & 0.5 & 60.4 & 340 & 0.18 & 49.0 & 9 & 15.2\\
 &  &  & Slit-L & 22 & 38.5 & 0.9 & 196.0 & 362 & 0.54 & 53.5 & 29 & 44.9\\
\hline
 &  &  & Drop-S & 34 & 38.5 & 0.4 & 38.9 & 369 & 0.11 & 50.2 & 6 & 10.0\\
1 & EPIQ Cvxi & X8-2T & Drop-M & 24 & 38.5 & 0.6 & 87.0 & 366 & 0.24 & 47.5 & 11 & 18.1\\
 & &  & Drop-L & 21 & 63.9 & 0.7 & 196.8 & 365 & 0.54 & 53.7 & 29 & 35.6\\
\hline
 &  &  & Circle-S & 17 & 42.8 & 0.4 & 43.2 & 447 & 0.10 & 68.9 & 7 & 5.59 \\
2 & EPIQ 7C & X8-2T & Circle-M & 18 & 47.0 & 0.6 & 106.2 & 461 & 0.23 & 72.5 & 17 & 22.43\\
 &  &  & Circle-L & 17 & 47.5 & 0.7 & 146.3 & 689 & 0.21 & 106.0 & 22 & 35.1\\
\hline
 &  &  & Slit-S & 17 & 12.1 & 0.7 & 37.3 & 413 & 0.09 & 56.3 & 5 & 9.1\\
2 & EPIQ 7C & X8-2T & Slit-M & 22 & 46.2 & 0.5 & 72.5 & 343 & 0.21 & 47.5 & 10 & 15.2\\
 &  & & Slit-L & 15 & 41.5 & 0.8 & 166.8 & 451 & 0.37 & 67.2 & 25 & 44.9\\
\hline
 &  &  & Drop-S & 20 & 46.2 & 0.5 & 72.5 & 354 & 0.20 & 55.7 & 11 & 10.0\\
2 & EPIQ 7C & X8-2T & Drop-M & 18 & 43.4 & 0.6 & 98.1 & 372 & 0.26 & 54.1 & 14 & 18.1\\
 &  &  & Drop-L & 22 & 47.2 & 0.7 & 145.4 & 381 & 0.38 & 65.9 & 25 & 35.6\\
\hline
 &  &  & Circle-S & 16 & 31.2 & 0.5 & 49.0 & 384 & 0.13 & 50.1 & 6 & 5.59 \\
3 & iE33 & X7-2t & Circle-M & 20 & 31.2 & 0.7 & 96.0 & 394 & 0.24 & 58.1 & 14 & 22.43\\
 &  & & Circle-L & 17 & 31.2 & 0.8 & 125.4 & 336 & 0.37 & 49.2 & 18 & 35.1\\
\hline
 &  &  & Slit-S & 19 & 31.2 & 0.5 & 49.0 & 286 & 0.17 & 40.7 & 7 & 9.1\\
3 & iE33 & X7-2t & Slit-M & 20 & 31.2 & 0.7 & 96.0 & 294 & 0.33 & 36.1 & 12 & 15.2\\
 &  &  & Slit-L & 19 & 31.2 & 0.7 & 96.0 & 279 & 0.34 & 33.0 & 11 & 44.9\\
\hline
 &  &  & Drop-S & 24 & 31.2 & 0.6 & 70.5 & 270 & 0.26 & 38.4 & 10 & 10.0\\
3 & iE33 & X7-2t & Drop-M & 31 & 31.2 & 0.7 & 96.0 & 317 & 0.30 & 43.4 & 13 & 18.1\\
 &  &  & Drop-L & 21 & 31.2 & 0.8 & 125.4 & 300 & 0.42 & 43.2 & 18 & 35.6\\
\bottomrule
\end{tabular}

\label{tab:ultrasound}

\end{table*}
\end{document}